\documentclass[opre,nonblindrev]{informs3_hide}
\DoubleSpacedXI 

\usepackage[english]{babel}
\usepackage[autostyle, english = american]{csquotes}
\MakeOuterQuote{"}
\usepackage{tablefootnote}
\usepackage{setspace}

\usepackage{bbm,xspace,multirow,multicol}
\usepackage[hypertexnames=false]{hyperref}
\usepackage{cleveref}
\crefname{subsection}{subsection}{subsections}
\usepackage[normalem]{ulem}

\newcommand{\eps}{\varepsilon}

\newcommand{\bE}{\mathbb{E}}

\newcommand{\OPT}{\mathsf{OPT}}

\usepackage[dvipsnames]{xcolor}

\NewDocumentEnvironment{myproof}{o}
{\IfNoValueTF{#1}{\paragraph{{Proof.} }} {\paragraph{{#1.} }} }
{\hfill$\Halmos$}

\usepackage{natbib}
 \bibpunct[, ]{(}{)}{,}{a}{}{,}%
 \def\bibfont{\normalsize}%

\usepackage{algorithm}
\usepackage[noend]{algpseudocode}

\usepackage{epsfig}
\usepackage{bm}
\usepackage{comment}

\usepackage{enumitem}

\numberwithin{equation}{section}

\newcommand{\Ber}{\textup{Ber}}

\newcommand{\Bin}{\mathrm{Bin}}

\usepackage{bbm}

\usepackage{algpseudocode}
\usepackage{datetime}

\usepackage[english]{babel}
\usepackage[autostyle, english = american]{csquotes}

\newcommand{\scr}{\mathcal}
\newcommand{\mb}{\mathbb}
\newcommand{\til}{\widetilde}

\newcommand{\E}{\mathbb E}

\newcommand{\obj}{\text{obj}}

\newcommand{\tpartial}{\til{\partial}}

\newcommand{\bP}{\mathbb{P}}
\newcommand{\cP}{\mathcal{P}}
\newcommand{\matched}{\mathsf{matched}}
\newcommand{\blocked}{\mathsf{blocked}}
\newcommand{\alone}{\mathsf{alone}}

\newcommand{\simple}{\mathsf{sbl}}
\newcommand{\blocker}{\mathsf{sbl}}
\newcommand{\candidate}{\mathsf{candidate}}
\newcommand{\advmin}{\mathrm{AdvMin}}
\newcommand{\advminaux}{\mathrm{AdvMinAux}}

\newcommand{\ol}[1]{\overline{#1}}

\newcommand{\fc}{f^{c}}
\newcommand{\Rf}{\scr{R}}
\newcommand{\sv}{s}

\def\Var{\mbox{{\bf Var}}}

 \newcommand{\CM}{}

\TheoremsNumberedThrough     
\ECRepeatTheorems

\EquationsNumberedThrough    

\MANUSCRIPTNO{OPRE-2023-06-0339} 

\begin{document}


\RUNAUTHOR{MacRury et al.}

\RUNTITLE{On Online Contention Resolution Schemes for the Matching Polytope of Graphs}

\TITLE{On (Random-order) Online Contention Resolution Schemes for the Matching Polytope of (Bipartite) Graphs}

\ARTICLEAUTHORS{
\AUTHOR{Calum MacRury}
\AFF{Graduate School of Business, Columbia University, New York, NY 10027, \EMAIL{cm4379@columbia.edu}}

\AUTHOR{Will Ma}
\AFF{Graduate School of Business and Data Science Institute, Columbia University, New York, NY 10027, \EMAIL{wm2428@gsb.columbia.edu}}

\AUTHOR{Nathaniel Grammel}
\AFF{Department of Computer Science, University of Maryland, College Park.
\EMAIL{ngrammel@umd.edu}}

}

\ABSTRACT{
Online Contention Resolution Schemes (OCRS's) represent a modern tool for selecting a subset of elements, subject to resource constraints, when the elements are presented to the algorithm sequentially.
OCRS's have led to some of the best-known competitive ratio guarantees for online resource allocation problems, with the added benefit of treating different online decisions---accept/reject, probing, pricing---in a unified manner.
This paper analyzes OCRS's for resource constraints defined by matchings in graphs, a fundamental structure in combinatorial optimization.
We consider two dimensions of variants: the elements being presented in adversarial or random order; and the graph being bipartite or general.
We improve the state of the art for all combinations of variants, both in terms of algorithmic guarantees and impossibility results.
Some of our algorithmic guarantees are best-known even compared to Contention Resolution Schemes that can choose the order of arrival or are offline.
All in all, our results for OCRS directly improve the best-known competitive ratios for online accept/reject, probing, and pricing problems on graphs in a unified manner.
}



\maketitle



\section{Introduction}

Contention Resolution Schemes (CRS's) are tools for selecting a subset of elements, subject to feasibility constraints that cause "contention" between the different elements to be selected.
The goal is to resolve this contention using randomization, and select each element with the same ex-ante probability conditional on it being "fit" for selection, or \textit{active}.
Only active elements can be selected, and this criterion is what distinguishes CRS's from other randomized rounding procedures that have been used for algorithm design in combinatorial optimization since \citet{raghavan1987randomized}.
In the CRS model, each element is active \textit{independently}, according to a known probability.

CRS's as originally introduced in \citet{chekuri2014submodular} were \textit{offline} in that they could observe whether every element was active before making any selection decisions, with applications in constrained submodular optimization.
Since then, CRS's have also been a tool of choice to design \textit{online} algorithms \citep{feldman2021online}, unlocking tight performance guarantees in online resource allocation problems \citep{jiang2022tight}, and generally not worsening the guarantees that are possible \citep{Lee2018}.
These CRS's sequentially discover whether each element is active; if so, they must immediately decide whether that element should be selected.
(An element cannot be selected if it would violate the feasibility constraints.)
Different CRS's can be designed depending on the order in which elements are processed.

The CRS literature is demarcated by the class of resource constraints being considered, and whether the processing is offline or online (and in the latter case, the processing order).  This paper focuses on resource constraints defined by \textit{matchings} in graphs which is a fundamental feasibility constraint in combinatorial optimization, and derives state-of-the-art results for all processing orders.
The definition of matching constraints is that each element is represented by an edge in a graph, and if selected, consumes two arbitrary resources in the form of its two incident vertices.
We derive online CRS's that work under arbitrary (adversarial) processing orders and CRS's with improved performance guarantees if the edges are processed in a uniformly random order.
The latter guarantees are best-known even compared to offline CRS's on general graphs.

One of the principle applications of online CRS's is to the \textit{prophet inequality} problem. We discuss this in detail in \Cref{sec:prophet_application}, as well as other applications specific to matching constraints.

\subsection{Formal Definitions of Contention Resolution Schemes for Matchings} \label{sec:problem_definition}

Let $G=(V,E)$ be a graph.  An edge $e=(u,v)$ is said to be \textit{incident} to vertices $u$ and $v$, and $v$ is said to be a \textit{neighbor} of $u$ (and vice versa).
For any vertex $v\in V$, let $\partial(v)\subseteq E$ denote the set of edges incident to $v$,
and for any $e =(u,v)\in E$, let $\partial(e) := \partial(u) \cup \partial(v) \setminus \{e\}$.
A \textit{matching} $M$ is a subset of edges no two of which are incident to the same vertex, i.e.\ satisfying $|M\cap\partial(v)|\le 1$ for all $v\in V$.  A vector $\bm{x}\in[0,1]^E$ lies in the \textit{matching polytope} of $G$ (which we denote $\cP_G$), if $\sum_{e\in\partial(v)}x_e\le 1$ for all $v\in V$. In this case,
we refer to $\bm{x}$ as a \textit{fractional matching} for $G$.

Fixing a fractional matching $\bm{x}=(x_e)_{e\in E}$ of $G$, each edge $e$ has an \textit{activeness state} $X_e$ that realizes to 1 with probability (w.p.) $x_e$ and 0 w.p.~$1-x_e$, independent of everything else.  We denote this random draw as $X_e\sim\Ber(x_e)$, where $\Ber(x)$ represents an independent Bernoulli random variable of mean $x$ for any $x\in[0,1]$.
Edges $e$ with $X_e=1$ are active.

A \textit{contention resolution scheme} (CRS) is passed $G$ and $\bm{x}$ as input and \textit{selects} a subset
of active edges, under the additional constraint that the selected subset must form a matching.
In the original \textit{offline setting},
the CRS is also passed $(X_e)_{e \in E}$, and thus learns the edge states prior to making
its selections.
A CRS is then said to be \textit{$c$-selectable} if for any graph $G$ and any vector $\bm{x}\in\cP_G$, it selects each edge $e$ with probability at least $cx_e$, where $c$ is a constant in $[0,1]$. 

A \textit{sequential} CRS is also passed $G$ and $\bm{x}$ as input, but initially does
not know the edge states $(X_e)_{e \in E}$. Instead, an ordering on $E$ is chosen,
and the edge states are presented one-by-one to the sequential CRS using this ordering. Upon
learning $X_e$, it makes an irrevocable decision on whether or not to select $e$.
Two types of sequential schemes have been defined in the literature depending on how the ordering
is generated: \textit{online} contention resolution schemes (OCRS), where this order is chosen by an adversary\footnote{
Like \citet{Ezra_2020} we assume that this adversary is \textit{oblivious}, in that it fixes the arrival order based on the algorithm and cannot change the order based on realizations.
}; and \textit{random-order} contention resolution schemes (RCRS),
where this order is chosen uniformly at random.

\subsection{Contributions}

To describe our results, we define \textit{selectability} as the maximum value of $c$ for which an OCRS or RCRS is $c$-selectable, evaluated on the worst case graph $G$ and vector $\bm{x}\in\cP_G$ for the algorithm.
Without further specification, selectability considers the best possible algorithm and takes a worst case over general graphs, although we also refer to the selectability of a specific algorithm or the selectability taken over bipartite graphs.
By definition, the selectability of a specific algorithm is worse (smaller) than that of the best algorithm; the selectability for general graphs is worse than that for bipartite graphs; and the selectability of OCRS is worse than that of RCRS.

\begin{table}[t]
\centering
\begin{tabular}{c|c|c} \hline
Selectability Bounds & General Graphs & Bipartite Graphs \tablefootnote{
Our algorithmic result for OCRS holds as long as the graph does not contain 3-cycles.
Our algorithmic result for RCRS holds as long as the graph does not contain 3-cycles or 5-cycles.
}\\ 
\hline
OCRS of [1] & $\ge 0.337$ [1]  $\to\ \ge \mathbf{0.344}$ [\S\ref{sec:ocrsGeneral}] & $\ge{0.337}$ [1] $\to \ge\mathbf{0.349}$ [\S\ref{sec:ocrsBipartite}]\\
& $\le\textbf{0.361}$ [\S\ref{sec:ocrsNegative}] & $\le 0.382$ $[\S\ref{sec:ocrsNegative},\text{ folklore}]$ \\
\hline
Any OCRS & $\le\textbf{0.4}$ [\S\ref{sec:ocrsNegative}] & $ \le 3/7$ [2] \\
\hline
Any RCRS \tablefootnote{Our random-order CRS's provide the best-known selectability guarantees even compared to sequential CRS's that can choose the order.  Compared to the yet more powerful offline CRS's, our $0.474$-selectable RCRS for general graphs is still best-known, but a $0.509$-selectable offline CRS is known for bipartite graphs \citep{nuti2023towards}.} & $\ge 0.45$ [3] $\to\ \ge\textbf{0.474}$ [\S\ref{sec:rcrsGeneral}] & $\ge0.456$ [3] $\to\ \ge \CM{\textbf{0.478}}$ [\S\ref{sec:rcrsBipartite}] \\
& & $\le 0.544$ [4] $\to\ \le\textbf{0.5}$ [\S\ref{sec:rcrsNegative}] \\
\hline
\end{tabular}
\caption{New results are \textbf{bolded}.  "$\ge$" refers to lower bounds on $c$ (algorithmic results), "$\le$" refers to upper bounds (impossibility results), and arrows indicate improvement from state of the art. [1], [2], [3], [4], [5] refer to \citet{Ezra_2020}, \citet{correa2022optimal}, \citet{pollner2022improved}, \citet{karp1981maximum} respectively.}
\label{tab:my_label}
\end{table}

Given this understanding, our results are summarized in \Cref{tab:my_label}.  We
improve algorithmic results for CRS's that select matchings in graphs, on all fronts.
We also derive many new impossibility results, and believe another contribution of this paper to lie in elucidating the limitations of different algorithms or analyses.
We now describe each new result individually, its significance, and sketch the techniques used to derive it. 

\subsection{Technical Comparison with Existing Work} \label{sec:technical}

Recall that the algorithm must select each edge $e$ with probability at least $cx_e$.  The algorithm is not rewarded for selecting $e$ with probability greater than $cx_e$, so a common idea behind both OCRS and RCRS is to \textit{attenuate} this probability, by only considering an edge $e$ for selection when its activeness state and another independent random bit $A_e$ both realize to 1.
In this case, we say that $e$ "survives", which occurs with a probability that can be calibrated to any value less than $x_e$.
The algorithms we study are all \textit{myopic} with respect to some appropriately-defined attenuation, i.e.\ they select any surviving edge that is feasible to select at its time of arrival. 

\textbf{Existing $c$-selectable OCRS.}
For OCRS the state of the art is a myopic OCRS that calibrates the survival probabilities so that every edge $e$ is selected with probability \textit{exactly} $cx_e$
\citep{Ezra_2020}.
For this OCRS to be valid, when any edge $e=(u,v)$ arrives, it must be feasible to select (i.e.\ neither vertices $u,v$ have already been matched) with probability at least $c$, so that there is the possibility of selecting $e$ with probability at least $cx_e$.
\citet{Ezra_2020} show that $c=1/3\approx0.333$ easily yields a valid algorithm.
Then by arguing that the bad events of $u$ being matched and $v$ being matched cannot be perfectly negatively correlated, or equivalently by providing a non-trivial lower bound on the probability of both $u$ and $v$ being matched (not to each other), \citet{Ezra_2020} show that the improved value of $c=0.337$ is also valid.

\textbf{Our improvements to OCRS.}
We consider the same OCRS as \citet{Ezra_2020}.
First we show that $c=0.349$ is valid for bipartite graphs, using a different analysis based on the FKG inequality.
Note that when edge $e=(u,v)$ arrives, $u$ is guaranteed to be matched if it has a neighbor $u'$ such that: (i) edge $(u,u')$ already arrived and survived; and (ii) no edge incident to $u'$ that arrived before $(u,u')$ survived.
A neighbor $v'$ of $v$ satisfying (i)--(ii) can be defined analogously.
We show that $u$ having such a neighbor $u'$ is positively correlated with $v$ having such a neighbor $v'$, by the FKG inequality.
Moreover, whether two neighbors $u_1,u_2$ of $u$ satisfy condition (ii) are independent (because there cannot be an edge between $u_1$ and $u_2$).
Ultimately this reveals that the worst case for the existence of both $u'$ and $v'$ occurs when $u,v$ are surrounded by edges with infinitesimally-small $x$-values, implying that $c=0.349$ yields a valid algorithm.
We note that it may be tempting to improve this guarantee by applying FKG directly on the events of $u$ and $v$ being matched; however we construct an example in \Cref{sec:bipartiteNegCorr} showing that these events, surprisingly, may not be positively correlated even on bipartite graphs.

The preceding argument breaks down for general graphs, both because $u'$ could be the same vertex as $v'$, and because satisfying condition~(ii) is no longer independent.
To rectify this argument, we take an approach motivated by \citet{Ezra_2020}---$u$ and $v$ will each randomly choose up to one neighbor satisfying~(i), and hope that they end up choosing distinct vertices that also satisfy~(ii), which would again certify both $u$ and $v$ to be matched.
Our choice procedure differs
from \citet{Ezra_2020} and is designed so that the probabilities of two good events ($u,v$ choosing any neighbors at all, and~(ii) being satisfied)
cannot be simultaneously minimized
in a worst-case configuration.
Interestingly, this leads to a "hybrid" worst case for general graphs, in which both endpoints $u,v$ of the arriving edge $e$ neighbor a "large" vertex $w$ with $x_{uw}=x_{vw}=1/2$, but otherwise $u,v$ are surrounded by edges with infinitesimally-small $x$-values.
To prove that this hybrid is the worst case, we bound an infinite-dimensional optimization problem using a finite one with vanishing loss, and solve the finite one numerically.
This worst case implies that $c=0.344$ is valid.
\CM{We also believe that our new procedure for $u$ and $v$ to randomly choose neighbors is simpler and more flexible than the original "witness" argument from \citet{Ezra_2020}. In \Cref{sec:recoverEzra}, we demonstrate this by showing how their bound of $c=0.337$ can be recovered through our procedure.}

\textbf{Impossibility results for OCRS.}
To complement our algorithmic results, we construct a simple example on which no OCRS can be more than 0.4-selectable, and the OCRS of \citet{Ezra_2020} in particular is no more than 0.361-selectable.
This example is related to the worst case from our analysis of general graphs above, in that it has an edge $e$ connected to two "large" vertices $w$.
\CM{Performance on this example also demonstrates the shortcoming of the OCRS of \citet{Ezra_2020}---it does not discriminate between different states in which an arriving edge could be feasibly selected.
This shortcoming is echoed in the example showing it to be no more than 0.382-selectable on bipartite graphs.}


\textbf{Existing $c$-selectable RCRS.}
For RCRS the state of the art also uses the attenuation framework, with the attenuation bit $A_e$ in this case being set a priori
to some value $a(e)\in[0,1]$, where $a$ is a function of the edge $e$.
The challenge again lies in lower-bounding the probability of an arriving edge $e$ being feasible for selection, in this case by $c/a(x_e)$.
\citet{brubach2021offline} lower-bound this probability using a condition similar to~(ii) above---when $e=(u,v)$ arrives, if there are no edges incident to $u$ or $v$ that arrived before $e$ and survived (i.e.\ are active with $A_e=1$), then $e$ must be feasible to select.
We refer to these bad edges incident to $u$ or $v$ as \textit{relevant}.
\citet{brubach2021offline} show for many attenuation functions $a$, in all of which $a(e)$ depends only on $x_e$,
that the probability of $e$ having no relevant edges is at least $c/a(x_e)$, with $c=(1-e^{-2})/2\approx0.432$.
\CM{\citet{pollner2022improved} later identify a barrier of $(1-e^{-2})/2$ for the analysis method of \citet{brubach2021offline}, and overcome it by deriving a lower bound on the probability of $e$ having exactly one relevant edge, say $f=(u,w)$, but $f$ being \textit{blocked}, in that $w$ was already matched when $f$ arrived.
Of course, this lower bound must be 0 if $w$ is only incident to $f$, so \citet{pollner2022improved} also use a more elaborate $a$ function that heavily attenuates $f$ in this case where $\partial(w)=\{f\}$.  Combining these ingredients, \citet{pollner2022improved} derive a $0.45$-selectable RCRS, that is $0.456$-selectable for bipartite graphs.}

\textbf{Our improvements to RCRS.}
We provide an improved 0.474-selectable RCRS for general graphs.
Our algorithm executes on the \textit{1-regularized} version of the graph $G$, which means that "phantom" edges and vertices are added to make $\sum_{e\in\partial(v)}x_e$ equal to $1$ for all $v$.
These phantom edges serve only the purpose of blocking relevant edges, and allow us to return to simpler attenuation functions based only on $x_e$ (which would have been stuck at $(1-e^{-2})/2$ without 1-regularity).

Restricting to these simple functions $a$ that map $x_e$ to a probability, our technique is to identify analytical properties of $a:[0,1]\to[0,1]$ that lead to characterizable worst-case configurations for the arriving $e=(u,v)$ having relevant edges and for these edges being blocked.
First, conditioning on the only relevant edge being say $f=(u,w)$, we formulate analytical constraints on function $a$ under which the worst case (minimum probability) for $f$ being blocked arises when $w$ is incident to a \textit{single} edge other than $(u,w)$ and $(v,w)$.
Given this worst case for $f$ being blocked, we can formulate further constraints on $a$ under which the worst case for $e$ having zero relevant edges or one blocked relevant edge arises when $u,v$ are surrounded by edges $f$ with infinitesimally-small $x_f$.
We show that there exist functions $a:[0,1]\to[0,1]$ satisfying both sets of constraints, and taking the best one yields a 0.474-selectable RCRS for general graphs.
\CM{For bipartite graphs our constraints on $a$ get looser (since the optimization for the worst case is more restricted), allowing us to push the envelope of feasible functions. Moreover, due to the lack of triangles and $5$-cycles, we are able to analyze when \textit{each} endpoint of $u$ and $v$ simultaneously has its own relevant edge. Taken together, these properties allow us to choose a different attenuation function which leads to a $0.478$-selectable RCRS. Notably, this surpasses the tight $0.476$ selectablity result of \cite{BruggmannZ22} for \textit{monotone} offline contention resolution on bipartite graphs, so our result shows that monotonicity is more constraining than having to process the edges online in a random order.

We note that in essence, our 1-regularity reduction achieves the same goals as the elaborate attenuation function from \citet{pollner2022improved}. They define an attenuation function which penalizes edges whose
endpoints have small fractional degree via an additional parameter $s_e$. The additional term forces the worst-case input for their RCRS to be 1-regular, in which case the term "$s_e$" equals $x_e$ (and so disappears). Afterwards,
their computations and our computations proceed similarly, and they also lower bound the probability that a single ``relevant'' edge adjacent to $e$ is ``blocked''. Thus, their approach can be thought of as \textit{implicitly} reducing to $1$-regular inputs, whereas we do this explicitly. This allows us to better "engineer" worst-case configurations through the design of $a:[0,1]\to[0,1]$, and is best exemplified in the case of bipartite graphs where we are able to analyze multiple relevant edges.} We also find it interesting that our technique leads to the best-known RCRS despite using attenuation functions that do not take arrival time into account (as is done in \citet{Lee2018,pollner2022improved}).
In fact, our 0.474-selectable RCRS based on these simple $a$ functions improves the state of the art even for \textit{offline} contention resolution schemes and \textit{correlation gaps} on general graphs (see the discussion in \citet{pollner2022improved}).

\textbf{Impossibility result for RCRS.}
We show that no RCRS can be more than 1/2-selectable, on a complete bipartite graph with $n$ vertices on each side and all edge values equal to $1/n$ as $n\to\infty$.
This is achieved by analyzing the more fundamental problem of online (unweighted) matching on random graphs: when the edges of this graph arrive in a uniformly random order, and active edges must be irrevocably accepted or rejected, what fraction of vertices can an optimal online algorithm match?
The main challenge here is that an arriving edge $(u,v)$ which is both active and feasible may not be optimal to accept, if many edges between $u$ or $v$ and another unmatched vertex are yet to arrive.  Nonetheless, we upper-bound the value that an online algorithm can gain through judiciously rejecting edges this way.

More precisely, we show that the greedy algorithm, which accepts any feasible edge, is suboptimal up to $o(n)$ terms as $n\to\infty$.
We do this by tracking the size of the matching constructed by
an arbitrary online algorithm after $t \ge 0$ edges arrive. Denoting this random variable by $M(t)$, we prove that $\mb{E}[ M(t+1) - M(t) \mid \scr{H}_t] \le \frac{1}{n} \left(1 - \frac{M(t)}{n^2} \right)^2$ for an appropriate choice of ``typical histories'' $\scr{H}_t$. \CM{On these histories, the online algorithm knowing which edges have already arrived is not particularly informative. By applying the ``one-sided'' differential equation method of \cite{bennett2023}, we conclude that the expected matching size of an arbitrary online algorithm is
at most $n/2 + o(n)$. We note that our approach can be thought of as implicitly reducing to a problem where each edge is drawn independently \textit{with replacement} uniformly from the $n^2$ possibilities.  In the problem with replacement, a greedy algorithm can be easily seen to be optimal for all $n$, and one can apply the (standard) differential equation method of \cite{de} to prove that it constructs
a matching of expected size $n/2 + o(n)$.


All in all, this represents a fundamental barrier for online matching on large random graphs when edges arrive in a uniformly random order.
In this setting, \citet{karp1981maximum} have shown that an \textit{offline} algorithm can match 54.4\% of the vertices as $n\to\infty$, and to our knowledge no smaller upper bounds have been previously shown for online algorithms.
We also remark that both our result and that of \citet{karp1981maximum} continue to hold if we consider large complete graphs instead of large complete bipartite graphs.
Finally, we mention that \citet{nuti2023towards} recently designed a $0.509$-selectable \textit{offline} contention resolution scheme for bipartite graphs. Combined with this, our $1/2$ impossibility result establishes a separation that offline contention resolution is strictly easier than random-order contention resolution on bipartite graphs.}

\

\section{Details of Online Contention Resolution Schemes} \label{sec:OCRS}
\begin{definition}[Terminology and Notation for OCRS]
Let $G=(V,E)$ be a graph with fractional matching $(x_e)_{e \in E}$ passed as input to an OCRS.
At the time an edge $e\in E$ arrives, we say that a vertex $v\in V$ is \textit{matched} if an edge incident to $v$ that has already arrived has been selected.
We denote this event using $\matched_v(e)$, noting that it depends on the random active states of edges arriving before $e$ and any randomness in the algorithm.
We say that an edge $e=(u,v)$ is \textit{blocked} if either $u$ or $v$ has been matched by the time $e$ arrives, and denote this event using $\blocked(e)$.  Blocked edges, even if active, cannot be selected.
\end{definition}
Our improved lower bound for OCRS is based on a new analysis of the algorithm of \citet{Ezra_2020}, which we restate in \Cref{alg:attenuate_aom} using our terminology.

\begin{algorithm}{H}[t]
\caption{OCRS of \citet{Ezra_2020}} 
\label{alg:attenuate_aom}
\begin{algorithmic}[1]
\SingleSpacedXI
\Require $G=(V,E)$, $\bm{x}=(x_e)_{e \in E}$, and $c\in[0,1]$ a constant to be determined later
\Ensure subset of active edges forming a matching $\scr{M}$
\State $\scr{M} \leftarrow \emptyset$
\For{arriving edges $e$}
\State Let $\alpha_e:=c/\bP[\overline{\blocked(e)}]$, where the denominator is the probability that edge $e$ is not blocked, taken over the randomness in the activeness of past edges and the algorithm
\State Draw $A_e \sim \Ber(\alpha_e)$
\If{$e$ is active, not blocked, and $A_e=1$}
\State $\scr{M} \leftarrow \scr{M} \cup \{e\}$
\EndIf
\EndFor
\State \Return $\scr{M}$
\end{algorithmic}
\end{algorithm}

\begin{remark}
In \Cref{alg:attenuate_aom}, $\alpha_e$ is a probability over the hypothetical scenarios that could have occurred, based on what the OCRS knows about the edges that have arrived so far.  Computing these probabilities exactly requires tracking exponentially many scenarios, but fortunately sampling these scenarios yields an $\eps$ loss in selectability given $O(1/\eps)$ runtime \citep{Ezra_2020}.

We also remark that the values of $\alpha_e$ used by \Cref{alg:attenuate_aom} are fixed once the graph and order of edge arrival are determined.
This is where the assumption that the adversary is \textit{oblivious} comes in---the order, and hence the values of $\alpha_e$, must be independent of any realizations.
\end{remark}

We now define some further concepts specific to \Cref{alg:attenuate_aom}. Recall that
$X_e$ is an indicator random variable for the event edge $e$ is active, and $A_e$ is defined
in \Cref{alg:attenuate_aom}.
We say that $e$ \textit{survives} if both $X_e$ and $A_e$ realize to 1, and we let $S_e=X_eA_e$ indicate this event, which is an independent Bernoulli random variable with mean $x_e \alpha_e$.
The OCRS of \citet{Ezra_2020} can then be concisely described as "select every surviving unblocked edge".
The survival probabilities are calibrated so that
\begin{align} \label{eqn:exactly_c}
\bP[e\in\scr{M}]=x_e\alpha_e\bP[\overline{\blocked(e)}]=cx_e && \forall e\in E
\end{align}
(by definition of $\alpha_e$), resulting in a $c$-selectable OCRS.

However, \Cref{alg:attenuate_aom} only defines a valid OCRS if $\alpha_e$ is a probability in [0,1] for all $e\in E$.
Put another way, constant $c$ must be small enough such that
\begin{align} \label{eqn:desired}
    \bP[\overline{\blocked(e)}] &\ge c
\end{align}
for every graph $G$, fractional matching $\bm{x}$, and arriving edge $e$ (which would ensure that $\alpha_e\le1$).
Following \citet{Ezra_2020}, validity can be inductively established by assuming~\eqref{eqn:desired} holds for all $e$ under a given $G$, $\bm{x}$, and arrival order, and then proving that it also holds for an arbitrary edge $e\notin E$ which could arrive next. \citet{Ezra_2020} further observe that if this newly arriving edge is $e=(u,v)$, then
\begin{align}
\bP[\overline{\blocked(e)}]
&=1-\bP[\matched_u(e)\cup\matched_v(e)] \nonumber
\\ &=1-\bP[\matched_u(e)]-\bP[\matched_v(e)]+\bP[\matched_u(e)\cap\matched_v(e)] \label{eqn:match_uv}
\\ &=\textstyle 1-c\sum_{f\in\partial(u)\setminus e}x_f-c\sum_{f\in\partial(v) \setminus e}x_f+\bP[\matched_u(e)\cap\matched_v(e)] \nonumber
\end{align}
(where the final equality holds by~\eqref{eqn:exactly_c} and the induction hypothesis).  Therefore, the real challenge and intricacy of the problem lies in bounding the term $\bP[\matched_u(e)\cap\matched_v(e)]$, and thus the correlation between $u$ and $v$ being matched (to different partners) in the past. 
\subsection{Analysis for General Graphs} \label{sec:ocrsGeneral}

We present a new way of analyzing, given a newly arriving edge $(u,v)$, the probability of both $u,v$ being matched.  This will allow us to show that \Cref{alg:attenuate_aom} remains valid for $c=0.344$.

We consider the following sufficient condition for both $u,v$ being matched.
Suppose $u$ inspects all its surviving incident edges, and chooses one (if any exist), and $v$ (independently) does the same.
If these chosen edges are $(u,u')$ and $(v,v')$, where $u'$ and $v'$ are vertices in $V\setminus\{u,v\}$, then we call $u'$ and $v'$ the \textit{candidates} of $u$ and $v$, respectively.
Now, if candidate $u'$ was \textit{alone}
in that it had no surviving incident edges at the time of arrival of $(u,u')$, then this guarantees vertex $u$ to be matched, either to $u'$, or via a surviving incident edge that arrived before $(u,u')$.  A similar argument can be made for candidate $v'$ of vertex $v$.
Therefore, if $u'$ and $v'$ are \textit{distinct} candidates, and both alone at the arrival times of $(u,u')$ and $(v,v')$ respectively, then this guarantees both $u$ and $v$ to be matched.

We note that \citet{Ezra_2020} take a similar approach, but our procedure for choosing candidates is quite different from their "sampler", and generally more likely to choose any candidate at all.
Let $u_1,\ldots,u_k$ be vertices in $V\setminus\{u,v\}$ such that $(u,u_1),\ldots,(u,u_k)$ are the edges in $E$ incident to $u$ (recall that $E$ does not include the newly arriving edge $(u,v)$).
If $u$ has multiple surviving edges $(u,u_i)$ it will prioritize choosing the one with the smallest index $i$; however, it adds some noise to reduce the likelihood that $v$ (after defining an analogous procedure) will choose the same candidate.  The ordering of vertices $u_1,\ldots,u_k$ will be specified later based on the analysis.

To add this noise, we define a random bit $R_{u,u_i}$ for each $i=1,\ldots,k$.  We couple $R_{u,u_i}$ with $S_{u,u_i}$ (the random bit for edge $(u,u_i)$ surviving)
so that $R_{u,u_i}$ and $S_{u,u_i}$ are \textit{perfectly positively correlated}. Vertex $u$ then chooses $u_i$ as its candidate if $i$ is the smallest index for which $R_{u,u_i}$ realizes to 1.
We let $\candidate^u_{u_i}$ denote this event, noting that $u$ can have at most one candidate, and possibly none.
Now, although the random bits $R_{u,u_i}$ are coupled with $S_{u,u_i}$, the bits $S_{u,u_i}$ are independent from everything else, so we can use independence to deduce that
\begin{align}
\bP[\candidate^u_{u_i}]
&=\bE[R_{u,u_i}]\prod_{i'<i}(1-\bE[R_{u,u_{i'}}]) &\forall i=1,\ldots,k. \label{eqn:candProb}
\end{align}
We define an analogous procedure for the edges $(v,v_1),\ldots,(v,v_\ell)$ incident to $v$.
We will specify the means of the random bits $R_{u,u_i}$ and $R_{v,v_j}$ later, after establishing some concepts that bound the probabilities of edges surviving.

\begin{definition}
Let $e=(u',v')$ be an edge that has already arrived, with $u',v'$ being generic vertices in $V\setminus\{u,v\}$ (not necessarily candidates).
Let $x_{u'}(e):=\sum_{f\in\partial(u'):f\prec e}x_f$, where $f\prec e$ indicates that the edge $f$ arrived before $e$ (the sum does not include edge $e$ itself).
Similarly, let $x_{v'}(e):=\sum_{f\in\partial(v'):f\prec e}x_f$.

Let $\alone_{u'}(u',v')$ (respectively $\alone_{v'}(u',v')$) denote the event that $u'$ (respectively $v'$) does not have any surviving incident edges at the time of arrival of edge $(u',v')$.
We note that \citet{Ezra_2020} use a similar notion in their definition of "witness", but without the qualifier "at time of arrival of $(u',v')$".  We need this qualifier in order to make our subsequent argument.
\end{definition}

\begin{proposition}[proven in \S\ref{pf:prop:survBound}] \label{prop:survBound}
For any edge $e=(u',v')$, the probability of it surviving satisfies
\begin{align*}
\frac{cx_e}{1-c\cdot\max\{x_{u'}(e),x_{v'}(e)\}}
\le\bP[S_e=1] \le\frac{cx_e}{1-cx_{u'}(e)-cx_{v'}(e)}.
\end{align*}
\end{proposition}

\begin{proposition}[proven in \S\ref{pf:prop:aloneBound}] \label{prop:aloneBound}
For any edge $e=(u',v')$, the probability of a vertex $u'$ being alone satisfies
$
\bP[\alone_{u'}(e)]\ge\frac{1-c-cx_{u'}(e)}{1-c}.
$
\end{proposition}


Having established these \namecrefs{prop:survBound}, we can now use the lower bound in \Cref{prop:survBound} to define the probabilities for the random bits $R_{u,u_i}$.  We would like to ensure that whenever $u$ chooses vertex $u_i$ as its candidate, edge $(u,u_i)$ actually survives.  This will be the case whenever $\bE[R_{u,u_i}]\le\bE[S_{u,u_i}]$, since the bits $R_{u,u_i},S_{u,u_i}$ are coupled using perfect positive correlation.
By the lower bound in \Cref{prop:survBound}, this is ensured if we set
\begin{align} \label{eqn:random_bit}
\bE[R_{u,u_i}]:=\frac{cx_{u,u_i}}{1-cx_{u_i}(u,u_i)}
\end{align}
for all $i=1,\ldots,k$,
and similarly set $\bE[R_{v,v_j}]:=\frac{cx_{v,v_j}}{1-cx_{v_j}(v,v_j)}$ for all $j=1,\ldots,\ell$.
These values are set so that if $x_{u_i}(u,u_i)$ is large, which worsens the lower bound of $\frac{1-c-cx_{u_i}(u,u_i)}{1-c}$ on the probability of $u_i$ being alone, then at least we have the consolation prize that $\bE[R_{u,u_i}]$ is large, making it more likely that $u$ has a candidate.
This will prevent a worst-case configuration from simultaneously minimizing the two good events of $u_i$ being alone and $u$ having a candidate, which is precisely the motivation behind our choice procedure and definition of $\alone_{u_i}(u,u_i)$ that differs from \citet{Ezra_2020}.
We note that the analysis of \citet{Ezra_2020} is recovered if instead of~\eqref{eqn:random_bit}, we set $\bE[R_{u,u_i}]:=\frac{cx_{u,u_i}}{1-cx_{u}(u,u_i)}$, which also validly satisfies $\bE[R_{u,u_i}]\le\bE[S_{u,u_i}]$ (by the lower bound in \Cref{prop:survBound}).  We explain this analysis in \Cref{sec:recoverEzra} as well as demonstrate why it gets stuck at a guarantee worse than ours.

Having defined these random bits, we are ready to state and prove our main result, which lower-bounds the selectability of \Cref{alg:attenuate_aom} using an elementary optimization problem.


\begin{definition} \label{def:adv_opt}
For any positive integer $k$ and non-negative real number $b$, let
\begin{align*}
\advmin_k(b):=\inf\ &b^2\left(\sum_{i=1}^k\frac{y_i-by_i+by_i^2}{1+by_i}\prod_{i'<i}\frac{1}{1+by_{i'}}\right)\left(\sum_{i=1}^k\frac{z_i-bz_i+bz_i^2}{1+bz_i}\prod_{i'<i}\frac{1}{1+bz_{i'}}\right)
\\ &-b^2\sum_{i=1}^k\frac{y_i-by_i+by_i^2}{1+by_i}\frac{z_i-bz_i+bz_i^2}{1+bz_i}\prod_{i'<i}\frac{1}{1+by_{i'}}\frac{1}{1+bz_{i'}}
\\ 
&\begin{aligned}
\text{s.t. } &\sum_{i=1}^k y_i=\sum_{i=1}^k z_i=1
\\ & y_{i-1} \ge y_i, z_{i-1} \ge z_i, y_i + z_i\le 1, && \forall i=1,\ldots,k
\\ & y_i, z_i\ge0  && \forall i=1,\ldots,k.
\end{aligned}
\end{align*}
\end{definition}

\begin{theorem}[proven in \S\ref{pf:thm:ocrsGeneral}] \label{thm:ocrsGeneral}
\leavevmode
\begin{enumerate}
\item[(i).] \Cref{alg:attenuate_aom} is $c$-selectable for any 
$c$ satisfying $1-3c+\inf_k\advmin_k(\frac{c}{1-c})\ge0$.
\item[(ii).] $c=0.3445$ satisfies $1-3c+\inf_k\advmin_k(\frac{c}{1-c})\ge0$.
\end{enumerate}
Therefore, \Cref{alg:attenuate_aom} provides a 0.3445-selectable OCRS for general graphs.
\end{theorem}
Note that for fixed $b$, $\advmin_k(b)$ is decreasing in $k$, so $\inf_k\advmin_k(b)=\lim_{k\to\infty}\advmin_k(b)$.

\begin{remark}
Part (ii) of \Cref{thm:ocrsGeneral} is proved with the aid of computational verification, after bounding the difference between $\lim_{k\to\infty}\advmin_k(b)$ and an optimization problem with $2K$ variables as $O(1/K)$.
We then use Non-Linear Programming (NLP) solver COUENNE, modeled with JuMP \citep{dunning2017jump}, providing a link to the code.
The NLP solver establishes a \textit{provable lower bound} on the infimum value of this finite-dimensional NLP, allowing us to finish the proof.
Interestingly, the optimal solution suggested by the solver for a large $K$ is a "hybrid" in which $y_1=z_1=1/2$, and all other values of $y_i,z_i$ are infinitesimally-small.
\end{remark}

\subsection{Improvement for Bipartite Graphs} \label{sec:ocrsBipartite}

We improve the analysis of \Cref{alg:attenuate_aom} in the special case where $G = (V,E)$ is a bipartite graph.
Adopting the same proof skeleton and terminology, our goal is to lower-bound, given a newly arriving edge $(u,v) \notin E$, the probability that vertices $u$ and $v$ have both been matched.

In \Cref{sec:ocrsGeneral}, we analyzed the probability of the sufficient condition that $u$ and $v$ "randomly chose" distinct candidates who were alone.
In this \namecref{sec:ocrsBipartite}, we can analyze the easier-to-satisfy condition of $u$ and $v$ both \textit{having} candidates who are alone.
The reason for this is twofold:
the neighbors $u_1,\ldots,u_k$ of $u$ (i.e.\ the potential candidates) are clearly distinct from the neighbors of $v$, because edge $(u,v)$ cannot form a 3-cycle; and,
the neighbors $u_1,\ldots,u_k$ being alone are independent events, because there cannot be any edges between them (which again would form a 3-cycle).

By lower-bounding the probability of this easier-to-satisfy condition, we show that \Cref{alg:attenuate_aom} is 0.349-selectable for all graphs without a $3$-cycle (which includes all bipartite graphs), improving upon the earlier guarantee of 0.344 for general graphs.

\begin{theorem}[proven in \S\ref{pf:thm:ocrsBipartite}] \label{thm:ocrsBipartite}
On bipartite graphs, \Cref{alg:attenuate_aom} provides a $c$-selectable OCRS for any value of $c\in[0,1/2]$ satisfying $1-3c+\left(1-\exp(-\frac{c(1-2c)}{(1-c)^2})\right)^2\ge0$.
Therefore, \Cref{alg:attenuate_aom} is 0.349-selectable.
\end{theorem}
In the proof of \Cref{thm:ocrsBipartite}, we use the FKG inequality to argue that the events of $u$ and $v$ satisfying the condition of having candidates who are alone are positively correlated.  It may be tempting to argue that the events of $u$ and $v$ being matched are also positively correlated, under the intuition that in a bipartite graph, $u$ and $v$ are competing for different partners.  However, we show in \Cref{sec:bipartiteNegCorr} that this is false, justifying why we are arguing for positive correlation on this sufficient condition instead.

\subsection{Impossibility Results for OCRS} \label{sec:ocrsNegative}


The first two impossibility results use the following construction, which to our knowledge is new.

\begin{example} \label{eg:4cycle}
Let $G$ be a complete graph on vertices $V=\{1,2,3,4\}$, and consider the fractional matching whose edge values are $x_{12}=x_{23}=x_{34}=x_{41}=(1-\eps)/2$ along a 4-cycle and $x_{13}=x_{24}=\eps$ on the diagonals.
$\eps$ is a small positive constant that we will take to 0.
The arrival order of edges, known in advance, is: $(1,2),(3,4)$ (a diametrically opposite pair of edges), followed by $(2,3),(4,1)$ (another diametrically opposite pair), followed by $(1,3),(2,4)$ (the diagonal edges).
\end{example}

\begin{proposition}[proven in \S\ref{pf:thm:generalOcrsUB}] \label{thm:generalOcrsUB}
On the $G,\bm{x}$ given in \Cref{eg:4cycle}, any OCRS is no more than 0.4-selectable.
\end{proposition}

\begin{proposition}[proven in \S\ref{pf:thm:ezraOcrsUB}] \label{thm:ezraOcrsUB}
On \Cref{eg:4cycle}, the OCRS of \citet{Ezra_2020} is no more than 0.361-selectable.
\end{proposition}


\begin{proposition}[proven in \S\ref{pf:prop:3path}]  \label{prop:3path}
The OCRS of \citet{Ezra_2020} is no more than 0.382-selectable for bipartite graphs.
\end{proposition}

\section{Details of Random-Order Contention Resolution Schemes} \label{sec:RCRS}

We reuse the terminology and notation about graphs and matching polytopes defined in \Cref{sec:problem_definition}
and add the following definitions below.

\begin{definition}[Terminology and Notation for RCRS]
Suppose the edges of $G=(V,E)$ arrive uniformly at random.
In our analysis, we will treat each edge $e$ as having an \textit{arrival time} $Y_e$ drawn independently and uniformly from $[0,1]$.
Edges then arrive in increasing order of arrival times.

Also, if $\bm{x} = (x_e)_{e \in E}$ satisfies constraints $\sum_{e\in\partial(v)}x_e \le 1$ for all $v\in V$ as equality,
then we then say that $G$ is $1$-\textit{regular} (with respect to $\bm{x}$), and refer to $(G, \bm{x})$ as a $1$-\textit{regular input}. 
\end{definition}
\begin{remark}
Reformulating the random arrivals in this way has become a standard technique in online matching (for instance, see \citet{Ehsani2017, huang2018online}), and was originally used to analyze the Ranking algorithm in \citet{DJK2013}.
For RCRS, it was first employed by \citet{Lee2018}, and has since been used by \citet{Fu2021}, \citet{brubach2021offline},\citet{pollner2022improved} and \citet{macrury2023induction}.
\end{remark}


\begin{lemma}[proven in \S\ref{pf:lem:one_regular}]
\label{lem:one_regular}
If there exists a $c$-selectable RCRS for all $1$-regular inputs, then
there exists a $c$-selectable RCRS for all inputs via a reduction to a $1$-regular input. Moreover, this reduction can be
computed efficiently, and preserves the absence of cycles of length $3$ and $5$.
\end{lemma}
\begin{remark}
Although not needed for our results, in \Cref{lem:one_regular_long_bipartite} of \Cref{sec:added_reduction} we also present a reduction that preserves bipartiteness.
\end{remark}

Let us now fix an arbitrary \textit{attenuation function} $a: [0,1] \rightarrow [0,1]$. Consider the template RCRS in \Cref{alg:attenuate_rom}, which is presented the edges of a graph $G=(V,E)$ in random order.
\begin{algorithm}[t]
\caption{Attenuate-ROM} 
\label{alg:attenuate_rom}
\begin{algorithmic}[1]
\SingleSpacedXI
\Require Graph $G=(V,E)$ and a fractional matching $\bm{x} =(x_e)_{e \in E}$.
\Ensure subset of active edges forming a matching $\scr{M}$.
\State $\scr{M} \leftarrow \emptyset$.
\For{arriving edges $e \in E$}
\State Draw $A_{e} \sim \Ber(a(x_e))$ independently. \Comment{attenuate with probability $a(x_e)$}
\If{$e$ is active, not blocked and $A_{e} =1$}
\State $\scr{M} \leftarrow \scr{M} \cup \{e\}$.
\EndIf
\EndFor
\State \Return $\scr{M}$
\end{algorithmic}
\end{algorithm}
We consider Algorithm \ref{alg:attenuate_rom} with a quadratic attenuation function $a_1(x):=(1 - (3- e)x)^2$ when working with general graphs, and a different attenuation function, $a_2(x):=(1 - x)^4/(e^x - e x)^2$ for $x \in [0,1)$
where $a_2(1):= \lim_{x \rightarrow 1^-} a(x)= 4/e^2 $, when working with bipartite graphs.

\begin{theorem}[proven in \S\ref{sec:rcrsGeneral}] \label{thm:rom_selectability_general}
If $a(x)= a_1(x)$, where $a_1(x) := (1 - (3- e)x)^2$, then \Cref{alg:attenuate_rom} is $\frac{ e^2 - 4 e^3 + e^4 + 20 e -22}{4 e^2}\ge 0.474035$ selectable for $1$-regular general graphs.
\end{theorem}

\begin{theorem}[proven in \S\ref{sec:rcrsBipartite}] \label{thm:rom_selectability_triangle_free}
If $a(x) = a_2(x)$, where $a_2(x) := (1 - x)^4/(e^x - e x)^2$, then \Cref{alg:attenuate_rom} is
$\frac{e^6 + e^4 -42 - 4 e^2}{2 e^6} \ge 0.478983$ selectable for $1$-regular graphs without cycles
of length $3$ or $5$.
\end{theorem}


As in the adversarial order setting, we define $S_e := X_e \cdot A_e$
and say that $e$ \textit{survives} (the attenuation function $a$) if $S_e =1$. Observe that each edge $e$ survives independently with probability $\sv(x_e):=x_e a(x_e)$. 
We say that $f \in \partial(e)$ is \textit{relevant} (for $e$), provided $Y_f < Y_e$
and $f$ survives (recall that $\partial(e) = \partial(u) \cup \partial(v) \setminus \{e\}$ if $e=(u,v)$). Otherwise, $f$ is \textit{irrelevant} (for $e$). Denote the relevant edges of $e$ by $\Rf_e$. 
Observe that if $e$
survives and $\Rf_e = \emptyset$, then $e$ is selected by \Cref{alg:attenuate_rom} (note that the latter
event is equivalent to $\alone_{u}(e) \cap \alone_{v}(e)$ in our OCRS terminology).
\citet{brubach2021offline} ~use a different
attenuation function $a$
to argue that $\mb{P}[\Rf_e = \emptyset \mid X_e =1] \ge (1 - e^{-2})/2 \ge 0.432$, and since $a(0)=1$, it is not hard
to see that their analysis is tight\footnote{Consider $G=(V,E)$ with $V=\{u_i,v_i\}_{i=0}^n$ and $E=\{(u_0,v_0)\}\cup\{(u_0,u_i),(v_0,v_i)\}_{i=1}^n$. The activeness probability $x_e$ of every $e\in E$ is $1/(n+1)$. When $(u_0,v_0)$ survives, the RCRS of \citet{brubach2021offline} selects $(u_0,v_0)$ if and only if $\Rf_{u_0,v_0} = \emptyset$.
Moreover, since $a(0)=1$, this occurs with probability \textit{exactly} $(1-e^{-2})/2$ as $n\to\infty$.}.  \CM{Our improvement comes from restricting to $1$-regular inputs, as this allows $e$ to be matched even when $\Rf_e \neq \emptyset$. Specifically, when $|\Rf_e|=1$,
we use the $1$-regularity of $G$ to lower bound the probability $f \in \Rf_e$ is \textit{not} matched. This yields a lower bound
on $\mb{P}[e \in \scr{M}, \Rf_e \neq \emptyset \mid X_e =1]$,  and combined with the previous lower bound on $\mb{P}[\Rf_e = \emptyset \mid X_e =1]$, allows us to surpass a selection guarantee of $(1 - e^{-2})/2$. We now provide some additional definitions and notation needed to formalize this argument. These are similar to what is used in \cite{pollner2022improved} after they implicitly reduce to $1$-regular inputs via vertex-based attenuation (see \Cref{sec:technical} for a more detailed discussion on this).

\begin{definition} \label{def:simple_blocker}
Fix $e=(u,v) \in E$, and suppose that $f \in \partial(e)$ has vertex $w$ not in $e$. We say that $h \in \partial(w) \setminus \{(u,w),(v,w)\}$
is a \textit{simple-blocker} for $f$, indicated by the event $\simple_{f}(h)$, if:
\begin{enumerate}
    \item $h$ is relevant for $f$ (i.e., $S_h = X_h \cdot A_h =1$, and $Y_h < Y_f$).
    \item Each $h' \in \partial(h) \setminus \partial(e)$ is irrelevant for $h$.
\end{enumerate}
We denote the event in which $f$
has \textit{some} simple-blocker by $\blocker_f$. 

\end{definition}
Observe the following basic properties of the simple-blocker definition:
\begin{proposition} \label{prop:simple_blocker_useful}
For any $f \in \partial(e)$:
\begin{enumerate}
    \item $f$ has at most one simple-blocker.
    \item The event $\blocker_{f}$ is independent from the random variables $S_f$
    and $(Y_{g}, S_{g})_{g \in \partial(e) \cup \{e\} \setminus \{f\}}$.
\end{enumerate}
\end{proposition}
Note that the first item of \Cref{prop:simple_blocker_useful} is true, since if $f$ had two simple blockers, then one of them must be relevant for the other, which contradicts \Cref{def:simple_blocker}. The second item follows immediately since $\blocker_f$ does
\textit{not} depend on the random variables $(Y_{g}, S_{g})_{g \in \partial(e) \cup \{e\} \setminus \{f\}}$ in \Cref{def:simple_blocker}. 


\begin{proposition} \label{prop:one_relevant_edge}
\label{prop:safe_early_edges}
If $e$ survives, $|\Rf_e| \le 1$, and each $f \in \Rf_e$ satisfies $\blocker_f$, then $e \in \scr{M}$. 
\end{proposition}
Suppose $|\Rf_e| =1$, where $f =(u,w) \in \Rf_e$. \Cref{prop:one_relevant_edge} follows by observing that if $\blocker_{f}(h)$ occurs for
some $h \in \partial(w) \setminus \{(u,w),(v,w)\}$, then $h$ must be matched \textit{prior}
to the arrival of $f$, and so $f$ is not matched. Since every remaining edge of $\partial(e) \setminus \{f\}$ is irrelevant for $e$,
and $S_e =1$, $e$ will be matched when it arrives.}

\subsection{Analysis for General Graphs} \label{sec:rcrsGeneral}
Throughout this section, we analyze Algorithm \ref{alg:attenuate_rom}
when executed with the quadratic attenuation function $a(x) = (1 - (3- e)x)^2$. However, we are careful
to isolate the required analytic properties of $a$ as we proceed through the argument. (See Propositions \ref{property:first_order}, \ref{property:second_order} and \ref{property:vertex_split}).


We consider the case when there is at most one relevant edge; that is, $|\Rf_e| \le 1$. Observe first that by \Cref{prop:safe_early_edges},
\begin{equation} \label{eqn:general_graph_lower_bound}
\mb{P}[e \in \scr{M} \mid S_e = 1] \ge \mb{P}[ |\Rf_e| = 0] + \sum_{f \in \partial(e)} \mb{P}[\text{$\blocker_f$ and $\Rf_e = \{f\}$}].
\end{equation}
In order to lower bound the r.h.s. of \eqref{eqn:general_graph_lower_bound},
it will be convenient to first condition on $Y_e = y$ for an arbitrary $y \in [0,1]$. 
The expression $\mb{P}[ |\Rf_e| = 0 \mid Y_e =y]$ is then
easy to control, since $|\Rf_e|$ is distributed as $\sum_{f \in \partial(e)} \Ber(y \sv(x_{f}))$ where the Bernoulli's are independent, and so
\begin{equation} \label{eqn:no_early_edges}
     \mb{P}[ |\Rf_e| = 0 \mid Y_e = y] = \prod_{f \in \partial(e)} \ell(x_{f},y),
\end{equation}
where $\ell(x_{f},y):=1 - y\sv(x_{f})$ is the probability that $f$ is irrelevant. For an edge $f\in \partial(e)$ with vertex $w$ not in $e$,
we now lower bound $\mb{P}[\blocker_f \mid \Rf_e = \{f\}, Y_e = y]$.
In order to do so, we first lower bound the likelihood that $h \in \partial(w) \setminus \{(u,w),(v,w)\}$ is a simple-blocker for $f$, conditional on $\Rf_e = \{f\}$. Note that if $f = (w,u)$ (or $f=(w,v)$),
then we define $\fc:=(w,v)$ (respectively, $\fc := (w,u)$) to
be the \textit{pair} of $f$ in the triangle $\{(u,v), (w,v),(w,u)\}$.
\begin{lemma}[First-order minimization: proven in \S\ref{pf:lem:first_order_block}] \label{lem:first_order_block}
If $f$ has vertex $w$ not in $e$, then for each $h \in \partial(w) \setminus \{f, \fc \}$,
\begin{equation}\label{eqn:first_order_block}
   \mb{P}[\simple_{f}(h) \mid \Rf_e = \{f\}, Y_e =y] \ge  \frac{\sv(x_h)}{2(1 - x_h) - x_f - x_{\fc}}\left(1 - \frac{1 - e^{-(2(1 - x_h) - x_f - x_{\fc})y}}{(2(1 - x_h) - x_f - x_{\fc})y}\right).
\end{equation}
\end{lemma}
In order to prove \Cref{lem:first_order_block}, we show that the minimum probability of
the event $\simple_{f}(h)$ corresponds to when all the edges $h' \in \partial(h) \setminus \partial(e)$
have infinitesimally small values. This is implied by the following analytic properties of the attenuation function $a$:
\begin{proposition}[First-order minimization: proven in \S \ref{pf:property:first_order}] \label{property:first_order}
 For all $y \in [0,1]$, the function $x \rightarrow \ln(1 - y x a(x))$ is convex. Moreover, $a(0) =1$, and $a$ is continuous and decreasing on $[0,1]$. 

\end{proposition}

Next, we lower bound the probability that $f$ has \textit{some} simple-blocker, conditional on $f \in \Rf_e$.
\begin{lemma}[Second-order minimization: proven in \S\ref{pf:lem:second_order_block}] \label{lem:second_order_block}
For each $f \in \partial(e)$,   
\[
\mb{P}[\blocker_f \mid  \Rf_e = \{f\}, Y_e =y] \ge \frac{\sv(1-x_f -x_{\fc} ) }{x_f + x_{\fc} } \left(1 - \frac{1- e^{-(x_f + x_{\fc})y}}{(x_f + x_{\fc} ) y}\right) =:T(x_f + x_{\fc},y).
\]
\end{lemma}
We prove Lemma \ref{lem:second_order_block} by characterizing the minimum probability of the event $\blocker_f$. This minimum occurs when $w$ of $f=(w,u)$ has a single neighbor (other than $u$ and $w$), and its corresponding edge value is $1 - x_f - x_{\fc}$. Note that this is the \textit{opposite} worst-case in comparison to Lemma \ref{lem:first_order_block}.
Our proof relies on the following property of $a$:
\begin{proposition}[Second-order minimization: proven in \S \ref{pf:property:second_order}] \label{property:second_order}
For all $x \in [0,1]$,
    $\frac{a'(x)}{a(x)} + \frac{4}{1 - x} - \frac{2 (1 - \exp(x-1))}{\exp(x-1) -x} \le 0.$
\end{proposition}

Recall that by \Cref{prop:simple_blocker_useful},
for each $f \in \partial(e)$ the event $\blocker_f$ is independent
from random variables $(S_{g},Y_{g})_{g \in \partial(e) \setminus \{f\}}$. We can therefore apply \Cref{lem:second_order_block}, to get that
\[
    \sum_{f \in \partial(e)} \mb{P}[ \blocker_f, \Rf_e = \{f\} \mid Y_e =y] \ge \sum_{f \in \partial(e)} T(x_f +x_{\fc} ,y) \cdot \sv(x_f)y \prod_{g \in \partial(e) \setminus \{f\}} \ell(x_{g},y).
\]
By combining this equation with \eqref{eqn:no_early_edges}, and using \eqref{eqn:general_graph_lower_bound},
we get that $\mb{P}[e \in \scr{M} \mid S_e = 1] \ge \int_{0}^{1} \obj_{G}(e,y) dy$,
where for $y \in [0,1]$,
\begin{equation} \label{eqn:general_graph_opt}
    \obj_{G}(e,y):= \prod_{g \in \partial(e)} \ell(x_{g},y) + \sum_{f \in \partial(e)} T(x_f +x_{\fc} ,y) \cdot \sv(x_f)y \prod_{g \in \partial(e) \setminus \{f\}} \ell(x_{g},y).
\end{equation}
We must now identify the infimum of \eqref{eqn:general_graph_opt} over graphs which
contain $e$, and whose fractional matching assigns $x_e$ to $e$. 
We claim that no matter the value of $x_e$, this occurs as $\max_{f \in \partial(e)} x_f \rightarrow 0$ (i.e.,
the \textit{Poisson regime}). In order to prove
this, we apply a \textit{vertex-splitting procedure}.
Specifically, fix any $k \ge 1$, and replace an \textit{arbitrary} vertex
$w \in N(u) \cup N(v) \setminus \{u,v\}$ with $k$ copies of itself,
say $w_1, \ldots ,w_k$. Let $G_k=(V_k,E_k)$ be the resulting graph,
whose fractional matching $\bm{x}'$ is constructed by splitting the values
of the edges incident to $w$ uniformly amongst $w_1, \ldots ,w_k$,
and keep the remaining edge values the same. That is,
$x'_{w_i,r} := x_{w,r}/k$ for each $i \in [k]$ and $r \in V \setminus \{w_1, \ldots ,w_k\}$,
and $x'_f := x_f$ for all other $f \in E$. We lower bound
$\int_{0}^{1}\obj_{G}(e,y) dy$ by the limiting value of $\int_{0}^{1} \obj_{G_k}(e,y)dy$
as $k \rightarrow \infty$.
\begin{lemma}[Vertex Splitting: proven in \S\ref{pf:lem:vertex_split}] \label{lem:vertex_split}
$\int_{0}^{1}\obj_{G}(e,y) dy \ge \lim_{k \rightarrow \infty} \int_{0}^{1} \obj_{G_k}(e,y)dy$.
\end{lemma}
Lemma \ref{lem:vertex_split} relies on the following technical properties of $a$:
\begin{proposition}[Vertex splitting:  proven in \S\ref{pf:property:vertex_split}] \label{property:vertex_split}
For all $x_1, x_2 \in [0,1]$,
\begin{enumerate}
\item $y \rightarrow \ell(x_1,y) \ell(x_2,y) - \exp(-(x_1 + x_2)y)$ is non-negative for $y \in [0,1]$. \label{property:late_function}
\item Recalling the definition of the function $T(x_1 + x_2, y)$ from \Cref{lem:first_order_block}, the function $$y \rightarrow \ell(x_1,y) \ell(x_2,y) + T(x_1 +x_2,y)(y \sv(x_1) \ell(x_2,y) + y \sv(x_2) \ell(x_1,y))
- e^{-(x_1 + x_2) y}\left(1 + \frac{(x_1 + x_2)a(1) y^2}{2} \right)$$
is initially non-negative for $ y \in [0,1]$, and changes sign at most once. Moreover, its
integral over $[0,1]$ is non-negative.\label{property:integral}
\end{enumerate}
\end{proposition}
We also make use of the following elementary lower bound on the integral of the product of two functions.

\begin{proposition} \label{lem:integral}
Suppose that $\lambda, \phi:[0,1] \rightarrow \mb{R}$ are integrable, $\lambda \ge 0$, and $\lambda$ is non-increasing. Moreover, assume that there exists $0 \le z_c \le 1$ such that $\phi(z) \ge 0$ for all $z \in [0, z_c]$, and $\phi(z) \le 0$ for all $z \in [z_c,1]$. Then,
\[
    \int_{0}^{1} \lambda(z) \phi(z)  \, dz \ge \lambda(z_c) \int_{0}^{1} \phi(z) \, dz
\]
\end{proposition}

\begin{myproof}[Proof of \Cref{thm:rom_selectability_general}]

Suppose $G^*_k$ is the graph formed after splitting \textit{each} vertex of $N_{G}(u) \cup N_{G}(v) \setminus \{u,v\}$ into $k \ge 1$ copies. In order to compare $\int_{0}^{1}\obj_{G}(e,y) dy$ with $\lim_{k \rightarrow \infty} \int_{0}^{1} \obj_{G^*_k}(e,y)$, we can consider the sequence of $|N_{G}(u) \cup N_{G}(v) \setminus \{u,v\}|$ graphs formed by starting with $G$, and splitting a (new) vertex of $N_{G}(u) \cup N_{G}(v) \setminus \{u,v\}$ in each step. By applying \Cref{lem:vertex_split} to the consecutive pairs of graphs in the sequence, we can combine all the inequalities to
get that $ \int_{0}^{1} \obj_{G}(e,y) dy \ge \lim_{k \rightarrow \infty} \int_{0}^{1} \obj_{G^*_k}(e,y) \, dy$.
On the other hand, for each $y \in [0,1]$,
\begin{equation} \label{eqn:limiting_full_split}
    \obj_{G^*_k}(e,y) = \prod_{g \in \partial(e)} (\ell(x_{g}/k,y))^k + \sum_{f \in \partial(e)} T( (x_f +x_{\fc})/k ,y) \cdot \sv(x_f/k)y \prod_{g \in \partial(e) \setminus \{f\}} (\ell(x_{g}/k,y) )^k.
\end{equation}
Now, by taking $k \rightarrow \infty$, and applying the same asymptotic computations
from the proof of Lemma \ref{lem:vertex_split}, 
\[
\lim_{k \rightarrow \infty} \obj_{G^*_k}(e,y) = \left( 1 + \sum_{f \in \partial(e)} \frac{a(1) x_f y^2}{2}\right) e^{-\sum_{f \in \partial(e)} x_f y} .
\]
Thus, since $\sum_{f \in \partial(e)} x_f = 2(1-x_e)$, $\lim_{k \rightarrow \infty} \obj_{G^*_k}(e,y) =  \exp\left( -2y (1- x_e) \right) (1 + a(1)(1-x_e) y^2)$,
and so 
\begin{equation} \label{eqn:integral_lower_bound}
\int_{0}^{1} \obj_{G}(e,y) dy \ge \int_{0}^{1} e^{-2(1-x_e)y} \left(1 + a(1)(1-x_e) y^2 \right) dy,
\end{equation}
where we have exchanged the order of integration and point-wise convergence by using the dominated convergence theorem. Now, if we multiply \eqref{eqn:general_graph_opt} by $a(x_e)$, then
$$\mb{P}[ e \in \scr{M} \mid X_e = 1] \ge a(x_e) \cdot \int_{0}^{1} \obj_{G}(e,y) dy.$$
Thus, after applying \eqref{eqn:integral_lower_bound},
\begin{equation}\label{eqn:sel_e}
a(x_e) \cdot \int_{0}^{1} \obj_{G}(e,y) dy \ge a(x_e) \int_{0}^{1} e^{-2y (1- x_e)} (1 + a(1)(1-x_e) y^2) \, dy.
\end{equation}
Upon evaluating the integral in \eqref{eqn:sel_e},
we get a function of $x_e$ whose minimum occurs at $x_e=0$
when it takes on the value $\frac{ e^2 - 4 e^3 + e^4 + 20 e -22}{4 e^2}\ge 0.474035$. \Cref{thm:rom_selectability_general} is thus proven.
\end{myproof}

\CM{\subsection{Improvement for Bipartite Graphs}} \label{sec:rcrsBipartite}
We now consider when \Cref{alg:attenuate_rom} is executed on
a bipartite graph $G=(V,E)$ using the attenuation function $a(x) = (1 - x)^4/(e^x - e x)^2$.
In fact, everything in this section holds for graphs without cycles of length $3$ or $5$. In particular, the $1$-regular reduction of \Cref{lem:one_regular_long} preserves the lack of triangles and $5$-cycles. However, to simplify the resulting terminology, we state everything for bipartite graphs with the understanding that our $\approx 0.4789$ selectability result holds slightly more generally.

Our proof follows the same structure as the general graph case in that after fixing
$e =(u,v) \in E$, we consider the relevant edges $\Rf_e$ of $e$. 
The proof that our new attenuation function satisfies the analytic properties of \Cref{property:first_order} follows in the same way as before, and so we omit the argument. It also satisfies \Cref{property:second_order} by definition, as $a$ is unique solution
to the differentiation equation $\frac{a'(x)}{a(x)} + \frac{4}{1 - x} - \frac{2 (1 - \exp(x-1))}{\exp(x-1) -x} = 0$ with initial condition $a(0)=1$. Recalling the definition of the function $T(x,y)$ as stated in \Cref{lem:first_order_block}, we can therefore apply the argument from the previous section to get that
\begin{equation} \label{eqn:triangle_free_lower_bound_bad}
\mb{P}[|\Rf_e| \le 1, \cap_{f \in \Rf_e} \blocker_f \mid Y_e = y]  \ge \prod_{g \in \partial(e)} \ell(x_{g},y) + \sum_{f \in \partial(e)} T(x_f,y) \cdot \sv(x_f)y \prod_{g \in \partial(e) \setminus \{f\}} \ell(x_{g},y),
\end{equation}
where we've used the fact that $x_{\fc} =0$ for each $f \in \partial(e)$, as $G$ has no triangles. After integrating over $y \in [0,1]$, it is possible to argue that the infinum of the r.h.s. of \eqref{eqn:triangle_free_lower_bound_bad} occurs as $\max_{f \in \partial(e)} x_f \rightarrow 0$, when it takes value $\approx 0.4761$. Combined with \Cref{prop:one_relevant_edge}, one can then prove a lower bound of $\approx 0.4761$ for RCRS selectability on bipartite graphs. However, since our goal
is to exceed the tight $\approx 0.4762$ selectability bound for monotone contention resolution schemes on biparite graphs,
we now go beyond
the case of $|\Rf_e| \le 1$ and handle when each of $u$ and $v$ has its own relevant
edge. Specifically, define $\Rf_u = \partial(u) \setminus \{e\} \cap \Rf_e$
and $\Rf_v = \partial(v) \setminus \{e\} \cap \Rf_e$ as the relevant edges (for $e$) incident to 
$u$ and $v$, respectively. The following variation of \Cref{prop:one_relevant_edge} is easily proven:
\begin{proposition} \label{prop:two_early_edges}
\label{prop:safe_early_edges}
If $e$ survives, $|\Rf_u| = |\Rf_v| =1$, and each $f \in \Rf_e$ satisfies $\blocker_f$, then $e \in \scr{M}$. 
\end{proposition}
Due to \Cref{prop:two_early_edges}, our goal is to lower bound the probability that $\blocker_f$
occurs for each $f \in \Rf_e$,
and $|\Rf_u| = |\Rf_v| = 1$.
We first show that since $G$ is bipartite, and thus has no $5$-cycles, we can handle $u$ and $v$ separately.
\begin{lemma}[proven in \S\ref{pf:lem:bipartite_positive_correlation}] \label{lem:bipartite_positive_correlation}
For each $y \in [0,1]$ and $f \in \partial(u) \setminus \{e\}$, $f' \in \partial(v) \setminus \{e\}$,
$$
\mb{P}[\blocker_f \cap \blocker_{f'} \mid Y_e = y, \scr{R}_u = \{f\}, \scr{R}_v = \{f'\}] \ge T(x_f,y) \cdot T(x_{f'},y),
$$
where $T(x,y)=\frac{\sv(1-x) }{x} \left(1 - \frac{1- e^{-x y}}{x y}\right)$, and $\sv(1-x) = (1-x) a(1-x)$.
\end{lemma}
By applying \Cref{lem:bipartite_positive_correlation}, $\mb{P}[|\Rf_u| =|\Rf_v| = 1, \cap_{f \in \Rf_e} \blocker_f \mid Y_e = y]$ is lower bounded by  
$$
\sum_{f \in \partial(u) \setminus \{e\}, f' \in \partial(v) \setminus \{e\}} T(x_f,y) T(x_{f'},y) \cdot \sv(x_f)\sv(x_{f'})y^2 \prod_{g \in \partial(e) \setminus \{f, f'\}} \ell(x_{g},y).
$$
Thus, if
$$
\obj_{G \setminus e}(u,y) := \prod_{g \in \partial(u) \setminus e} \ell(x_{g},y) + \sum_{f \in \partial(u) \setminus e} T(x_f,y) \cdot \sv(x_f)y \prod_{g \in \partial(u) \setminus \{f,e\}} \ell(x_{g},y),
$$
(where $\obj_{G \setminus e}(v,y)$ is defined analogously) then combined with \eqref{eqn:triangle_free_lower_bound_bad}, we get that
$$
\mb{P}[|\Rf_u| \le 1, |\Rf_v| \le 1, \cap_{f \in \Rf_e} \blocker_f \mid Y_e = y]  \ge \obj_{G \setminus e}(u,y) \cdot \obj_{G \setminus e}(v,y). 
$$
As a result, Propositions \ref{prop:one_relevant_edge} and \ref{prop:two_early_edges} imply that $\mb{P}[e \in \scr{M} \mid S_e =1]$ is lower bounded by
\begin{equation} \label{eqn:new_obj}
\int_{0}^{1} \obj_{G \setminus e}(u,y) \cdot \obj_{G \setminus e}(v, y) dy 
\end{equation}
Our goal is now to identify the infimum of \eqref{eqn:new_obj}  restricted to \textit{bipartite} graphs which assign fractional value $x_e$ to $e$.
For $k \ge 1$, let us consider the same vertex splitting procedure as in the general graph case, where we select an arbitrary vertex
of $N_{G}(u) \cup N_{G}(v) \setminus \{u,v\}$ to construct $G_k$. Without loss, let us assume that the vertex split is from $N_{G}(u) \setminus \{v\}$. Note that the resulting graph $G_k$ will not have any cycles of length $3$ or $5$, as $G$ has none. 
Before stating the bipartite version of \Cref{lem:vertex_split}, we need an additional lemma which says that $\obj_{G \setminus e}(v,y)$ is non-increasing as a function of $y$. Roughly speaking, we prove this by considering an alternative input $G^*$ which contains the edge $e=(u,v)$, as well as $N_{G}(u) \cup N_{G}(v) \setminus \{u,v\}$. If one executes \Cref{alg:attenuate_rom} on $G^*$ conditional on $Y_e =y$, then $\obj_{G \setminus e}(v,y)$ is the probability that $v$ has at most one relevant edge \textit{and} is matched by time $y \in [0,1]$. This probability gets smaller the later $e$ arrives, which is exactly what we wish to prove.
\begin{lemma}[proven in \S\ref{pf:lem:decreasing_probability}]  \label{lem:decreasing_probability}
The function 
$
y \rightarrow \obj_{G\setminus e}(v,y)
$
is non-negative and non-increasing for $y \in [0,1]$.
\end{lemma}

\begin{lemma}[Bipartite-free vertex splitting: proven in \S \ref{pf:lem:triangle_free_vertex_split}] \label{lem:triangle_free_vertex_split}
$$
\int_{0}^{1} \obj_{G \setminus e}(u,y) \cdot \obj_{G \setminus e}(v, y) dy \ge \lim_{k \rightarrow \infty} \int_{0}^{1} \obj_{G_k \setminus e}(u,y) \cdot \obj_{G_k \setminus e}(v, y) dy 
$$
\end{lemma}
\Cref{lem:triangle_free_vertex_split} relies on the following properties of $a$, which we note are a special case (i.e, weakening) of those presented in \Cref{property:vertex_split}. The proof follows identically to the proof of \Cref{property:vertex_split}, and so we omit the argument.
\begin{proposition} \label{property:triangle_free_vertex_split}
For all $x \in [0,1]$:
\begin{enumerate}
\item $y \rightarrow \ell(x,y) - \exp(-xy)\ge 0$ is a non-negative function for $y \in [0,1]$ \label{property:triangle_free_late_function}
\item The function $y \rightarrow \ell(x,y) + y \sv(x) T(x,y) - e^{-x y}\left(1 + \frac{x a(1) y^2}{2} \right)$
is initially non-negative for $ y \in [0,1]$, and changes sign at most once. Moreover,  its integral over $[0,1]$ is non-negative. \label{property:triangle_free_integral}
\end{enumerate}
\end{proposition}
The proof of \Cref{thm:rom_selectability_triangle_free} now follows similarly to the proof of \Cref{thm:rom_selectability_general},
and so we just provide an outline where we the indicate differences in the asymptotic computations.
\begin{myproof}[Proof of \Cref{thm:rom_selectability_triangle_free}]
We can use \Cref{lem:triangle_free_vertex_split} to conclude that no matter the value of $x_e$, the infimum of \eqref{eqn:new_obj} occurs as $\max_{f \in \partial(e)} x_f \rightarrow 0$. Moreover, the same asymptotic computation
used in \eqref{eqn:integral_lower_bound} can be applied to $\obj_{G \setminus e}(u,y)$ and $\obj_{G \setminus e}(v,y)$ individually,
with the difference being that the $\sum_{f \in \partial(e)} x_f =2(1- x_e)$ term is replaced by $\sum_{f \in \partial(u) \setminus e} x_f = \sum_{f \in \partial(v) \setminus e} x_f = 1 - x_e$. Thus, applied to \eqref{eqn:new_obj},
we get that
\begin{equation} \label{eqn:triangle_free_integral_lower_bound}
\mb{P}[e \in \scr{M} \mid S_e =1] \ge \int_{0}^{1} \left( e^{-(1-x_e)y} \left(1 + \frac{a(1)(1-x_e)}{2} y^2 \right) \right)^2 \, dy,
\end{equation}
and so
\[
\mb{P}[ e \in \scr{M} \mid X_e =1] \ge a(x_e) \int_{0}^{1} \left( e^{-(1-x_e)y} \left(1 + \frac{a(1)(1-x_e)}{2} y^2 \right) \right)^2 \, dy.
\]
After evaluating the above integral,
we get a function of $x_e$ whose minimum occurs at $x_e=0$
when it takes on the value $\frac{e^6 + e^4 -42 - 4 e^2}{2 e^6} \ge 0.478983$. The proof
is thus complete.
\end{myproof}





\subsection{Impossibility Result for RCRS} \label{sec:rcrsNegative}

\begin{theorem} \label{thm:hardness_edge_rom}
No RCRS is better than $1/2$-selectable on bipartite graphs.
\end{theorem}

In order to prove \Cref{thm:hardness_edge_rom},
we again analyze the complete 1-regular bipartite graph with
$2 n$ vertices and uniform edge values, except instead of adversarially chosen edge arrivals, we work with random order edge arrivals. 
Let $G=(U_1,U_2,E)$ where $E = U_1 \times U_2$, and $|U_1|= |U_2| =n$ for $n \ge 1$,
and set $x_e = 1/n$ for all $e \in E$. Once again, we work in the asymptotic setting as $n \rightarrow \infty$. We say that a sequence of events $(\scr{E}_n)_{n \ge 1}$ occurs \textit{with high probability} (w.h.p.), provided $\mb{P}[\scr{E}_n] \rightarrow 1$ as $n \rightarrow \infty$.

\begin{lemma}\label{lem:expectation_negative}
For any RCRS which outputs matching $M$ on $G$, $\E[|M|] \le \frac{(1 + o(1))n}{2}$.
\end{lemma}
Assuming \Cref{lem:expectation_negative}, \Cref{thm:hardness_edge_rom} then follows immediately.

To prove \Cref{lem:expectation_negative}, we consider an algorithm for maximizing $\bE[|M|]$ and show that the cardinality of the matching cannot exceed $\frac{(1 + o(1))n}{2}$ in expectation.  Without loss of generality, we can assume such an algorithm is deterministic (this follows via an averaging argument), even though we will refer to it colloquially as an "RCRS".

For each $1 \le t \le n^2$, let $F_t$ be the $t^{th}$ edge of $G$ presented to the RCRS,
and denote its state by $X_{F_t}$ (clearly, $X_{F_t} \sim \Ber(1/n)$).
Observe that if $E_t := \{F_1, \ldots , F_t\}$, then conditional on $E_t$, $F_{t+1}$ is distributed
u.a.r. amongst $E \setminus E_t$ for $0 \le t \le n^2 -1$. If $\scr{M}_t$ is the matching
constructed by the RCRS after $t$ rounds, then since the RCRS is deterministic,
$\scr{M}_t$ is a function of $(F_i, X_{F_i})_{i=1}^{t}$. Thus, $\scr{M}_t$
is measurable with respect to $\scr{H}_t$, the sigma-algebra generated from $(F_i, X_{F_i})_{i=1}^{t}$
(here $\scr{H}_{0}:= \{ \emptyset, \Omega\}$, the trivial sigma-algebra).
We refer to $\scr{H}_t$ as the \textit{history} after $t$ steps.
It will be convenient to define $W(t):=  n \cdot |\scr{M}_t|$ for each $0 \le t \le n^2$. We can think of $W(t)$ as indicating the weight of the matching $\scr{M}_t$, assuming each edge of $G$ has weight $n$.

Let $w(z):=z/(1+z)$ for each real $z \ge 0$. Note that $w$ is the unique solution
to the differential equation $w' = (1 -w)^2$
with initial condition $w(0)=0$. By applying the differential equation method \citep{de}, one can show the greedy algorithm returns a matching
of weight
$(1 + o(1)) w(t/n^2) n^2$ after $0 \le t \le n^2$ steps. We implicitly prove
that greedy is asymptotically optimal by arguing that w.h.p. the random variable $W(t)/n^2$ is upper bounded by $(1 + o(1)) w(t/n^2)$.

\begin{proposition} \label{prop:de_dominance}
For each constant $0 \le \eps < 1$,
$W(t) \le (1 + o(1)) w(t/n^2)n^2$
for all $0 \le t \le \eps n^2$ with probability at least $1 - o(1/n^2)$.
\end{proposition}

We emphasize that in \Cref{prop:de_dominance}, a constant $\eps$ is fixed first, and $n$ is taken to $\infty$ afterward.
As we take constant $\eps$ to be arbitrarily close to $1$, the $1/2$ upper bound in \Cref{lem:expectation_negative} is established.
We now provide the proof of this fact.
The rest of this section is then devoted to proving \Cref{prop:de_dominance}.

\begin{myproof}[Proof of \Cref{lem:expectation_negative} using \Cref{prop:de_dominance}]
Fix $0 \le \eps < 1$. Observe that \Cref{prop:de_dominance} implies
\begin{equation}
\E[ W(\eps n^2)] \le  (1 - o(1/n^2))(1 + o(1)) w(\eps) n^2 + o(1/n^2)\eps n^2 = (1 + o(1)) w(\eps) n^2,
\end{equation}
where the $o(1/n^2)\eps n^2$ term uses the bound that $W(\eps n^2)$ cannot exceed the expected weight of active edges up to time $\eps n^2$, which is $\eps n^2$.
Moreover, the same bound yields $\E[ W(n^2) - W(\eps n^2)] \le (1 - \eps) n^2$. Thus,
$
\E[W(n^2)] \le (1 + o(1))\frac{\eps}{1+\eps} n^2 + (1 - \eps) n^2,
$
and so after dividing by $n^2$, $\E[|\scr{M}_{n^2}|]/n \le (1 + o(1))\frac{\eps}{1+\eps} + (1 - \eps)$.
Since this holds for each $0 \le \eps < 1$, and $\frac{\eps}{1+\eps} + (1 - \eps)
\rightarrow 1/2$ as $\eps \rightarrow 1$, we get that
\[
\frac{\E[ |\scr{M}_{n^2}|]}{n} \le (1 + o(1)) \frac{1}{2}.
\]
As $\E[ |\scr{M}_{n^2}|]$ is an upper bound on the expected
size of any matching created by an RCRS, the proof is complete.
\end{myproof}

In order to prove \Cref{prop:de_dominance}, for each constant $0 \le \eps < 1$,
and $0 \le t \le \eps n^2$, 
we first upper bound the expected one-step changes of $W(t)$, conditional
on the current history $\scr{H}_t$. More formally, we upper bound
$\E[ \Delta W(t) \mid \scr{H}_t]$, where $\Delta W(t):= W(t+1) - W(t)$.
Our goal is to show that 
$$\E[ \Delta W(t) \mid \scr{H}_t] \le (1 + o(1)) \left(1 - \frac{W(t)}{n^2} \right)^2.$$
It turns out that this upper bound only holds for \textit{most} instantiations of the random variables
$(F_i)_{i=0}^{t}$ (upon which the history $\scr{H}_t$ depends). We quantify this
by defining a sequence of events, $(Q_t)_{t=0}^{\eps n^2}$, which occur w.h.p.,
and which help ensure the upper bound holds. 

Fix a pair of vertex subsets $(S_1,S_2)$,
where $S_j \subseteq U_j$ for $j=1,2$. We say that
$(S_1, S_2)$ is \textit{large}, provided $|S_j| \ge  n/2$ for $j=1,2$.
Given $0 \le t \le n^2$, we say that $(S_1,S_2)$
is \textit{well-controlled} at time $t$, provided
\begin{equation}\label{prop:controlled_subsets}
|L_t \cap  S_1 \times S_2| \le (1 + n^{-1/3}) |S_1| |S_2| \left(1 - \frac{t}{n^2}\right),
\end{equation}
where $L_t:= E \setminus E_t$ denotes the edges
which have yet to arrive after $t$ rounds. 
We define the event $Q_t$ to occur, 
provided \textit{each} pair of \textit{large} vertex subsets is well-controlled
at time $t$. Observe that the event $Q_t$ is $\scr{H}_t$-measurable. 

\begin{lemma} \label{lem:well_controlled_probability}
For any constant $0 \le \eps < 1$, $\mb{P}[\cap_{i=0}^{\eps n^2 } Q_{i}] \ge 1 - o(1/n^2)$.
\end{lemma}
\begin{myproof}[Proof of \Cref{lem:well_controlled_probability}]
We shall prove that for each $0 \le i \le \eps n^2$, $Q_i$ holds with probability
at least $1 - o(1/n^4)$. Since there are $\eps n^2 \le n^2$ rounds, this will imply
that $\mb{P}[\cap_{i=0}^{\eps n^2 } Q_{i}] \ge 1 - o(1/n^2)$ after applying
a union bound.

Observe first that $L_i = E \setminus E_i$ is a uniformly random subset of $E$ of size $n^2 -i$.
Thus, $|L_i \cap  S_1 \times S_2|$ is
distributed as a hyper-geometric random variable on a universe of size
$n^2$ with success probability $|S_1| |S_2|/(n^2 - i)$
(we denote this by $|L_i \cap  S_1 \times S_2| \sim \text{Hyper}(n^2, |S_1||S_2|, n^2 -i)$.
Now, the distribution $\text{Hyper}(n^2, |S_1||S_2|, n^2 -i)$
is at least as concentrated about its expectation
as the binomial distribution, $\Bin(n^2, |S_1| |S_2|/(n^2 - i))$  (see Chapter~21 in~\cite{Frieze2015} for details). As such, by
standard Chernoff bounds, if $\mu := |S_1| |S_2| \left(1 - \frac{i}{n^2}\right)$,
then for each $0 < \lambda < 1$,
\[
\mb{P}[ |L_i \cap  S_1 \times S_2| \ge (1 + \lambda ) \mu ] \le \exp\left(\frac{- \lambda^2   \mu}{3} \right).
\]
By assumption, $|S_1||S_2| \ge n^2/4$. Thus, since $0 \le i \le \eps n^2$,
$\mu \ge n^2 (1 - \eps)/4$. By taking $\lambda = n^{-1/3}$, we get that 
\[
|L_i \cap  S_1 \times S_2| \ge (1 + \lambda) |S_1| |S_2| \left(1 - \frac{i}{n^2}\right)
\]
with probability at most $\exp\left(-\frac{n^{4/3}(1-\eps)}{12} \right)$ which is
$\exp\left(-\Omega(n^{4/3}) \right)$ because $\eps<1$ is a constant.
Now, 
after union bounding over at most $4^n$ subsets, we get
that $Q_t$ does \textit{not} occur with probability
at most $4^n \exp(-\Omega(n^{4/3})) = o(1/n^4)$. The proof is thus complete.
\end{myproof}

Upon conditioning on the history $\scr{H}_t$ for $0 \le t \le \eps n^2$, if $Q_t$ occurs
and $W(t) \le (1 + o(1)) w(t/n^2) n^2$,
then we can upper bound $\E[ \Delta W(t) \mid \scr{H}_t]$.

\begin{lemma}\label{lem:matching_expected_difference}
For each $0 \le t \le \eps n^2$, if $Q_t$ occurs
and $W(t) \le (1 + o(1)) w(t/n^2) n^2$, then
\begin{equation}\label{eqn:one_step_expected_change}
\E[ \Delta W(t) \mid \scr{H}_t] \le (1 + n^{-1/3} ) \left(1 - \frac{W(t)}{n^2} \right)^2.
\end{equation}
\end{lemma}
\begin{myproof}[Proof of \Cref{lem:matching_expected_difference}] \label{pf:lem:matching_expected_difference}
Suppose $0 \le t \le \eps n^2$ is such that $Q_t$ occurs and
$W(t) \le (1 + o(1)) w(t/n^2) n^2$. Observe that since $W(t) = n |\scr{M}_t|$,
it suffices to show that
$$\E[ |\scr{M}_{t+1}|- |\scr{M}_t|\mid \scr{H}_t] \le \frac{1}{n} (1 + n^{-1/3}) \left(1 - \frac{|\scr{M}_t|}{n} \right)^2.$$
For $j=1,2$, let $U_{j,t}$ denote the vertices of $U_j$ which are \textit{not} selected
by the RCRS after edges $E_t=\{F_1,\ldots , F_t\}$ arrive, where $U_{j,0}:=U_j$. Since the graph is bipartite, we have $|U_{j,t}| = n - |\scr{M}_t|$. Observe that a necessary condition for the RCRS to match $F_{t+1}$ is that it must be an edge of
$U_{1,t} \times U_{2,t}$. On the other hand, conditional on $\scr{H}_t$, $F_{t+1}$
is distributed u.a.r. amongst $L_t:= E \setminus E_t$. Thus,
\begin{equation} \label{eqn:arriving_edge}
\mb{P}[ F_{t+1} \in U_{1,t} \times U_{2,t} \mid \scr{H}_t] = \frac{| (U_{1,t} \times U_{2,t}) \cap L_t|}{|E \setminus L_t|} 
= \frac{| (U_{1,t} \times U_{2,t}) \cap L_t|}{n^2 - t},
\end{equation}
where the equality follows since $|E \setminus L_t| = n^2 -t$.
In order to simplify \eqref{eqn:arriving_edge}, 
we make use the upper bound
on $W(t)$, and the occurrence of the event $Q_t$.
First, $W(t) \le (1 + o(1)) w(t/n^2) n^2$,
where we note that $w(t/n^2)\le w(\eps)=\frac{\eps}{1+\eps}<1/2$, and hence for a sufficiently large $n$ we have $W(t)\le n^2/2$ and hence
$|\scr{M}_t| \le n/2$.
Thus,  $|U_{j,t}| = (n - |\scr{M}_t|) \ge n/2$,
and so we can apply \eqref{prop:controlled_subsets}
to subsets $U_{1,t}$ and $U_{2,t}$ to ensure that 
\[
| (U_{1,t} \times U_{2,t}) \cap L_t| \le (1 + n^{-1/3}) (n - |\scr{M}_t|)^2 \left(1 - \frac{t}{n^2}\right).
\]
Combined with \eqref{eqn:arriving_edge}, this implies that
\begin{equation} \label{eqn:arriving_edge_simplified}
\mb{P}[ F_{t+1} \in U_{1,t} \times U_{2,t} \mid \scr{H}_t] \le \frac{ (1 + n^{-1/3}) (n - |\scr{M}_t|)^2}{n^2} = (1 + n^{-1/3}) \left(1 - \frac{|\scr{M}_t|}{n} \right)^2.
\end{equation}
Now, a second necessary condition for the RCRS to match $F_{t+1}$ is that $F_{t+1}$ must be active (i.e., $X_{F_{t+1}} =1$).
This event occurs with probability $1/n$, independently of the event $F_{t+1} \in U_{1,t} \times U_{2,t}$ and the history $\scr{H}_t$.
By combining both necessary conditions, and \eqref{eqn:arriving_edge_simplified},
\begin{align*}
\E[ |\scr{M}_{t+1}|- |\scr{M}_t|\mid \scr{H}_t] &\le \frac{1}{n} (1 + n^{-1/3}) \left(1 - \frac{|\scr{M}_t|}{n} \right)^2,\\
\end{align*}
and so the proof is complete.
\end{myproof}

Lemmas \ref{lem:well_controlled_probability} and \ref{lem:matching_expected_difference}  imply that $W(t)/n^2$ satisfies
the \textit{differential inequality}, $r' \le (1-r)^2$ with $r(0)=0$. Intuitively, since $w(t)$ satisfies the corresponding differential \textit{equation}, $w' = (1 -w)^2$ with $w(0)=0$,
this suggests that $W(t)/n^2$ ought to be dominated by $w(t/n^2)$. This is
precisely the statement of \Cref{prop:de_dominance}, and assuming the above lemmas,
follows from the general purpose ``one-sided'' differential equation method of \citet{bennett2023}.
We provide the details in \S\ref{pf:prop:de_dominance}.

\CM{\section{Conclusion and Discussion}}
While our paper improved on the state of the art for RCRS and OCRS, determining the tight selectability
bounds for these problems remains open. 

For OCRS, recall that the OCRS of \citet{Ezra_2020} (\Cref{alg:attenuate_aom}) is $c \approx 0.382$-selectable on trees, where $c \in (0,1)$ is the smaller root of  $c = (1 - c)^2$. This is easily proven, since for any edge $e=(u,v)$ of $G=(V,E)$, $u$ and $v$ are matched independently prior to the arrival of $e$. In contrast, our $0.361$ upper bound shows $0.382$ is not attainable by this OCRS on general graphs. Deriving an OCRS that can surpass our upper bound of $0.361$ for \Cref{alg:attenuate_aom} on general graphs (if it is even possible) would be intriguing.
The input we used to prove the upper bound of $0.361$ (i.e., \Cref{eg:4cycle}) shows that the limitation of \Cref{alg:attenuate_aom} is due to the fact that it is \textit{non-adaptive}: it samples a bit $A_e$
independently for each edge $e \in E$, and then greedily accepts arriving edges with $A_e X_e =1$. This suggests that in order to beat $0.361$, one should design an OCRS whose bits $(A_e)_{e \in E}$ are correlated in an intelligent way.


For RCRS, our $1/2$ upper bound provides a fundamental separation from offline contention resolution,
where a $0.509$ lower bound is known for bipartite graphs \citep{nuti2023towards}. Our new $0.4789$ lower bound surpasses the tight $0.4762$ bound for \textit{monotone} contention resolution schemes on bipartite graphs, and thus shows that with respect to designing a CRS, monotonicity is more constraining than random order. On a technical level, our positive results first reduce to $1$-regular inputs and thus avoid more complicated forms of attenuation, as required by \citet{pollner2022improved}. The main question left open here is whether $1/2$ is the tight selectability bound for RCRS. Unlike our results for OCRS, we do not have an improved upper bound specific to our RCRS, and so it is unclear whether the gap between our upper and lower bounds is due to our analysis, or our choice of algorithm. Resolving this question is a natural first step towards proving (or disproving) a tight selectability bound of $1/2$.



\subsection*{Acknowledgements}The first author would like to thank Patrick Bennett for suggesting
the critical interval approach used to complete the proof of \Cref{prop:de_dominance} in the conference version of the paper.
The second author would like to thank Joey Huchette for suggesting to use the COUENNE package with JuMP.
An earlier version of this work
appeared in the ACM-SIAM Symposium on Discrete Algorithms (SODA) \citep{macrury2023random},
where a $0.4761$ lower bound on RCRS selectability for bipartite graphs was presented.
We improved this to $0.4789$ in this version. Finally, we would like to thank an anonymous reviewer for
suggesting the simplified $1$-regularity reduction of \Cref{lem:one_regular_long}.

\bibliographystyle{informs2014} 

\begingroup
    \bibliography{bibliography}

\begin{thebibliography}{26}
\providecommand{\natexlab}[1]{#1}
\providecommand{\url}[1]{\texttt{#1}}
\providecommand{\urlprefix}{URL }

\bibitem[{Adamczyk et~al.(2015)Adamczyk, Grandoni, \protect\BIBand{}
  Mukherjee}]{Adamczyk15}
Adamczyk M, Grandoni F, Mukherjee J (2015) Improved approximation algorithms
  for stochastic matching. Bansal N, Finocchi I, eds., \emph{Algorithms - {ESA}
  2015 - 23rd Annual European Symposium, Patras, Greece, September 14-16, 2015,
  Proceedings}, volume 9294 of \emph{Lecture Notes in Computer Science}, 1--12
  (Springer).

\bibitem[{Bansal et~al.(2012)Bansal, Gupta, Li, Mestre, Nagarajan,
  \protect\BIBand{} Rudra}]{BansalGLMNR12}
Bansal N, Gupta A, Li J, Mestre J, Nagarajan V, Rudra A (2012) When {LP} is the
  cure for your matching woes: Improved bounds for stochastic matchings.
  \emph{Algorithmica} 63(4):733--762,
  \urlprefix\url{http://dx.doi.org/10.1007/s00453-011-9511-8}.

\bibitem[{Baveja et~al.(2018)Baveja, Chavan, Nikiforov, Srinivasan,
  \protect\BIBand{} Xu}]{BavejaBCNSX18}
Baveja A, Chavan A, Nikiforov A, Srinivasan A, Xu P (2018) Improved bounds in
  stochastic matching and optimization. \emph{Algorithmica} 80(11):3225--3252,
  ISSN 1432-0541, \urlprefix\url{http://dx.doi.org/10.1007/s00453-017-0383-4}.

\bibitem[{Bennett \protect\BIBand{} MacRury(2023)}]{bennett2023}
Bennett P, MacRury C (2023) Extending wormald's differential equation method to
  one-sided bounds. \emph{CoRR} abs/2302.12358,
  \urlprefix\url{http://dx.doi.org/10.48550/arXiv.2302.12358}.

\bibitem[{Brubach et~al.(2021)Brubach, Grammel, Ma, \protect\BIBand{}
  Srinivasan}]{brubach2021offline}
Brubach B, Grammel N, Ma W, Srinivasan A (2021) Improved guarantees for offline
  stochastic matching via new ordered contention resolution schemes.
  \emph{Advances in Neural Information Processing Systems} 34:27184--27195.

\bibitem[{Bruggmann \protect\BIBand{} Zenklusen(2022)}]{BruggmannZ22}
Bruggmann S, Zenklusen R (2022) An optimal monotone contention resolution
  scheme for bipartite matchings via a polyhedral viewpoint. \emph{Math.
  Program.} 191(2):795--845,
  \urlprefix\url{http://dx.doi.org/10.1007/s10107-020-01570-6}.

\bibitem[{Chekuri et~al.(2014)Chekuri, Vondr{\'a}k, \protect\BIBand{}
  Zenklusen}]{chekuri2014submodular}
Chekuri C, Vondr{\'a}k J, Zenklusen R (2014) Submodular function maximization
  via the multilinear relaxation and contention resolution schemes. \emph{SIAM
  Journal on Computing} 43(6):1831--1879.

\bibitem[{Correa et~al.(2022)Correa, Cristi, Fielbaum, Pollner,
  \protect\BIBand{} Weinberg}]{correa2022optimal}
Correa J, Cristi A, Fielbaum A, Pollner T, Weinberg SM (2022) Optimal item
  pricing in online combinatorial auctions. \emph{International Conference on
  Integer Programming and Combinatorial Optimization}, 126--139 (Springer).

\bibitem[{Devanur et~al.(2013)Devanur, Jain, \protect\BIBand{}
  Kleinberg}]{DJK2013}
Devanur NR, Jain K, Kleinberg RD (2013) Randomized primal-dual analysis of
  ranking for online bipartite matching. \emph{Proceedings of the Twenty-fourth
  Annual ACM-SIAM Symposium on Discrete Algorithms}, 101--107, SODA '13
  (Philadelphia, PA, USA: Society for Industrial and Applied Mathematics), ISBN
  978-1-611972-51-1,
  \urlprefix\url{http://dl.acm.org/citation.cfm?id=2627817.2627824}.

\bibitem[{Dunning et~al.(2017)Dunning, Huchette, \protect\BIBand{}
  Lubin}]{dunning2017jump}
Dunning I, Huchette J, Lubin M (2017) Jump: A modeling language for
  mathematical optimization. \emph{SIAM review} 59(2):295--320.

\bibitem[{Ehsani et~al.(2018)Ehsani, Hajiaghayi, Kesselheim, \protect\BIBand{}
  Singla}]{Ehsani2017}
Ehsani S, Hajiaghayi M, Kesselheim T, Singla S (2018) Prophet secretary for
  combinatorial auctions and matroids. \emph{Proceedings of the Twenty-Ninth
  Annual ACM-SIAM Symposium on Discrete Algorithms}, 700–714, SODA ’18
  (USA: Society for Industrial and Applied Mathematics), ISBN 9781611975031.

\bibitem[{Ezra et~al.(2022)Ezra, Feldman, Gravin, \protect\BIBand{}
  Tang}]{Ezra_2020}
Ezra T, Feldman M, Gravin N, Tang ZG (2022) Prophet matching with general
  arrivals. \emph{Mathematics of Operations Research} 47(2):878--898.

\bibitem[{Feldman et~al.(2021)Feldman, Svensson, \protect\BIBand{}
  Zenklusen}]{feldman2021online}
Feldman M, Svensson O, Zenklusen R (2021) Online contention resolution schemes
  with applications to bayesian selection problems. \emph{SIAM Journal on
  Computing} 50(2):255--300.

\bibitem[{Frieze \protect\BIBand{} Karoński(2015)}]{Frieze2015}
Frieze A, Karoński M (2015) \emph{Introduction to Random Graphs} (Cambridge
  University Press),
  \urlprefix\url{http://dx.doi.org/10.1017/CBO9781316339831}.

\bibitem[{Fu et~al.(2021)Fu, Tang, Wu, Wu, \protect\BIBand{} Zhang}]{Fu2021}
Fu H, Tang ZG, Wu H, Wu J, Zhang Q (2021) {Random Order Vertex Arrival
  Contention Resolution Schemes for Matching, with Applications}. Bansal N,
  Merelli E, Worrell J, eds., \emph{48th International Colloquium on Automata,
  Languages, and Programming (ICALP 2021)}, volume 198 of \emph{Leibniz
  International Proceedings in Informatics (LIPIcs)}, 68:1--68:20 (Dagstuhl,
  Germany: Schloss Dagstuhl -- Leibniz-Zentrum f{\"u}r Informatik), ISBN
  978-3-95977-195-5, ISSN 1868-8969,
  \urlprefix\url{http://dx.doi.org/10.4230/LIPIcs.ICALP.2021.68}.

\bibitem[{Gravin \protect\BIBand{} Wang(2019)}]{gravin2019prophet}
Gravin N, Wang H (2019) Prophet inequality for bipartite matching: Merits of
  being simple and non adaptive. \emph{Proceedings of the 2019 ACM Conference
  on Economics and Computation}, 93--109.

\bibitem[{Huang et~al.(2018)Huang, Tang, Wu, \protect\BIBand{}
  Zhang}]{huang2018online}
Huang Z, Tang Z, Wu X, Zhang Y (2018) Online vertex-weighted bipartite
  matching: Beating 1-1/e with random arrivals. \emph{ACM Transactions on
  Algorithms} 15, \urlprefix\url{http://dx.doi.org/10.1145/3326169}.

\bibitem[{Jiang et~al.(2022)Jiang, Ma, \protect\BIBand{}
  Zhang}]{jiang2022tight}
Jiang J, Ma W, Zhang J (2022) Tight guarantees for multi-unit prophet
  inequalities and online stochastic knapsack. \emph{Proceedings of the 2022
  Annual ACM-SIAM Symposium on Discrete Algorithms (SODA)}, 1221--1246 (SIAM).

\bibitem[{Karp \protect\BIBand{} Sipser(1981)}]{karp1981maximum}
Karp RM, Sipser M (1981) Maximum matching in sparse random graphs. \emph{22nd
  Annual Symposium on Foundations of Computer Science (sfcs 1981)}, 364--375
  (IEEE).

\bibitem[{Lee \protect\BIBand{} Singla(2018)}]{Lee2018}
Lee E, Singla S (2018) {Optimal Online Contention Resolution Schemes via
  Ex-Ante Prophet Inequalities}. Azar Y, Bast H, Herman G, eds., \emph{26th
  Annual European Symposium on Algorithms (ESA 2018)}, volume 112 of
  \emph{Leibniz International Proceedings in Informatics (LIPIcs)}, 57:1--57:14
  (Dagstuhl, Germany: Schloss Dagstuhl--Leibniz-Zentrum fuer Informatik), ISBN
  978-3-95977-081-1, ISSN 1868-8969,
  \urlprefix\url{http://dx.doi.org/10.4230/LIPIcs.ESA.2018.57}.

\bibitem[{MacRury \protect\BIBand{} Ma(2023)}]{macrury2023induction}
MacRury C, Ma W (2023) Random-order contention resolution via continuous
  induction: Tightness for bipartite matching under vertex arrivals.
  \emph{CoRR} abs/2310.10101,
  \urlprefix\url{http://dx.doi.org/10.48550/ARXIV.2310.10101}.

\bibitem[{MacRury et~al.(2023)MacRury, Ma, \protect\BIBand{}
  Grammel}]{macrury2023random}
MacRury C, Ma W, Grammel N (2023) On (random-order) online contention
  resolution schemes for the matching polytope of (bipartite) graphs.
  \emph{Proceedings of the 2023 Annual ACM-SIAM Symposium on Discrete
  Algorithms (SODA)}, 1995--2014 (SIAM).

\bibitem[{Nuti \protect\BIBand{} Vondr{\'a}k(2023)}]{nuti2023towards}
Nuti P, Vondr{\'a}k J (2023) Towards an optimal contention resolution scheme
  for matchings. \emph{International Conference on Integer Programming and
  Combinatorial Optimization}, 378--392 (Springer).

\bibitem[{Pollner et~al.(2022)Pollner, Roghani, Saberi, \protect\BIBand{}
  Wajc}]{pollner2022improved}
Pollner T, Roghani M, Saberi A, Wajc D (2022) Improved online contention
  resolution for matchings and applications to the gig economy.
  \emph{Proceedings of the 23rd ACM Conference on Economics and Computation},
  EC '22 (New York, NY, USA: Association for Computing Machinery), ISBN
  9781450391504.

\bibitem[{Raghavan \protect\BIBand{} Tompson(1987)}]{raghavan1987randomized}
Raghavan P, Tompson CD (1987) Randomized rounding: a technique for provably
  good algorithms and algorithmic proofs. \emph{Combinatorica} 7(4):365--374.

\bibitem[{Wormald et~al.(1999)}]{de}
Wormald NC, et~al. (1999) The differential equation method for random graph
  processes and greedy algorithms. \emph{Lectures on approximation and
  randomized algorithms} 73:155.

\end{thebibliography}
\endgroup


\ECSwitch
\ECDisclaimer
\ECHead{E-Companion}


%
%
%

\begin{APPENDICES}
\crefalias{section}{appendix}

\section{Applications of Contention Resolution} \label{sec:prophet_application}

\textbf{Example application: prophet matching.}
In the prophet inequality (resp. prophet secretary) matching problem \citep{gravin2019prophet, Ezra_2020}, an online algorithm
is given a graph $G=(V,E)$ with each edge $e$ having weight $W_e$ drawn
independently from a known distribution; for simplicity, assume each distribution
is continuous. Initially, the instantiations of the
edge weights are \textit{unknown} to the algorithm
and are instead revealed to it in adversarial order (resp.\ random-order). Upon learning the weight
of an edge, the algorithm must make an irrevocable decision on whether to include it in its current matching. Its goal is to maximize the expected weight of the matching it outputs,
and it is bench-marked against $\OPT(G)$, the expected weight of an optimal matching of $G$ (optimized knowing all the weight realizations in advance). An algorithm
is said to be $c$-\textit{competitive} on $G$, provided the expected weight of the matching 
it returns is at least $c\cdot \OPT(G)$.

One way to handle the complexity of the prophet matching problem
is to first consider a relaxation  of $\OPT$: if $x_e$ is the probability that the optimal matching contains $e$, then $(x_e)_{e \in E}$ is a fractional matching. Moreover,
if $q(x_e) := \mb{E}[W_e \mid \text{$W_e$ in its top $x_e$-fraction of realizations}]$, then  $\OPT(G) \le  \sum_{e \in E} x_e q(x_e)$, and $(x_e)_{e \in E}$
can be computed by the online algorithm in advance (see \citet{feldman2021online}  for details).
\citet{feldman2021online} observed the following reduction to designing a $c$-selectable OCRS (RCRS). Upon learning the weight $W_e$ of an edge $e$, check if $W_e$ is in its top $x_e$-fraction of realizations. If the answer is yes, then refer to $e$ as \textit{active}, and use the OCRS (RCRS) to determine whether to add $e$ to the current matching. Since the OCRS (RCRS) is guaranteed to select each active edge with probability $c$, the expected weight of the matching returned is at least $c \cdot \sum_{e \in E} x_e q(x_e) \ge c \cdot \OPT(G)$, and so the algorithm is $c$-competitive.




\textbf{General application to online algorithms.}
Beyond the preceding example, our OCRS's and RCRS's also imply state-of-the-art results for other online problems on graphs, including stochastic probing \citep{BansalGLMNR12,Adamczyk15,BavejaBCNSX18,brubach2021offline} and a recent application of sequentially pricing jobs for gig workers \citep{pollner2022improved}.  In these problems, the algorithm makes a probing/pricing decision, after which there is stochasticity in whether an edge $e$ actually gets matched.  Nonetheless, by querying a CRS, and when the CRS says "select" calibrating the probing/pricing decision so that the probability of getting matched is exactly $x_e$, one also recovers online algorithms for these problems whose total reward is at least $c$ fraction of an optimal algorithm.  This provides the analogous desideratum of $c$-competitiveness.

All in all, given arbitrary resource constraints, OCRS's and RCRS's can be viewed as a parsimonious abstraction that allows different online decisions---accept/reject, probing, pricing, etc.---to be viewed under the same lens.
Philosophically, they also make it easier to identify worst cases for online algorithms, by imposing the stronger condition that it must extract $c$ fraction of an optimal algorithm's reward from \textit{every} element $e$ (instead of only in total).
It has been shown in some cases \citep{Lee2018} that this stronger condition does not worsen the competitiveness $c$ attainable.

\section{Deferred Proofs from \Cref{sec:OCRS}}

\subsection{Proof of \Cref{prop:survBound}} \label{pf:prop:survBound}
The proof follows easily from~\eqref{eqn:match_uv}: the induction hypothesis implies that
$$
1-cx_{u'}(e)-cx_{v'}(e)\le\bP[\overline{\blocked(e)}]\le 1-\max\{cx_{u'}(e),cx_{v'}(e)\},
$$
where we note that $\bP[\matched_{u'}(e)\cup\matched_{v'}(e)]\ge\max\{\bP[\matched_{u'}(e)],\bP[\matched_{v'}(e)]\}$.
Recalling that $\bP[S_e=1]=x_e\alpha_e$ with $\alpha_e$ defined to equal $c/\bP[\overline{\blocked(e)}]$, this completes the proof.

\subsection{Proof of \Cref{prop:aloneBound}} \label{pf:prop:aloneBound}
Let $(u',w_1),\ldots,(u',w_m)$ be the edges incident to $u'$ arriving before $(u',v')$, in that order.  Note that $x_{u'}(e)=\sum_{i=1}^mx_{u',w_i}$.  We can use independence to derive
\begin{align*}
\bP[\alone_{u'}(e)]
&=\prod_{i=1}^m(1-\bP[S_{u',w_i}])
\\ &\ge\prod_{i=1}^m \left(1-\frac{cx_{u',w_i}}{1-c x_{u'}(u',w_i)-c x_{w_i}(u',w_i)} \right)
\\ &\ge\prod_{i=1}^m \left(1-\frac{cx_{u',w_i}}{1-c\sum_{j<i}x_{u',w_j}-c} \right)
\\ &=\prod_{i=1}^m\frac{1-c-c\sum_{j\le i}x_{u',w_j}}{1-c-c\sum_{j<i}x_{u',w_j}}
\\ &=\frac{1-c-cx_{u'}(e)}{1-c}
\end{align*}
where the first inequality uses the upper bound in~\Cref{prop:survBound}, and
the second inequality uses the definition that $x_{u'}(u',w_i)=\sum_{j<i}x_{u',w_j}$ and the fact that $x_{w_i}(u',w_i)\le1$.  This leads to the desired result.

\subsection{Proof of \Cref{thm:ocrsGeneral}} \label{pf:thm:ocrsGeneral}
We prove parts (i) and (ii) of \Cref{thm:ocrsGeneral} in order.

Recall from~\eqref{eqn:desired} and~\eqref{eqn:match_uv} that it suffices to show that $\bP[\matched_u(u,v)\cup\matched_v(u,v)]\le 1-c$ for the newly arriving edge $(u,v)$.
Note that if $\sum_{f\in\partial(u)}x_f<1$, then $\bP[\matched_u(u,v)\cup\matched_v(u,v)]$ can only be increased after adding a dummy edge between $u$ and a new vertex that is active with probability $1-\sum_{f\in\partial(u)}x_f$, which arrives right before $(u,v)$.  The same argument can be made if $\sum_{f\in\partial(v)}x_f<1$.
Therefore, we can without loss of generality assume that $\sum_{f\in\partial(u)}x_f=\sum_{f\in\partial(v)}x_f=1$, which represents the hardest case for $\bP[\matched_u(u,v)\cup\matched_v(u,v)]\le 1-c$ to be satisfied.
Rewriting $\bP[\matched_u(u,v)\cup\matched_v(u,v)]$ following~\eqref{eqn:match_uv}, it suffices for $c$-selectability to show that $$0\le1-3c+\bP[\matched_u(u,v)\cap\matched_v(u,v)].$$

Therefore, we must show $\bP[\matched_u(u,v)\cap\matched_v(u,v)]\ge \inf_k\advmin_k(\frac{c}{1-c})$, where we have assumed that $\sum_{f\in\partial(u)}x_f=\sum_{f\in\partial(v)}x_f=1$.
Recall that $\matched_u(u,v)\cap\matched_v(u,v)$ occurs whenever all four events $\candidate^u_{u_i}$, $\candidate^v_{v_j}$, $\alone_{u_i}(u,u_i)$, and $\alone_{v_j}(v,v_j)$ occur, for any choice of indices $i\in\{1,\ldots,k\},j\in\{1,\ldots,\ell\}$ such that the vertices $u_i,v_j$ do not coincide.
This is because $\candidate^u_{u_i}$ implies $R_{u,u_i}=1$, which implies edge $(u,u_i)$ survives (see~\eqref{eqn:random_bit}), and this in conjunction with $\alone_{u_i}(u,u_i)$ ensures that $\matched_u(u,v)$ occurs.
An analogous argument ensures that $\matched_v(u,v)$ occurs, assuming $u_i$ is not the same vertex as $v_j$.
Finally, we note that since $u$ and $v$ each choose at most one candidate, the events $\candidate^u_{u_i}\cap \candidate^v_{v_j}\cap\alone_{u_i}(u,u_i)\cap\alone_{v_j}(v,v_j)$ are disjoint across the different combinations of $i,j$.  Therefore, we can derive
\begin{multline} \label{eqn:disjoint}
\bP[\matched_u(u,v)\cap\matched_v(u,v)]\\ \ge \sum_{i,j:u_i\neq v_j}\bP[\candidate^u_{u_i}\cap \candidate^v_{v_j}\cap\alone_{u_i}(u,u_i)\cap\alone_{v_j}(v,v_j)].
\end{multline}

The next step consists in showing that for any combination of $i,j$ such that $u_i\neq v_j$, the probability term on the r.h.s. is lower-bounded by the independent case, i.e.\
\small
\begin{multline} \label{eqn:OCRSindep}
\bP[\candidate^u_{u_i}\cap \candidate^v_{v_j}\cap\alone_{u_i}(u,u_i)\cap\alone_{v_j}(v,v_j)] \\
\ge
\bP[\candidate^u_{u_i}]\bP[\candidate^v_{v_j}]\bP[\alone_{u_i}(u,u_i)]\bP[\alone_{v_j}(v,v_j)].
\end{multline}
\normalsize
We argue this using the FKG inequality.  Consider the bits $\{S_e:e\in E\}$ about the survival of the edges.  Note that all four events $\candidate^u_{u_i},\candidate^v_{v_j},\alone_{u_i}(u,u_i),\alone_{v_j}(v,v_j)$ are fully determined by these bits, and moreover are increasing in the bits $S_{u,u_i}$, $S_{v,v_j}$ (they must necessarily be 1 for $\candidate^u_{u_i},\candidate^v_{v_j}$ to be 1, and note that this does not adversely affect $\alone_{u_i}(u,u_i),\alone_{v_j}(v,v_j)$ since $u_i\neq v_j$), and decreasing in all bits $S_e$ when $e$ is not $(u,u_i)$ or $(v,v_j)$.
Since the bits $S_e$ are independent across $e$, we have that~\eqref{eqn:OCRSindep} holds, for any $i,j$ such that $u_i\neq v_j$.

Now, we can use~\eqref{eqn:candProb} and \Cref{prop:aloneBound} to lower-bound $\bP[\candidate^u_{u_i}]$ and $\bP[\alone_{u_i}(u,u_i)]$ respectively.  Therefore, we derive
\begin{align}
\bP[\candidate^u_{u_i}]\bP[\alone_{u_i}(u,u_i)]
&\ge\frac{1-c-cx_{u_i}(u,u_i)}{1-c}\frac{cx_{u,u_i}}{1-cx_{u_i}(u,u_i)}\prod_{i'<i}(1-\frac{cx_{u,u_{i'}}}{1-cx_{u_{i'}}(u,u_{i'})}) \nonumber
\\ &\ge\frac{1-c-c(1-x_{u,u_i})}{1-c}\frac{cx_{u,u_i}}{1-c(1-x_{u,u_i})}\prod_{i'<i}(1-\frac{cx_{u,u_{i'}}}{1-c(1-x_{u,u_{i'}})}) \nonumber
\\ &=\frac{1-2c+cx_{u,u_i}}{1-c+cx_{u,u_i}}\frac{cx_{u,u_i}}{1-c}\prod_{i'<i}\frac{1-c}{1-c+cx_{u,u_{i'}}} \label{eqn:sub_in}
\end{align}
where the second inequality holds because the first expression is decreasing in both $x_{u_i}(u,u_i)$ and $x_{u_{i'}}(u,u_{i'})$, which must satisfy $x_{u_i}(u,u_i)\le 1-x_{u,u_i}$ and $x_{u_{i'}}(u,u_{i'})\le 1-x_{u,u_{i'}}$ respectively.

Combining the derivations in~\eqref{eqn:disjoint}, \eqref{eqn:OCRSindep}, and~\eqref{eqn:sub_in} (and lower bounding the analogous expression $\bP[\candidate^v_{v_j}]\bP[\alone_{v_j}(v,v_j)]$), we see that $\bP[\matched_u(u,v)\cap\matched_v(u,v)]$ is at least
\begin{align}
\sum_{i,j:u_i\neq v_j}
\left(\frac{1-2c+cx_{u,u_i}}{1-c+cx_{u,u_i}}\frac{cx_{u,u_i}}{1-c}\prod_{i'<i}\frac{1-c}{1-c+cx_{u,u_{i'}}}\right)
\left(\frac{1-2c+cx_{v,v_j}}{1-c+cx_{v,v_j}}\frac{cx_{v,v_j}}{1-c}\prod_{j'<j}\frac{1-c}{1-c+cx_{v,v_{j'}}}\right). \label{eqn:extant}
\end{align}

Finally, to relate to $\inf_k\advmin_k(\frac{c}{1-c})$, let $U(i)$ denote the first expression in large parentheses in~\eqref{eqn:extant}, and let $V(j)$ denote the second expression in large parentheses in~\eqref{eqn:extant}.
We can assume without loss that $k=\ell=|V|-2$, by adding edges with $x_{u,u_i}=0$ or $x_{v,v_j}=0$ as necessary, in which case $U(i)=0$ or $V(j)=0$ respectively.
This allows us to rewrite~\eqref{eqn:extant} as
\begin{align} \label{eqn:extantSimp}
\sum_{i=1}^k U(i) \sum_{j=1}^k V(j) - \sum_{i,j:u_i=v_j}U(i)V(j).
\end{align}

This is where we specify the ordering of the vertices $u_1,\ldots,u_k$ and $v_1,\ldots,v_k$ in a way that aids our analysis.
We specify $u_1$ so that $x_{u,u_1}=\max_i x_{u,u_i}$, and similarly specify $v_1$ so that $x_{v,v_1}=\max_j x_{v,v_j}$.
We let $v_2=u_1$, and similarly $u_2=v_1$; if $u_1=v_1$ then we instead let $u_2=v_2$ be any other vertex in $V\setminus\{u,v,u_1\}$.
We have completed the specification of $u_1,u_2,v_1,v_2$ in a way such that $\{u_1,u_2\}=\{v_1,v_2\}$.
Hence, both $u_3,\ldots,u_k$ and $v_3,\ldots,v_k$ must be orderings of the vertices in $V\setminus\{u,v,u_1,u_2\}$.
We define these orderings in such a way so that $x_{u,u_3}\ge\cdots\ge x_{u,u_k}$ and $x_{v,v_3}\ge\cdots\ge x_{v,v_k}$. This implies $U(3)\ge\cdots\ge U(k)$ and $V(3)\ge\cdots\ge V(k)$.

We have completed the specification of the orderings $u_1,\ldots,u_k$ and $v_1,\ldots,v_k$.
Now, consider an adversary trying to design the values of $x_{u,u_1},\ldots,x_{u,u_k},x_{v,v_1},\ldots,x_{v,v_k}$ to minimize expression~\eqref{eqn:extantSimp}, subject to all aforementioned constraints.
By the rearrangement inequality, the sum being subtracted is maximized if the largest values of $U(i)$ are paired with the largest values of $V(j)$.  That is, the adversary wants $u_i=v_i$ for all $i=3,\ldots,k$. Moreover, this assignment of vertices is guaranteed to feasibly satisfy $x_{u,u_i}+x_{v,v_i}\le 1$, since both $x_{u,u_i}$ and $x_{v,v_i}$ must be at most 1/2 (recall that $x_{u,u_i}\le x_{u,u_1}$ and $x_{u,u_i}+x_{u,u_1}\le 1$).
Therefore, if we assume that $u_i=v_i$ for all $i=3,\ldots,k$, then this only provides a lower bound on expression~\eqref{eqn:extantSimp}.

To finish, let $b:=\frac{c}{1-c}$.  We define shorthand notation $y_i:=x_{u,u_i}$ and $z_i:=x_{v,v_i}$ for all $i=3,\ldots,k$, as well as $y_1,y_2,z_1,z_2$ such that $y_1$ and $z_1$ correspond to the same vertex (and $y_2$ and $z_2$ correspond to the same vertex).
We drop the constraint that at least one of $y_1,y_2$ must correspond to a maximal value of $x_{u,u_i}$ (and similarly for $z_1,z_2$).
Noting that $U(i)$ can be rewritten as $\frac{1-b+by_i}{1+by_i}by_i\prod_{i'<i}\frac{1}{1+by_{i'}}$ under the new notation (and similarly for $V(j)$), we can express the adversary's optimization problem as minimizing
\begin{align*}
&\left(\sum_{i=1}^{k}\frac{1-b+by_i}{1+by_i}by_i\prod_{i'<i}\frac{1}{1+by_{i'}}\right)
\left(\sum_{i=1}^{k}\frac{1-b+bz_i}{1+bz_i}bz_i\prod_{i'<i}\frac{1}{1+bz_{i'}}\right)
\\ &-\sum_{i=1}^{k}\frac{1-b+by_i}{1+by_i}by_i\frac{1-b+bz_i}{1+bz_i}bz_i\prod_{i'<i}\frac{1}{1+by_{i'}}\frac{1}{1+bz_{i'}}
\end{align*}
subject to constraints $\sum_{i=1}^k y_i=\sum_{i=1}^k x_{u,u_i}=1=\sum_{i=1}^k z_i=\sum_{i=1}^k x_{v,v_i}$ (recall the assumption that $\sum_{f\in\partial(u)}x_f = \sum_{f\in\partial(v)}x_f = 1$), constraint $y_i+z_i\le 1$ for all $i=1,\ldots,k$, and constraints $y_3\ge\cdots\ge y_k,z_3\ge\cdots\ge z_k$ (due to the ordering chosen by the algorithm) as well as non-negativity constraints.
This is a lower bound on the original expression in~\eqref{eqn:extant}, and is exactly the $\advmin_k(b)$ optimization problem.
Therefore, $\bP[\matched_u(u,v)\cap\matched_v(u,v)]\ge\advmin_k(b)$ where $k$ denotes the number of vertices in $V\setminus\{u,v\}$.
To ensure $1-3c+\bP[\matched_u(u,v)\cap\matched_v(u,v)]\ge0$, it suffices to ensure $1-3c+\inf_k\advmin_k(b)$, completing the proof of \Cref{thm:ocrsGeneral}, part (i).

Fix a large positive integer "cutoff" $K$ and consider any $k\ge K$.
Since any term subtracted in the latter sum in the objective of $\advmin_k(b)$ also appears when the first two large parentheses are expanded, the objective can only be reduced if we reduce the term
\begin{align} \label{eqn:1080}
\frac{y_i-by_i+by_i^2}{1+by_i}\prod_{i'<i}\frac{1}{1+by_{i'}}
\end{align}
for any index $i$.  To reduce this term, note that $\frac{y_i-by_i+by_i^2}{1+by_i}=\frac{1-b+by_i}{1+by_i}y_i\ge(1-b)y_i$ and $\frac{1}{1+by_{i'}}\ge1-by_{i'}$, which allows us to reduce~\eqref{eqn:1080} to $(1-b)y_i\prod_{i'<i}(1-by_{i'})$.
We can similarly lower bound $\frac{z_i-bz_i+bz_i^2}{1+bz_i}\prod_{i'<i}\frac{1}{1+bz_{i'}}$ by $(1-b)z_i\prod_{i'<i}(1-bz_{i'})$.
Therefore, the objective of $\advmin(k)$ is lower-bounded by the following:
\small
\begin{align}
&b^2\left(\sum_{i=1}^K\frac{y_i-by_i+by_i^2}{1+by_i}\prod_{i'<i}\frac{1}{1+by_{i'}}+\sum_{i>K}(1-b)y_i\prod_{i'<i}(1-by_{i'})\right)\cdot \nonumber
\\ &\left(\sum_{i=1}^K\frac{z_i-bz_i+bz_i^2}{1+bz_i}\prod_{i'<i}\frac{1}{1+bz_{i'}}+\sum_{i>K}(1-b)z_i\prod_{i'<i}(1-bz_{i'})\right) \nonumber
\\ &-b^2\sum_{i=1}^K\frac{y_i-by_i+by_i^2}{1+by_i}\frac{z_i-bz_i+bz_i^2}{1+bz_i}\prod_{i'<i}\frac{1}{1+by_{i'}}\frac{1}{1+bz_{i'}}-b^2\sum_{i>K}(1-b)^2y_iz_i\prod_{i'<i}(1-by_{i'})(1-bz_{i'}) \nonumber
\\ &\ge b^2\left(\sum_{i=1}^K\frac{y_i-by_i+by_i^2}{1+by_i}\prod_{i'<i}\frac{1}{1+by_{i'}}+\frac{1-b}{b}\prod_{i'=1}^K(1-by_{i'})\sum_{i=K+1}^k by_i\prod_{i'=K+1}^{i-1}(1-by_{i'})\right) \nonumber
\\ &\cdot\left(\sum_{i=1}^K\frac{z_i-bz_i+bz_i^2}{1+bz_i}\prod_{i'<i}\frac{1}{1+bz_{i'}}+\frac{1-b}{b}\prod_{i'=1}^K(1-bz_{i'})\sum_{i=K+1}^k bz_i\prod_{i'=K+1}^{i-1}(1-bz_{i'})\right) \nonumber
\\ &-b^2\sum_{i=1}^K\frac{y_i-by_i+by_i^2}{1+by_i}\frac{z_i-bz_i+bz_i^2}{1+bz_i}\prod_{i'<i}\frac{1}{1+by_{i'}}\frac{1}{1+bz_{i'}}-b^2\sum_{i>K}(1-b)^2\frac{1}{(i-2)^2} \nonumber
\\ &\ge b^2\left(\sum_{i=1}^K\frac{y_i-by_i+by_i^2}{1+by_i}\prod_{i'<i}\frac{1}{1+by_{i'}}+\frac{1-b}{b}\prod_{i=1}^K(1-by_{i})\left(1-\prod_{i=K+1}^k(1-by_i)\right)\right) \nonumber
\\ &\cdot\left(\sum_{i=1}^K\frac{z_i-bz_i+bz_i^2}{1+bz_i}\prod_{i'<i}\frac{1}{1+bz_{i'}}+\frac{1-b}{b}\prod_{i=1}^K(1-bz_{i})\left(1-\prod_{i=K+1}^k(1-bz_i)\right)\right) \nonumber
\\ &-b^2\sum_{i=1}^K\frac{y_i-by_i+by_i^2}{1+by_i}\frac{z_i-bz_i+bz_i^2}{1+bz_i}\prod_{i'<i}\frac{1}{1+by_{i'}}\frac{1}{1+bz_{i'}}-b^2(1-b)^2\int_{K-2}^\infty \frac{1}{x^2}dx \nonumber
\\ &\ge \left(\sum_{i=1}^K by_i \left(1-\frac{b}{1+by_i}\right)\prod_{i'<i}\frac{1}{1+by_{i'}}+\frac{1-b}{b}\prod_{i=1}^K(1-by_{i})\left(1-\exp(-b(1-\sum_{i=1}^Ky_i))\right)\right) \label{eqn:long1}
\\ &\cdot\left(\sum_{i=1}^K bz_i \left(1-\frac{b}{1+bz_i}\right)\prod_{i'<i}\frac{1}{1+bz_{i'}}+\frac{1-b}{b}\prod_{i=1}^K(1-bz_{i})\left(1-\exp(-b(1-\sum_{i=1}^Kz_i))\right)\right) \label{eqn:long2}
\\ &-\sum_{i=1}^K\left(1-\frac{b}{1+by_i}\right)\left(1-\frac{b}{1+bz_i}\right)\prod_{i'<i}\frac{1}{1+by_{i'}}\frac{1}{1+bz_{i'}}-\frac{b^2(1-b)^2}{K-2}. \label{eqn:long3}
\end{align}
\normalsize
We explain each inequality.
The first inequality rewrites terms in the first two lines and applies the bounds $y_i\le \frac{1}{i-2}$ and $z_i\le \frac{1}{i-2}$ on the final subtracted term, which hold because $\sum_{i=1}^k y_k=1$ and $y_3\ge\cdots\ge y_k\ge0$ (and similarly for the $z_i$'s).
For the second inequality, note that $\sum_{i=K+1}^k by_i\prod_{i'=K+1}^{i-1}(1-by_{i'})$ is equivalent to the probability that at least one of independent Bernoulli random variables with means $by_i$ for $i=K+1,\ldots,k$ realizes to 1 (similarly for the $z_i$'s).
Moreover, we have $\sum_{i>K}\frac{1}{(i-2)^2}\le\int_{K-2}^\infty \frac{1}{x^2} dx$ by Riemann sums.
For the final inequality, we have applied the fact $1-by_i\le\exp(-by_i)$ and the constraint that $\sum_{i=1}^K y_i=1$ (similarly for the $z_i$'s) and evaluated the integral.

Since this holds for all $k\ge K$, we have proven that for any positive integer $K>2$, $\inf_k \advmin_k(b)$ is lower-bounded by the auxiliary optimization problem defined by
\begin{align*}
\\ \advminaux_K(b):=\inf\ &\text{\eqref{eqn:long1}--\eqref{eqn:long3}}
\\ \text{s.t. } &\sum_{i=1}^K y_i \le 1
\\ &\sum_{i=1}^K z_i \le 1
\\ & y_i + z_i\le 1 &\forall i=1,\ldots,K
\\ &y_i, z_i\ge0 &\forall i=1,\ldots,K
\end{align*}
(note that we have relaxed the constraints $y_3\ge\cdots\ge y_K$ and $z_3\ge\cdots\ge z_K$ on the adversary).
That is, we have $1-3c+\inf_k\advmin_k(\frac{c}{1-c})\ge1-3c+\advminaux_K(\frac{c}{1-c})$. The proof of \Cref{thm:ocrsGeneral}, part~(ii) is then completed by computationally verifying that for $c=0.3445$ and $K=80$ (a finite optimization problem), $1-3c+\advminaux_K(\frac{c}{1-c})\ge 0$. (Code can be found at \url{https://github.com/Willmasaur/OCRS_matching/blob/main/ocrs.jl}, which uses the JuMP \citep{dunning2017jump} and COUENNE packages).

\subsection{Recovering the Analysis of \citet{Ezra_2020}} \label{sec:recoverEzra}

We show why the analysis of \citet{Ezra_2020} that yields $c=0.337$ is recovered if we set $\bE[R_{u,u_i}]:=\frac{cx_{u,u_i}}{1-cx_{u}(u,u_i)}$ for all neighbors $u_i$ of $u$ other than $v$ (and analogously, set $\bE[R_{v,v_j}]:=\frac{cx_{v,v_j}}{1-cx_v(v,v_j)}$ for all neighbors $v_j$ of $v$ other than $u$).
Following the proof of \Cref{thm:ocrsGeneral}, we derive
\begin{align*}
\bP[\matched_u(u,v)\cap\matched_v(u,v)] &\ge\sum_{i,j:u_i\neq v_j}\bP[\candidate^u_{u_i}]\bP[\candidate^v_{v_j}]\bP[\alone_{u_i}(u,u_i)]\bP[\alone_{v_j}(v,v_j)]
\\ &\ge\sum_{i,j:u_i\neq v_j}\bP[\candidate^u_{u_i}]\bP[\candidate^v_{v_j}]\frac{1-c-cx_{u_i}(u,u_i)}{1-c}\frac{1-c-cx_{v_j}(v,v_j)}{1-c}
\end{align*}
where the latter inequality applies \Cref{prop:aloneBound}.

Now, the probabilities of $\candidate^u_{u_i}$ and $\candidate^v_{v_j}$ will change under the new way in which we set the probabilities of the bits $R_{u,u_i}$ and $R_{v,v_j}$.
Contrasting the proof of \Cref{thm:ocrsGeneral}, in this analysis we index $i=1,\ldots,k$ so that the neighbors $u_1,\ldots,u_k$ of $u$ arrive in that order (we similarly index $j=1,\ldots,\ell$).
Then, we can substitute into~\eqref{eqn:candProb} and get
\begin{align*}
\bP[\candidate^u_{u_i}]
&=\frac{cx_{u,u_i}}{1-cx_u(u,u_i)}\prod_{i'<i}\left(1-\frac{cx_{u,u_{i'}}}{1-cx_u(u,u_{i'})} \right)
\\ &=\frac{cx_{u,u_i}}{1-cx_u(u,u_i)}\prod_{i'<i}\frac{1-cx_u(u,u_{i'})-cx_{u,u_{i'}}}{1-cx_u(u,u_{i'})}
\\ &=\frac{cx_{u,u_i}}{1-cx_u(u,u_i)}\prod_{i'<i}\frac{1-cx_u(u,u_{i'+1})}{1-cx_u(u,u_{i'})}
\\ &=cx_{u,u_i}
\end{align*}
where $x_u(u,u_{i'})+x_{u,u_{i'}}=x_u(u,u_{i'+1})$ by definition of the ordering, allowing for the telescoping product.  After similarly deriving that $\bP[\candidate^v_{v_j}]=cx_{v,v_j}$, we get that
\begin{align}
\bP[\matched_u(u,v)\cap\matched_v(u,v)]
&\ge\sum_{i,j:u_i\neq v_j}cx_{u,u_i}cx_{v,v_j}\frac{1-c-cx_{u_i}(u,u_i)}{1-c}\frac{1-c-cx_{v_j}(v,v_j)}{1-c} \label{eqn:129048}
\\ &\ge\sum_{i,j:u_i\neq v_j}cx_{u,u_i}cx_{v,v_j}\left(\frac{1-2c}{1-c}\right)^2. \nonumber
\end{align}
The final bound corresponds exactly to the expression being analyzed in \citet[Lem.~3]{Ezra_2020}, which has a minimum value of $c^2(\frac{1-2c}{1-c})^2/2$, resulting in a constraint of $c\le1-2c+c^2(\frac{1-2c}{1-c})^2/2$ and leading to $c\approx0.337$.

We remark that even if one tries to improve the analysis by lower-bounding the RHS of~\eqref{eqn:129048} using the tighter expression
\begin{align*}
\sum_{i,j:u_i\neq v_j}cx_{u,u_i}cx_{v,v_j}\frac{1-c-c(1-x_{u,u_i})}{1-c}\frac{1-c-c(1-x_{v,v_j})}{1-c},
\end{align*}
the expression can be as small as $c^2(\frac{1-2c}{1-c})^2$ when the values of $x_{u,u_i},x_{v,v_j}$ become infinitesimally small and satisfy $\sum_{i=1}^k x_{u,u_i}=\sum_{j=1}^\ell x_{v,v_j}=1$.  This would result in a constraint of $c\le1-2c+c^2(\frac{1-2c}{1-c})^2$ and lead to $c\approx0.342$, which is still worse than the selection guarantee achieved in our \Cref{thm:ocrsGeneral}.

\subsection{Proof of \Cref{thm:ocrsBipartite}} \label{pf:thm:ocrsBipartite}
By the same argument as in the start of the proof of \Cref{thm:ocrsGeneral} part (i), we can without loss of generality assume that $\sum_{f\in\partial(u)}x_f=\sum_{f\in\partial(v)}x_f=1$, after which it suffices to show that
$1-3c+\bP[\matched_u(u,v)\cap\matched_v(u,v)]\ge0.$
We will show that $\bP[\matched_u(u,v)\cap\matched_v(u,v)]\ge\left(1-\exp(-\frac{c(1-2c)}{(1-c)^2})\right)^2.$
To do so, recall that $\matched_u(u,v)\cap\matched_v(u,v)$ occurs whenever $u$ and $v$ both have a neighbor that survives (i.e.\ can be a candidate) and is alone.
Letting $u_1,\ldots,u_k$ be the vertices in $V\setminus\{u,v\}$ such that $\{(u,u_i):i=1,\ldots,k\} = \partial(u)$ are the edges in $E$ incident to $u$, and respectively $v_1,\ldots,v_\ell$ be the vertices (which are distinct from $u_1,\ldots,u_k$) such that $\{(v,v_j):j=1,\ldots,\ell\}=\partial(v)$, we have that
\begin{multline*}
\bP[\matched_u(u,v)\cap\matched_v(u,v)] \\ \ge
\bP\left[\left(\bigcup_i(S_{u,u_i}\cap\alone_{u_i}(u,u_i))\right)
\bigcap
\left(\bigcup_j(S_{v,v_j}\cap\alone_{v_j}(v,v_j))\right)\right].
\end{multline*}
We argue that the r.h.s. of the preceding inequality is lower-bounded by the independent case, i.e.
\begin{align} \label{eqn:0289}
\bP[\matched_u(u,v)\cap\matched_v(u,v)]\ge
\bP\left[\bigcup_i(S_{u,u_i}\cap\alone_{u_i}(u,u_i))\right]
\bP\left[\bigcup_j(S_{v,v_j}\cap\alone_{v_j}(v,v_j))\right],
\end{align}
again using the FKG inequality.
To see this, consider the bits $\{S_e:e\in E\}$, and note that the events $S_{u,u_i}\cap\alone_{u_i}(u,u_i)$ and $S_{v,v_j}\cap\alone_{v_j}(v,v_j)$ are fully determined by these bits, and moreover are increasing in the bits $\{S_e:e\in\partial(u)\cup\partial(v)\}$ (such bits affect only $S_{u,u_i}$ and $S_{v,v_j}$) and decreasing in the bits $\{S_e:e\notin\partial(u)\cup\partial(v)\}$ (such bits affect only $\alone_{u_i}(u,u_i)$ and $\alone_{v_j}(v,v_j)$).
Since the bits $S_e$ are independent across $e$, we have that~\eqref{eqn:0289} holds.

Now, we can derive that
\begin{align*}
\bP\left[\bigcup_i(S_{u,u_i}\cap\alone_{u_i}(u,u_i))\right]
&= 1- \prod_i\left(1-\bP[S_{u,u_i}]\bP[\alone_{u_i}(u,u_i)]\right)
\\ &\ge 1- \prod_i \left(1-\frac{cx_{u,u_i}}{1-cx_{u_i}(u,u_i)}\frac{1-c-cx_{u_i}(u,u_i)}{1-c} \right)
\\ &\ge 1- \prod_i(1-\frac{c(1-2c)}{(1-c)^2}x_{u,u_i})
\\ &\ge 1- \exp\left(-\frac{c(1-2c)}{(1-c)^2}\sum_i x_{u,u_i}\right).
\end{align*}
To explain the equality, note that event $\alone_{u_i}(u,u_i)$ depends only on the independent bits $\{S_e:e\in\partial(u_i)\setminus(u,u_i)\}$, which must be disjoint from $\{S_e:e\in\partial(u_{i'})\setminus(u,u_{i'})\}$ for any $i'\neq i$, since otherwise $u_i$ and $u_{i'}$ would form a 3-cycle with $u$.
Therefore, the $2k$ events $S_{u,u_1}, \cdots, S_{u,u_k}$, $\alone_{u_1}(u,u_1), \ldots, \alone_{u_k}(u,u_k)$ are mutually independent, allowing us to decompose the probability $\bP\left[\bigcup_i(S_{u,u_i}\cap\alone_{u_i}(u,u_i))\right]$ into the product in the first line.
After that, the first inequality holds by \Cref{prop:survBound,prop:aloneBound},
the second inequality holds because $x_{u_i}(u,u_i)\le1$ and $c\le1/2$,
and the final inequality holds elementarily.
Finally, applying the assumption that $\sum_{i=1}^k x_{u,u_i}=1$, we conclude that $\bP\left[\bigcup_i(S_{u,u_i}\cap\alone_{u_i}(u,u_i))\right]\ge1-\exp(-\frac{c(1-2c)}{(1-c)^2})$.

After an analogous lower bound for $\bP\left[\bigcup_j(S_{v,v_j}\cap\alone_{v_j}(v,v_j))\right]$ and substituting into~\eqref{eqn:0289}, we have shown that $\bP[\matched_u(u,v)\cap\matched_v(u,v)]\ge(1-\exp(-\frac{c(1-2c)}{(1-c)^2}))^2$.  It can be numerically verified that $c=0.349$ satisfies $1-3c+(1-\exp(-\frac{c(1-2c)}{(1-c)^2}))^2\ge0$, completing the proof that \Cref{alg:attenuate_aom} is 0.349-selectable.

\CM{\subsection{Negative Correlation between $\matched_u(e)$ and $\matched_v(e)$ on Bipartite Graphs}} \label{sec:bipartiteNegCorr}

Consider a bipartite graph between vertices $u_1,u_2,u_3$ and $v_1,v_2,v_3$.  Let the edges be $e_1=(u_3,v_2)$, $e_2=(u_2,v_3)$, $e_3=(u_2,v_2)$, $e_4=(u_2,v_1)$, $e_5=(u_1,v_2)$, and $e_6=(u_1,v_1)$, arriving in that order, with $x_{e_1}=\cdots=x_{e_6}=1/3$ (the constant 1/3 is not important for our argument).  Let the edge in question be $e=e_6$, with $u=u_1$ and $v=v_1$.  Our claim is that under the execution of \Cref{alg:attenuate_aom}, we have $\Pr[\matched_u(e)\matched_v(e)]<\Pr[\matched_u(e)]\Pr[\matched_v(e)]$, i.e.\ the events of $u$ and $v$ being matched upon the arrival of the final edge are negatively correlated.

Note that when the final edge arrives, the algorithm would have conducted greedy matching (selecting every feasible edge) on the subsequence of $(e_1,\ldots,e_5)$ of edges that survive.  Conditional on $e_3$ not surviving, $u$ is disconnected from $v$ and hence $\matched_u(e)$ and $\matched_v(e)$ are independent events, i.e.
\begin{align}\label{eqn:198809}
\Pr[\matched_u(e)\matched_v(e)|S_3=0]=\Pr[\matched_u(e)|S_3=0]\Pr[\matched_v(e)|S_3=0].
\end{align}

Meanwhile, conditional on $S_3=1$, vertices $u$ and $v$ cannot both be matched, because if both vertices $u_2$ and $v_2$ are not blocked when edge 3 arrives, then $e_3$ would be selected, blocking $u_2$ and $v_2$ and preventing vertices $u$ and $v$ from being matched.  Yet, it is clear that both $\Pr[\matched_u(e)|S_3=1]$ and $\Pr[\matched_v(e)|S_3=1]$ are positive, noting that all edges in this example have a positive probability of surviving.  Therefore, we get that
\begin{align}\label{eqn:169500}
\Pr[\matched_u(e)\matched_v(e)|S_3=1]=0<\Pr[\matched_u(e)|S_3=1]\Pr[\matched_v(e)|S_3=1].
\end{align}
Combining~\eqref{eqn:198809} and~\eqref{eqn:169500} establishes that $\Pr[\matched_u(e)\matched_v(e)]<\Pr[\matched_u(e)]\Pr[\matched_v(e)]$, as desired.

\subsection{Proof of \Cref{thm:generalOcrsUB}} \label{pf:thm:generalOcrsUB}
Since edge (3,4) comes after (1,2), the probability of it being selected conditional on (1,2) being selected is at most $x_{34}=\frac{1-\eps}{2}$.  That is, $\bP[(1,2)\in\scr{M}\cap(3,4)\in\scr{M}]\le\frac{1-\eps}{2}\bP[(1,2)\in\scr{M}]$.
Thus,
\begin{align*}
\bP[(1,2)\in\scr{M}\cup(3,4)\in\scr{M}]
&\ge\bP[(1,2)\in\scr{M}]+\bP[(3,4)\in\scr{M}]-\frac{1-\eps}{2}\bP[(1,2)\in\scr{M}]
\\ &=\frac{1+\eps}{2}\bP[(1,2)\in\scr{M}]+\bP[(3,4)\in\scr{M}].
\\ &\ge\left(\frac{1+\eps}{2}+1\right)c\frac{1-\eps}{2}
\end{align*}
where the final inequality must hold if we were to have a $c$-selectable OCRS.
We can similarly derive that $\bP[(2,3)\in\scr{M}\cup(4,1)\in\scr{M}]\ge\frac{3+\eps}{2}c\frac{1-\eps}{2}$.
Now, note that $(1,2)\in\scr{M}\cup(3,4)\in\scr{M}$ and $(2,3)\in\scr{M}\cup(4,1)\in\scr{M}$ are disjoint events.
Hence, the probability that any of the edges (1,2),(2,3),(3,4),(4,1) is selected is at least $\frac{(3+\eps)(1-\eps)}{2}c$.
If any such edges are selected, then the diagonal edges (1,3),(2,4) cannot be selected.
Therefore, the probability that (1,3) can be selected is at most $(1-\frac{(3+\eps)(1-\eps)}{2}c)\eps$, which must be at least $c\eps$ in order to have a $c$-selectable OCRS.
Consequently we have $1-\frac{(3+\eps)(1-\eps)}{2}c\ge c$, and taking $\eps\to0$ implies $c\le 0.4$.

\subsection{Proof of \Cref{thm:ezraOcrsUB}} \label{pf:thm:ezraOcrsUB}
 First, note that when the first two edges $(1,2)$ and $(3,4)$ arrive, they
  cannot be blocked. Therefore,
  $\bP[\ol{\blocked(1,2)}] = \bP[\ol{\blocked(3,4)}] = 1$. Therefore,
  $\alpha_{(1,2)} = \alpha_{(3,4)} = c$, and matched with probability $c(1-\epsilon)/2$ (and
  independently of each other).

  Next, each of the next two edges (i.e., $(2,3)$ and $(4,1)$) are blocked if \emph{either} of the first two edges was matched. We have:
  \begin{align*}
    \bP[\ol{\blocked(2,3)}] = \bP[\ol{\blocked(4,1)}]
    &= \bP[(1,2) \notin \scr{M} \cap (3,4) \notin \scr{M}] \\
    &= (1-c(1-\epsilon)/2)^{2}
  \end{align*}
  which further gives that $\alpha_{(2,3)} = \alpha_{(4,1)} = c(1 - c(1-\epsilon)/2)^{-2}$.

  Finally, consider the final two arrivals, the diagonal edges. Edge $(1,3)$ is
  not blocked as long as none of the previous arrivals was matched. That is,
  both of edges $(1,2)$ and $(3,4)$ must have been left unmatched (each with
  probability $1-c(1-\epsilon)/2$), and then each of the next two edges (i.e., $(2,3)$
  and $(4,1)$) must have failed to survive (which occurs with probability
  $1 - c(1-\epsilon)(1-c(1-\epsilon)/2)^{2}/2$). The same holds for edge $(2,4)$. This gives us:
  \begin{align*}
    \bP[\ol{\blocked(1,3)}] = \bP[\ol{\blocked(2,4)}]
    &= \left(1 - c\frac{1-\epsilon}{2}\right)^{2} \left( 1 - \frac{c(1-\epsilon)}{2(1 - c(1-\epsilon)/2)^{2}} \right)^{2}
  \end{align*}
  As per~\eqref{eqn:desired}, we want this probability to be at least $c$. As $\epsilon\to 0$, we get
  \[
    \left(1 - \frac{c}{2}\right)^{2} \left( 1 - \frac{c}{2(1 - c/2)^{2}} \right)^{2} \ge c
  \]
  which is satisfied only if $c\le 0.3602$. Thus, on this graph, we must have
  $c<0.361$, and the OCRS cannot be better than $0.361$-selectable.

\subsection{Proof of \Cref{prop:3path}} \label{pf:prop:3path}
Let $G$ be a graph on vertices $V=\{1,2,3,4\}$ that is a path of three edges (1,2),(2,3),(3,4), and consider the fractional matching whose edge values are $x_{12}=1-\eps,x_{23}=\eps,x_{34}=1-\eps$, with
$\eps$ being a small positive constant.
The arrival order of edges is (1,2),(3,4),(2,3), where the middle edge arrives last.

  Notice that the first two edges, $(1,2)$ and $(3,4)$ cannot be blocked and so
  $\bP[\ol{\blocked(1,2)}] = \bP[\ol{\blocked(3,4)}] = 1$. This means
  $\alpha_{(1,2)} = \alpha_{(3,4)} = c$. Each of these edges will therefore be matched with probability $c(1-\epsilon)$. When the middle edge arrives, then, the probability it is not blocked is $(1 - c(1-\epsilon))^{2}$. Applying~\eqref{eqn:desired}, we get
  \[
    (1 - c(1-\epsilon))^{2} \ge c
  \]
  and for $\epsilon\to 0$ this gives $c \le 0.3819$. This means that the OCRS cannot be better than $0.382$-selectable for this graph.

\section{Deferred Proofs from Section \ref{sec:RCRS}} \label{sec:defPfs_rom}

\CM{\subsection{Proof of \Cref{lem:one_regular}}} \label{pf:lem:one_regular}

Below we state a detailed version of \Cref{lem:one_regular}. Note that this reduction also works for
OCRS, however we do not use this anywhere in the paper.

\begin{lemma}[Reduction to $1$-Regular Inputs -- Long Version] \label{lem:one_regular_long}
Given $G=(V,E)$ with fractional matching $\bm{x}=(x_e)_{e \in E}$,
there exists $G'=(V', E')$ and $\bm{x}' = (x'_e)_{e \in E'}$ with the following properties:
\begin{enumerate}
    \item $(G', \bm{x}')$ can be computed in time polynomial in the size of $(G,\bm{x})$
    \item $(G', \bm{x}')$ is $1$-regular. \label{property:one_regular}
    \item If $G$ has no cycles of length $3$ or $5$, then neither does $G'$.
    \item If there is an $\alpha$-selectable RCRS (respectively, OCRS) for $G'$ and $\bm{x}'$, then there exists an $\alpha$-selectable
    RCRS (respectively, OCRS) for $G$ and $\bm{x}$. \label{property:reduction}
\end{enumerate}

\end{lemma}

\begin{myproof}[Proof of Lemma \ref{lem:one_regular_long}]
The graph $G'=(V',E')$ and fractional matching is constructed as follows. For $v_0 \in V$, we add $6$ additional vertices,
say $v_1, \ldots ,v_6$, and create a cycle of length $7$ of the form $v_0,v_1, \ldots ,v_6, v_0$. 
If $x_{v_0}:= \sum_{e \in \partial_{G}(v_0)} x_{e}$, we define
$x'_{v_i, v_{i+1}}= (1-x_{v_0})/2$ if $i \in \{0,1, \ldots , 6\}$ is even,
and $x'_{v_i, v_{i+1}}= (1+x_{v_0})/2$ if $i$ is odd. We repeat this construction
for \textit{each} $v_0 \in V$, creating an additional $6$ vertices each time. Finally,
we set $x'_{e}:= x_e$ for each $e \in E$. Observe that $(G', \bm{x}')$ can be computed efficiently,
and it is clearly $1$-regular. Moreover, each cycle we created has length $7$, and so if $G$ has no cycles
of length $3$ or $5$, then neither will $G'$.

Finally, suppose we are given an $\alpha$-selectable RCRS for $(G', \mathbf{x'})$. We now show
how to produce an $\alpha$-selectable RCRS for $(G, \mathbf{x})$. For each
edge $e\in E' \setminus E$, generate a uniformly random arrival time
$Y'_{e} \in [0,1]$. Then, we run the RCRS for $G'$, allowing the edges of
$E'\setminus E$ to arrive in the order of $Y'_e$ (letting $Y'_e$ for an edge
$e\in E$ be equal to its original arrival time in $G$).

Clearly, the arrivals of $E'$ are uniformly random, so the RCRS for $G'$
selects each edge $e\in E'$ with probability at least $\alpha x_{e}$ (since
the RCRS is $\alpha$-selectable). Since the RCRS further processes the edges
of $E$ in their original (random) order, this is also an RCRS for $G$, and it
is clearly $\alpha$-selectable.

The reduction for OCRS proceeds similarly, and so the proof is complete.
\end{myproof}




\subsection{Proof of \Cref{lem:first_order_block}} \label{pf:lem:first_order_block}
Let us assume that $f = (u,w)$, and $h \in \partial(w) \setminus \{f,\fc\}$. We then condition on $Y_e =y$, $S_f =1$, $Y_f = y_f$, and $Y_h = y_h$, where $y_f, y_h \in [0, y]$
satisfy $y_h < y_f$. Our goal is to first derive a lower bound on $\mb{P}[ \simple_{f}(h) \mid Y_h = y_h, Y_f = y_f]$. Observe that since $y_h < y_f$, 
$h$ is a simple-blocker for $f$ if and only if each $h' \in \partial(h) \setminus \partial(e)$ is irrelevant for $h$. Thus,
\begin{equation} \label{eqn:first_order_blocker}
\mb{P}[ \simple_{f}(h) \mid Y_h = y_h, Y_f = y_f, Y_e =y] = x_h a(x_h) \prod_{h' \in \partial(h) \setminus \partial(e)} \ell(x_{h'},y_h)
\end{equation}  
where we recall that $\ell(x_{h'},y_h):= 1 - y_h \sv(x_{h'})$. Now, \eqref{eqn:first_order_blocker} is when minimized when $\partial(h) \setminus \partial(e)$ has as many edges as possible, so we hereby assume without loss of generality that $\partial(h) \cap \partial(e) = \{f, \fc \}$. In order to minimize \eqref{eqn:first_order_blocker},
we analyze
\begin{equation}\label{eqn:first_order_block_exponent}
    \sum_{h' \in \partial(h) \setminus \{f, \fc \}} \log \ell(x_{h'},y_h),
\end{equation}
subject to $\sum_{h' \in \partial(h) \setminus \{f, \fc \}} x_{h'} = 2 - 2x_h -x_f - x_{\fc}$.
The convexity of $x_{h'} \rightarrow \log \ell(x_{h'},y_h)$ guaranteed by \Cref{property:first_order} allows
us to conclude that \eqref{eqn:first_order_block_exponent} is minimized when $\max_{h' \in \partial(h) \setminus \{f, \fc \}} x_{h'} = o(1)$
and $|\partial(h) \setminus \{f, \fc \}| \rightarrow \infty$.
Thus, 
\begin{equation} \label{eqn:first_order_convex_minimization}
    \mb{P}[ \simple_{f}(h) \mid Y_h = y_h, Y_f = y_f, Y_e =y] \ge x_h a(x_h) \exp\left( -(2(1 - x_h) -x_f - x_{\fc} )y_h \right).
\end{equation}
(We omit the full details, as this part of the argument is due to \cite{brubach2021offline}).
Using \eqref{eqn:first_order_convex_minimization}, we integrate over $y_h \in [0,y_f]$, followed by $y_f \in [0,y]$,
to get that
\begin{align*}
    \mb{P}[ \text{$Y_f < y$ and $\simple_{f}(h)$} \mid Y_e =y] &\ge x_h a(x_h) \int_{0}^{y} \int_{0}^{y_f} \exp\left( -(2(1 - x_h) -x_f - x_{\fc} ) y_h \right) \, dy_h  \, dy_f \\
                        &= \frac{x_h a(x_h)}{{2(1 - x_h) -x_f - x_{\fc}} } \int_{0}^{y} \left(1 - \exp(-(2(1 - x_h) -x_f - x_{\fc})y_f\right) dy_f \\
                        & =\frac{x_h a(x_h)}{(2(1 - x_h) -x_f - x_{\fc} )}\left(y - \frac{1 - \exp(-(2(1 - x_h) -x_f - x_{\fc} ) y}{2(1 - x_h) -x_f - x_{\fc}}\right).
\end{align*}
Finally, after dividing both sides by $\mb{P}[Y_f  \le y] = y$, and observing that $S_f$ is independent of
$\simple_{f}(h)$, we get the claimed inequality, and so the proof is complete.

\subsection{Proof of Proposition~\ref{property:first_order}} \label{pf:property:first_order}

    First observe that $a_1(0)=1$, and $a_1$ is continuous and decreasing on $[0,1]$. Consider the second derivative $\frac{d^{2}}{dx^{2}}\ln \ell(x,y)$; minimizing this over all $x\in[0,1],y\in[0,1]$ gives a minimum of $0$ at $x=y=0$. Thus,
  if we fix any $y\in[0,1]$, the function
  $x\mapsto \frac{d^{2}}{dx^{2}}\ell(x,y)$ is nonnegative for $x\in[0,1]$,
  implying that for this fixed value of $y$, $x\mapsto \ln \ell(x,y)$ is convex
  on the interval $[0,1]$.

\subsection{Proof of \Cref{lem:second_order_block}} \label{pf:lem:second_order_block}
Let us assume that $f = (w,u)$ for some $w \in N(u) \setminus \{v\}$. 
We shall assume that $1 - x_f - x_{\fc} < 1$, as otherwise
the statement follows immediately. In this case, $\sum_{h \in \partial(w) \setminus \{f, \fc \}} x_h > 0$
since $w$ has fractional degree $1$ (as $G$ is $1$-regular). Observe that by definition, $\blocker_f= \cup_{h \in \partial(w) \setminus \{f, \fc\}} \simple_{f}(h)$.
On the other hand, $f$ has at most one simple-blocker by \Cref{prop:simple_blocker_useful}. Thus, after applying 
Lemma \ref{lem:first_order_block}, if $z_h:=2(1 - x_h) - x_f - x_{\fc}$,
then
\begin{equation}\label{eqn:second_order_optimization}
 \mb{P}[\blocker_f \mid f \in \Rf_e, Y_e = y] \ge \sum_{h \in \partial(w) \setminus \{f, \fc\}} \frac{x_h a(x_h)}{z_h}\left(1 - \frac{1-e^{-z_h y}}{z_hy}\right),
\end{equation}
subject to the constraint, $\sum_{h \in \partial(w) \setminus \{f, \fc\}} x_h = 1 - x_f - x_{\fc}$. Fix $y,x_f$ and $x_{\fc}$.
We claim that the worst-case for \eqref{eqn:second_order_optimization} occurs when 
$|\partial(w) \setminus \{f, \fc\}| =1$, and the single edge $h$ within this set
satisfies $x_h = 1- x_f - x_{\fc}$. In this case, the r.h.s of \eqref{eqn:second_order_optimization}
is $T(x_f + x_{\fc},y)$ so this will complete the proof.

Define $A(x_h):= \frac{a(x_h)}{z_h}\left(1 - \frac{1-e^{-z_h y}}{z_hy}\right)$.
Observe that if
we can show that $A(x_h)$ is decreasing as a function of $x_h$ on 
the interval $[0, 1 -x_f -x_{\fc}]$, then this will imply the claimed worst-case.
Now, setting $B(x_h) := \log A(x_h)$, we have that
$
    B(x_h) = \log a(x_h) - 2 \log z_h - \log y + \log( z_h y + e^{-yz_h} -1),
$
and so after differentiating $B$ with respect to $x_h$,
\begin{equation} \label{eqn:log_deriv}
B'(x_h) = \frac{a'(x_h)}{a(x_h)} + \frac{4}{z_h} - \frac{2y (1 - \exp(-y z_h)}{z_h y  + \exp(-y z_h -1)}.
\end{equation}
Our goal is to show that $B'(x_h) \le 0$ for all $x_h \in [0,1]$.
First, since $z_h \in [0,2]$, the function
$y \rightarrow \frac{2y (1 - \exp(-y z_h))}{z_h y  + \exp(-y z_h -1)}$
is decreasing, and so \eqref{eqn:log_deriv} is minimized at $y=1$, when it is equal to
$
\frac{a'(x_h)}{a(x_h)} + \frac{4}{z_h} - \frac{2(1 - \exp(-z_h))}{z_h  + \exp(-z_h -1)}.
$
Similarly, the function
$
    z_h \rightarrow \frac{4}{z_h} - \frac{2(1 - \exp(-z_h))}{z_h  + \exp(-z_h -1)}
$
is decreasing in $z_h$, and thus \textit{increasing} in $x_h$ (as $z_h=2 - 2 x_h - x_f - x_{\fc}$). Its maximum therefore
occurs at $x_h = 1 - x_f - x_{\fc}$, and so \eqref{eqn:log_deriv}
is upper-bounded by
 $$\frac{a'(x_h)}{a(x_h)} + \frac{4}{1 - x_h} - \frac{2 (1 - \exp(x_h-1))}{\exp(x_h-1) -x_h},$$
 which is at most $0$ by \Cref{property:first_order}. Thus, $B'(x_h) \le 0$ for all $x_h \in [0,1]$,
 and so $B(x_h)$ is decreasing as a function of $x_h$.
By exponentiating, the same statement is true for $A(x_h)$, and so the proof is complete.

\subsection{Proof of Proposition \ref{property:second_order}} \label{pf:property:second_order}
Recall that $a_1(x)=(1 - (3- e)x)^2$. The function
\[
x \mapsto \frac{a_1'(x)}{a_1(x)} + \frac{4}{1 - x} - \frac{2 (1 - \exp(x-1))}{\exp(x-1) -x}
\]
is decreasing on the interval $[0,1]$, as can be seen by examining its first derivative.
Thus, it is maximized at $x=0$, where it takes on value $0$.


\subsection{Proof of Lemma \ref{lem:vertex_split}} \label{pf:lem:vertex_split}
Recall that $w \in N_{G}(u) \cup N_{G}(v) \setminus \{u,v\}$ 
is the vertex which is copied $k \ge 1$ times in the construction $G_k$, yielding $w_1, \ldots ,w_k$ instead of $w$.
For convenience, we define $\tpartial(e): = \partial_{G}(e) \setminus \{(w,u),(w,v)\}$.
In order to relate $\obj_{G_k}(e,y)$ to $\obj_{G}(e,y)$,
it will be convenient to use
$$
T(x_1 +x_2,y) =\frac{\sv(1-x_1 -x_{2} ) }{x_1 + x_{2} } \left(1 - \frac{1- e^{-(x_1 + x_{2})y}}{(x_1 + x_{2} ) y}\right),$$
and $Q(x_1,x_2,y)= T(x_1 +x_2,y)(y \sv(x_1) \ell(x_2,y) + y \sv(x_2) \ell(x_1,y))$.
Observe then that $\obj_{G_k}(e,y)$ is equal to
\begin{align*}
\sum_{f \in \tpartial(e)} T(x_f +x_{\fc} ,y) \cdot y \sv(x_f)
    (\ell(x_{w,u}/k,y) \ell(x_{w,v}/k,y))^k  \prod_{g \in \tpartial(e) \setminus \{f\}}  \ell(x_{g},y) \\
   +  k  Q(x_{u,w}/k,x_{v,w}/k,y) \cdot (\ell(x_{w,u}/k,y) \ell(x_{w,v}/k,y))^{k-1}  \prod_{g \in \tpartial(e)} \ell(x_{g},y) \\
+ (\ell(x_{w,u}/k,y) \ell(x_{w,v}/k,y))^k \prod_{g \in \tpartial(e)} \ell(x_{g},y).
\end{align*}
(Here the middle term groups the contributions of the edges $(u,w_i)$ and $(v,w_i)$ for each $1 \le i \le k$).
Instead of working directly with $\obj_{G_k}(e,y)$, we consider
its point-wise limit as $k \rightarrow \infty$.
First, using the continuity of $a$, and the fact that $a(0)=1$,
\[
    \lim_{k \rightarrow \infty} (\ell(x_{w,u}/k,y) \ell(x_{w,v}/k,y))^k = 
    \lim_{k \rightarrow \infty} (\ell(x_{w,u}/k,y) \ell(x_{w,v}/k,y))^{k-1} = e^{-(x_{w,u} + w_{w,v})y},
\]
and $\lim_{k \rightarrow \infty} k (y \sv(x_{u,w}/k) \ell(x_{v,w}/k,y) + y \sv(x_{v,w}/k) \ell(x_{u,w}/k,y))= (x_{u,w} + x_{v,w})y$.
Moreover, it is not hard to show
that $\lim_{x \rightarrow 0^+} T(x,y)$ exists, and
is equal to $a(1) y/2$. Thus,
\[
 \lim_{k \rightarrow \infty} k Q(x_{u,w}/k,x_{v,w}/k,y) = \frac{a(1) (x_{u,w} +x_{v,w}) y^2}{2}.
\]
By combining all these expressions, $\lim_{k \rightarrow \infty} \obj_{G_k}(e,y)$
is equal to
\begin{align*}
 \sum_{f \in \tpartial(e)} T(x_f +x_{\fc} ,y) \cdot \sv(x_f)y e^{-(x_{w,u} + x_{w,v})y} \prod_{g \in \tpartial(e) \setminus \{f\}} \ell(x_{g},y) \\ + \left(\frac{a(1)(x_{u,w} + x_{v,w})y^2}{2} + 1 \right) e^{-(x_{w,u} + x_{w,v})y} \prod_{g \in \tpartial(e)} \ell(x_{g},y).
\end{align*}
Let us now compare $\lim_{k \rightarrow \infty} \obj_{G_k}(e,y)$ with $\obj_{G}(e,y)$, the latter of which we rewrite in the following way:
\[
  \sum_{f \in \tpartial(e)} T(x_f +x_{\fc} ,y) \cdot \sv(x_f)y \prod_{g \in \partial(e) \setminus \{f\}} \ell(x_{g},y) +  \prod_{g \in \partial(e)} \ell(x_{g},y) +  Q(x_{u,w},x_{v,w} ,y) \prod_{g \in \tpartial(e)} \ell(x_{g},y).
\]
Define $D_1(y)$ to be the difference of each expression's first term:
\[
D_1(y) := \sum_{f \in \tpartial(e)}
\left( \ell(x_{u,w},y) \ell(x_{v,w},y) -  e^{-(x_{w,u} + w_{w,v})y} \right) T(x_f +x_{\fc} ,y) y \sv(x_f) \prod_{g \in \tpartial(e) \setminus \{f\}} \ell(x_{g},y). 
\]
Similarly, let $D_{2}(y)$ be the difference of each expression's remaining terms:
\[
 \left(\ell(x_{u,w},y) \ell(x_{v,w},y) + Q(x_{u,w},x_{v,w} ,y) -  \left( 1 +\frac{a(1)(x_{u,w} + x_{v,w})y^2}{2} \right) e^{-(x_{w,u} + x_{w,v})y} \right) \prod_{g \in \tpartial(e)} \ell(x_{g},y).
\]
Observe now that  using the dominated convergence theorem, we can exchange the order of integration and point-wise convergence so 
that
\[
\lim_{k \rightarrow \infty} \int_{0}^{1} \obj_{G_k}(e,y) dy =  \int_{0}^{1} \lim_{k \rightarrow \infty} \obj_{G_k}(e,y) \, dy.
\]
Thus, to complete the proof it suffices to show that $\int_{0}^{1} D_i(y) \, dy \ge 0$ for each $i \in [2]$.
We start with $D_1$. Observe that $\ell(x_{u,w},y) \ell(x_{v,w},y) -  e^{-(x_{w,u} + x_{w,v})y} \ge 0$
for all $y \in [0,1]$ by the first property of \Cref{property:vertex_split}. Moreover, the remaining terms in each summand of $D_1$ are non-negative, so $\int_0^1 D_1 \ge 0$. Consider now $D_2$. Observe that the function $y \rightarrow \prod_{g \in \tpartial(e)} \ell(x_{g},y)$
is non-negative and non-increasing for $y \in [0,1]$, as $\ell(x_g,y)= 1 - y \sv(x_g)$, and $\sv(x_g) \in [0,1]$ for $x_g \in [0,1]$.
Moreover, by the second property of \Cref{property:vertex_split}, since $Q(x_{u,w},x_{v,w} ,y) = T(x_{u,w}+x_{v,w},y)(y\sv(x_{u,w}) \ell(x_{v,w},y) + y\sv(x_{v,w}) \ell(x_{u,w},y) )$, the function 
$$y \rightarrow \left(\ell(x_{u,w},y) \ell(x_{v,w},y) + Q(x_{u,w},x_{v,w} ,y) -  \left( 1 +\frac{a(1)(x_{u,w} + x_{v,w})y^2}{2} \right) e^{-(x_{w,u} + x_{w,v})y} \right),$$
is initially non-negative, changes sign at most once, and has a non-negative integral.
Thus, we can apply \Cref{lem:integral} (with $\lambda$ as the first function, and $\phi$ as the second),
to conclude that $\int_0^1 D_2 \ge 0$.


\subsection{Proof of Proposition \ref{property:vertex_split}} 
\label{pf:property:vertex_split}

We verify \Cref{property:vertex_split} for the attenuation function $a_1(x) = (1 - (3- e)x)^2$. 

Recall that $\ell(x,y) = 1 - y x a_{1}(x)$. The first property can be checked easily: the minimum value of $\ell(x_1,y) \ell(x_2,y) - \exp(-(x_1 + x_2)y)$ over
  $x_1,x_2, y \in[0,1]$ occurs either when of $x_1=x_2=0$ or $y=0$, for which $\ell(x_1,y)\ell(x_2,y) - \exp(-(x_1+x_2)y) = 0$. 
  
The second property is more complicated to verify. Recall that $$
T(x_1 +x_2,y) =\frac{\sv(1-x_1 -x_{2} ) }{x_1 + x_{2} } \left(1 - \frac{1- e^{-(x_1 + x_{2})y}}{(x_1 + x_{2} ) y}\right),$$
and $Q(x_1,x_2,y)= T(x_1 +x_2,y)(y \sv(x_1) \ell(x_2,y) + y \sv(x_2) \ell(x_1,y))$.
  Set $I(x_1,x_2) := \int_{0}^{1} \ell(x_1,z) \ell(x_2,z) + Q(x_1,x_2,s)
- e^{-(x_1 + x_2) z}\left(1 + \frac{(x_1 + x_2)a(1) z^2}{2} \right) \, dz$.
  This function has a closed form, and its minimum occurs when $x_1 = x_2 = 0$. Next, let
  $F_{x_1,x_2}(z) := \ell(x_1,z) \ell(x_2,z) + Q(x_1,x_2,z)
- e^{-(x_1 + x_2) z}\left(1 + \frac{(x_1 + x_2)a(1) z^2}{2} \right)$. It can
  be observed that for $x_1,x_2\in[0,1]$, $F_{x_1,x_2}(0)=0$, $F_{x_1,x_2}'(0)\ge 0$, and
  $F_{x_1,x_2}''(z)\le 0$ for all $z\in(0,1)$. These can be checked easily by e.g.\ numerically minimizing $F_{x_1,x_2}'(0)$ over $x_1,x_2\in [0,1]$ and numerically maximizing $F_{x_1,x_2}''(z)$ over $x_1,x_2,z\in[0,1]$ (we find that the minimum of $F_{x_1,x_2}'(0)$ is $F_{x_1,x_2}'(0)=0$, and the maximum of $F_{x_1,x_2}''(z)$ occurs for $x_1=x_2=0$ where $F_{0,0}''(z)=0$ for all $z$).
  Thus, $F_{x_1,x_2}(z)$ is initially
  nonnegative and increasing. Now, suppose for sake of contradiction that for a given $x_1,x_2$, there exist two points
  $z_{1} < z_{2}$ for which $F_{x_1,x_2}$ changes sign, and without loss of
  generality, assume these are the first two such points.
  Since $F_{x_1,x_2}$ is initially nonnegative and increasing, it must be the case
  that $F_{x_1,x_2}$ is positive on the interval $(0,z_{1})$ with
  $F_{x_1,x_2}'(z_{1}) < 0$ and negative on the interval $(z_{1},z_{2})$ with
  $F_{x_1,x_2}'(z_{2}) > 0$. Therefore, it must be the case that $F_{x_1,x_2}'(z)$ is
  increasing on the interval $(z_{1},z_{2})$, but this means that
  $F_{x_1,x_2}''(z) > 0$ somewhere on this interval; however, it was observed
  previously that $F_{x_1,x_2}''(z) \le 0$ on the entire interval $(0,1)$.

\CM{\subsection{Proof of \Cref{lem:bipartite_positive_correlation}}} \label{pf:lem:bipartite_positive_correlation}
Fix $f=(u,w) \in \partial(u) \setminus\{e\}$ and $f' =(v,w') \in \partial(v) \setminus \{e\}$, where we observe that $w \neq w'$, as $G$ is triangle-free. It suffices to prove
that
$$
\mb{P}[\blocker_f \cap \blocker_{f'} \mid  \max\{Y_f, Y_{f'}\} < y] \ge T(x_f,y) \cdot T(x_{f'},y),
$$
where $T(x,y)=\frac{\sv(1-x) }{x} \left(1 - \frac{1- e^{-x y}}{x y}\right)$. Now, $\mb{P}[\simple_{f}\cap \simple_{f'} \mid \max\{Y_f, Y_{f'}\} < y]$ is equal to:
\begin{align} \label{eqn:two_blocker_critical}
 \mb{P}[\simple_{f}(w,w') \cap \simple_{f'}(w,w') \mid   \max\{Y_f, Y_{f'}\} < y] + \sum_{\substack{h \in \partial(w) \setminus \{f, (w,w')\}, \\ h' \in \partial(w') \setminus \{f', (w,w')\}}} \mb{P}[\simple_{f}(h) \cap \simple_{f'}(h') \mid  \max\{Y_f, Y_{f'}\} < y].
\end{align}
We first lower bound the left-most term of \eqref{eqn:two_blocker_critical}, which corresponds to when $f$ and $f'$ are simultaneously blocked by the \textit{same} edge $(w,w')$.
In this case, for any $z < y$, $\mb{P}[\min\{Y_f,Y_{f'}\} \le z \mid \max\{Y_f,Y_{f'}\} < y] = 2z/y - (z/y)^2$, and so the probability density function of $\min\{Y_f,Y_{f'}\}$ (conditional on $\max\{Y_f,Y_{f'}\} < y$) is $z \rightarrow 2/y - 2z/y^2$. Thus, 
$\mb{P}[\simple_{f}(w,w') \cap \simple_{f'}(w,w') \mid \max\{Y_f, Y_{f'}\} < y]$ is equal to
\begin{align*}
 & \int_{0}^{y} \mb{P}[\simple_{f}(w,w') \cap \simple_{f'}(w,w') \mid \min\{Y_f,Y_{f'}\} = z, \max\{Y_f,Y_{f'}\} < y] \left(\frac{2}{y}- \frac{2z}{y^2} \right) dz \\
 &= \int_{0}^{y}  \left(\frac{2}{y}- \frac{2z}{y^2} \right) \int_{0}^{z} \sv(x_{w,w'}) \prod_{g \in \partial(w,w') \setminus \{f,f'\}} \left(1 - \sv(x_g) y_{w,w'} \right) dy_{w,w'} dz \\
&\ge \int_{0}^{y}  \sv(x_{w,w'})  \left(\frac{2}{y}- \frac{2z}{y^2} \right) \frac{1 - e^{-z (2 (1 - x_{w,w'}) - x_f - x_{f'})}}{2 (1 - x_{w,w'}) - x_f - x_{f'}}   dz,
\end{align*}
where we defer the details of the inequality, as it follows from the exact same computations used in the proof of \Cref{lem:first_order_block} (see \eqref{eqn:first_order_convex_minimization}, and the integral that follows immediately afterwards). Define
$Q(x_f,x_{f'},x_{w,w'},y)$ to be the final integral above, which has the closed form expression:
\begin{equation*}
   \frac{ \sv(x_{w,w'}) (2 - 
    2 e^{- (2(1 - x_{w,w'}) - x_f - x_{f'}) y} - (2(1 - x_{w,w'}) - x_f  - x_{f'}) y (2 - (2(1 - x_{w,w'}) - x_f  - x_{f'}) y)))}{(2(1- x_{w,w'}) -
     x_{f} - x_{f'})^3 y^2},
\end{equation*}
so that $\mb{P}[\simple_{f}(w,w') \cap \simple_{f'}(w,w') \mid \max\{Y_f, Y_{f'}\} < y] \ge Q(x_f,x_{f'},x_{w,w'},y)$.

We now consider the right-most term of \eqref{eqn:two_blocker_critical}.
Let $h \in \partial(w) \setminus \{f, (w,w')\}$ and $h' \in \partial(w') \setminus \{f',(w,w')\}$. We first prove a form of \textit{positive correlation} between 
the events $\simple_{f}(h)$ and $\simple_{f'}(h')$.
Specifically, 
\begin{equation} \label{eqn:blocker_positive_correlation}
 \mb{P}[\simple_{f}(h) \cap \simple_{f'}(h') \mid \max\{Y_f, Y_{f'}\} < y] \ge  \mb{P}[\simple_{f}(h)\mid Y_f < y] \cdot \mb{P}[\simple_{f'}(h')\mid  Y_{f'} < y].
\end{equation}
In order to see this, let us write $h=(w,r)$ and $h'=(w',r')$. Observe that due to the lack of $5$-cycles, $r \neq r'$. As a result, the events $\simple_{f}(h)$ and $\simple_{f'}(h')$ depend on different independent random variables, \textit{except} for the (possible) random variables associated with the edges $(w,w')$ and $(r,r')$. 
Thus, if $y_h < y_f < y$ and $y_{h'} < y_{f'} < y$, then 
$\mb{P}[\simple_{f}(h) \cap \simple_{f'}(h') \mid Y_{f} = y_f, Y_{f'} = y_{f'} , Y_{h} = y_h, Y_{h'} = y_{h'}]$
is equal to
\begin{align} \label{eqn:triple_product}
\prod_{g^* \in \{(w,w'),(r,r')\}} (1 - \max\{y_{h'},y_{h}\} \sv(x_{g^*})) \prod_{g \in \partial(h) \setminus \{f, (w,w'), (r,r')\}} (1 - y_{h} \sv(x_g)) \prod_{g' \in \partial(h') \setminus \{f', (w,w'), (r,r')\}} (1 - y_{h'} \sv(x_{g'}))
\end{align}
The product over $g^* \in \{(w,w'),(r,r')\}$ follows since we require $(w,w')$ and $(r,r')$ to be irrelevant for both $h$ \textit{and} $h'$:  this occurs precisely when $S_{g^*} = 0$ or $Y_{g^*} > \max\{Y_h,Y_{h'}\}$. The other two products use the fact that the sets $\partial(h) \setminus \{f, (w,w'), (r,r')\}$ and $\partial(h') \setminus \{f, (w,w'), (r,r')\}$ are disjoint, and so the relevant random variables for $h$ and $h'$ are independent. Now, for any $y_h, y_h' \in [0,y)$ and $g^* \in  \{(w,w'),(r,r')\}$,
$\mb{P}[\text{$S_{g^*} = 0$ or $Y_{g^*} > \max\{y_h, y_{h'}\}$} \mid Y_h = y_h, Y_{h'} = y_{h'}]$ is lower bounded by
\begin{equation} \label{eqn:overlapping_edges}
 (1 - y_{h} \sv(x_{g^*})) ( 1 - y_{h'} \sv(x_{g^*})),
\end{equation}
as $\mb{P}[Y_{g^*} > \max\{y_h , y_{h'}\} \mid Y_h = y_h, Y_{h'} = y_{h'}] = 1- \max\{y_{h},y_{h}\} \ge (1 - y_h)(1 - y_{h'})$. Observe that \eqref{eqn:overlapping_edges} is equal to
 $\mb{P}[\text{$Y_{g^*} > y_h$ or $S_{g^*} = 0$} \mid Y_h = y_h] \cdot \mb{P}[\text{$Y_{g^*} > y_{h'}$ or $S_{g^*} = 0$} \mid Y_{h'} = y_{h'}]$, and so 
 \begin{align*}
 \eqref{eqn:triple_product} &\ge \prod_{g \in \partial(h) \setminus \{f\}} (1 - y_{h} \sv(x_g)) \prod_{g' \in \partial(h') \setminus \{f'\}} (1 - y_{h'} \sv(x_{g'})) \\
 &= \mb{P}[\simple_{f}(h) \mid Y_{f} = y_f, Y_{h} = y_h] \cdot \mb{P}[\simple_{f'}(h') \mid  Y_{f'} = y_{f'}, Y_{h'} = y_{h'}].
 \end{align*}
 By integrating over $y_f, y_{f'} \in [0,y)$ and $y_{h} \in [0,y_f)$, $y_{h'} \in [0,y_{f'})$,
 \eqref{eqn:blocker_positive_correlation} follows.

Now, 
\Cref{lem:first_order_block} implies
 that $\mb{P}[\simple_{f}(h) \mid Y_f< y] \cdot \mb{P}[\simple_{f}(h') \mid Y_{f'} < y]$ is lower bounded by
\begin{equation} \label{eqn:sum_correlated}
 \frac{\sv(x_h)}{2(1 - x_h) - x_f}\left(1 - \frac{1 - e^{-(2(1 - x_h) - x_f)y}}{(2(1 - x_h) - x_f)y}\right)  \frac{\sv(x_{h'})}{2(1 - x_{h'}) - x_{f'}}\left(1 - \frac{1 - e^{-(2(1 - x_{h'}) - x_{f'})y}}{(2(1 - x_{h'}) - x_{f'})y}\right)
\end{equation}
Moreover, when summing \eqref{eqn:sum_correlated} over $h \in \partial(w) \setminus \{f, (w,w')\}$ and  $h' \in \partial(w) \setminus \{f', (w,w')\}$, \Cref{lem:second_order_block} implies that the minimum occurs when $|\partial(w) \setminus \{f, (w,w')\}| = |\partial(w') \setminus \{f', (w,w')\}| =1$,
 and if $h$ and $h'$ are the respective edges of these sets, then $x_h = 1 - x_f - x_{w,w'}$ and $x_{h'} = 1 - x_{f'} - x_{w,w'}$. In this case, the sum simplifies to 
 $$
 \frac{\sv(1-x_f -x_{w,w'} ) }{x_f + 2x_{w,w'} } \left(1 - \frac{1- e^{-(x_f + 2x_{w,w'})y}}{(x_f + 2x_{w,w'} ) y}\right)  \frac{\sv(1-x_{f'} -x_{w,w'} ) }{x_{f'} + 2x_{w,w'} } \left(1 - \frac{1- e^{-(x_{f'} + 2x_{w,w'})y}}{(x_{f'} + 2x_{w,w'} ) y}\right).
$$
and so $\sum_{\substack{h \in \partial(w) \setminus \{f, (w,w')\}, \\ h' \in \partial(w') \setminus \{f', (w,w')\}}} \mb{P}[\simple_{f}(h) \cap \simple_{f'}(h') \mid \max\{Y_f, Y_{f'}\} < y]$
is no greater than
\begin{equation} \label{eqn:almost_function_T}
 \frac{\sv(1-x_f -x_{w,w'} ) }{x_f + 2x_{w,w'} } \left(1 - \frac{1- e^{-(x_f + 2x_{w,w'})y}}{(x_f + 2x_{w,w'} ) y}\right)  \frac{\sv(1-x_{f'} -x_{w,w'} ) }{x_{f'} + 2x_{w,w'} } \left(1 - \frac{1- e^{-(x_{f'} + 2x_{w,w'})y}}{(x_{f'} + 2x_{w,w'} ) y}\right).
\end{equation}

Putting everything together, $\mb{P}[\simple_{f}\cap \simple_{f'} \mid \max\{Y_f, Y_{f'}\} < y]$ is lower bounded by
\begin{equation} \label{eqn:final_lower_bound}
      Q(x_f,x_{f'},x_{w,w'},y) +   \frac{\sv(1-x_f -x_{w,w'} ) }{x_f + 2x_{w,w'} } \left(1 - \frac{1- e^{-(x_f + 2x_{w,w'})y}}{(x_f + 2x_{w,w'} ) y}\right)  \frac{\sv(1-x_{f'} -x_{w,w'} ) }{x_{f'} + 2x_{w,w'} } \left(1 - \frac{1- e^{-(x_{f'} + 2x_{w,w'})y}}{(x_{f'} + 2x_{w,w'} ) y}\right).
\end{equation}
For any choice of $x_f, x_{f'}, y \in [0,1]$, the r.h.s. of \eqref{eqn:final_lower_bound}
is minimized when $x_{w,w'} =0$. Since $Q(x_f,x_f',0,y)=0$, and \eqref{eqn:almost_function_T} is equal 
to $T(x_f,y) T(x_{f'},y)$ when $x_{w,w'}=0$, this completes the proof.




\CM{\subsection{Proof of \Cref{lem:decreasing_probability}}} \label{pf:lem:decreasing_probability}
Our goal is to prove that $y \rightarrow \obj_{G\setminus e}(v,y)$
is non-negative and non-increasing for $y \in [0,1]$, where
\begin{equation} \label{eqn:obj_repeat}
\obj_{G\setminus e}(v,y) = \prod_{g \in \partial(v) \setminus e} \ell(x_{g},y) + \sum_{f \in \partial(v) \setminus e} T(x_f,y) \cdot \sv(x_f)y \prod_{g \in \partial(v) \setminus \{f,e\}} \ell(x_{g},y).
\end{equation}
By individually considering the various terms of \eqref{eqn:obj_repeat}, it is easy to see that $\obj_{G \setminus e}(v,y)$ is non-negative, so we focus on proving that $\obj_{G \setminus e}(v,y)$ is non-increasing.

Fix $k \ge 1$. We first construct an input $G^*_k =(V^*_k, E^*_k)$ with fractional matching
$\bm{x}^* = (x^*_f)_{f \in E^*_k}$. The first part of the construction copies 
the neighborhood structure of $e=(u,v)$ in $G$. Specifically, we add
the vertices of $N_{G}(u) \cup N_{G}(v)$ to $V^*_k$, as well as the edges of
$\partial_{G}(u) \cup \partial_{G}(v)$ to $E^*_k$ (this includes the
edge $e=(u,v)$). The fractional edge values also remain the same: i.e.,  $x^*_{f} = x_{f}$ for each $f \in \partial_{G}(u) \cup \partial_{G}(v)$.
The next part of the construction modifies the second and third neighbors
of $v$.
For each $w \in N_{G}(v) \setminus \{u\}$, we add a new vertex $w'$, and add the edge $(w,w')$
to $E^*_k$ while setting $x^*_{w,w'}:= 1 - x_{v,w}$. Denote the set of these vertices by
$W'$. Finally, for each $w' \in W'$, we add $k$ distinct vertices, say $w''_1, \ldots , w''_k$,
and set $x^*_{w',w''_i} = x_{v,w}/k$ for each $i=1, \ldots ,k$.


Let us now consider executing \Cref{alg:attenuate_rom} on $G^*_k$ using
the random arrival times $(Y_f)_{f \in E^*_k}$, as well as $(X_f)_{f \in E^*_k}$ and $(S_f)_{f \in E^*_k}$. (For notational convenience, we can assume that the same random variables are used for the edges which are in both $G$ and $G^*_k$).
For each $y \in [0,1]$, let $\scr{M}_{k}(y)$ be the matching returned by executing \Cref{alg:attenuate_rom} on $G^*_k$ when restricted to the edges which arrive before time $y$.
Recall that $\scr{R}_{e}$ denotes the relevant edges for $e$, where $f \in \partial_{G}(e)$ is relevant provided $Y_f < Y_e$ and $S_f =1$. Setting $\Rf_v = \scr{R}_e \cap \partial(v) \setminus \{u\}$, we claim that
\begin{equation} \label{eqn:coupling}
    y \rightarrow \mb{P}[v \in V(\scr{M}_{k}(y)), |\Rf_v| \le 1 \mid Y_e = y]
\end{equation}
is a non-decreasing function of $y$. (Here $V(\scr{M}_k(y))$ is the vertices matched by \Cref{alg:attenuate_rom} before time $y$). This follows by taking $y_1 < y_2$, and coupling the executions of \Cref{alg:attenuate_rom}, conditional on $Y_e = y_1$ and $Y_e = y_2$, respectively. Once this is done, it is easy to show that if $v \in V(\scr{M}_{k}(y_1))$ and $|\Rf_v| \le 1$ both occur, then $v \in V(\scr{M}_{k}(y_2))$ and $|\Rf_v| \le 1$ both occur. This implication implies the function
specified in \eqref{eqn:coupling} is non-decreasing.

To complete the proof, it suffices to show that for each $y \in [0,1]$, 
$$
     \mb{P}[v \notin V(\scr{M}_{k}(y)), |\Rf_v| \le 1 \mid Y_e = y] \rightarrow \obj_{G\setminus e}(v,y),
$$
as $k \rightarrow \infty$.
To prove this convergence, first observe that $
    \mb{P}[v \notin V(\scr{M}_{k}(y)), |\Rf_v| =0 \mid Y_e = y] = \prod_{g \in \partial(v) \setminus \{e\}} \ell(x_{g},y).$
Thus, since $T(x,y) = \frac{\sv(1-x ) }{x }\left(1 - \frac{1- e^{-x y}}{x y}\right)$ in \eqref{eqn:obj_repeat}, and
for $w \in N_{G}(v) \setminus \{u\}$ we have that $\mb{P}[\Rf_v = \{(v,w)\} \mid Y_e = y] = \sv(x_{v,w})y \prod_{g \in \partial(v) \setminus \{(v,w),e\}} \ell(x_{g},y)$, it remains to prove that as $k \rightarrow \infty$,
$$
\mb{P}[v \notin V(\scr{M}_{k}(y))  \mid \Rf_v = \{(v,w)\}, Y_e =y] \rightarrow \frac{\sv(1-x_{v,w} ) }{x_{v,w} } \left(1 - \frac{1- e^{-x_{v,w} y}}{x_{v,w} y}\right).
$$
Due to the neighborhood structure of $w  \in N_{G}(v) \setminus \{u\}$,
if we condition on $Y_e = y$ and $\Rf_v = \{(v,w)\}$, then $v \notin V(\scr{M}_{k}(y))$ occurs if and only if $(w,w')$ is a simple-blocker for $(v,w)$ with respect to the execution on $G^*_k$ (see \Cref{def:simple_blocker} for a review of this terminology). If we denote the latter event by $\simple^{(k)}_{(v,w)}(w,w')$ then this implies that
\begin{equation} \label{eqn:simple_blocker_matched_vertex}
    \mb{P}[ v \notin V(\scr{M}_{k}(y)) \mid \Rf_v = \{(v,w)\}, Y_e =y] = \mb{P}[\simple^{(k)}_{(v,w)}(w,w') \mid \Rf_v = \{(v,w)\},  Y_e =y].
\end{equation}
Now, $w'$ has neighbors $w''_1, \ldots ,w''_k$ apart from $w$. Thus, by applying
the same computations from the proof of \Cref{lem:first_order_block} and taking $k \rightarrow \infty$,
\begin{align*}
    \mb{P}[\simple^{(k)}_{(v,w)}(w,w') \mid  \{\Rf_v = \{(v,w)\}, Y_e =y] &= \sv(x^*_{w,w'}) \int_{0}^{y} \int_{0}^{y_{v,w}} \prod_{i=1}^k (1 - \sv(x^*_{w',w''_i})y_{w,w'}) dy_{w,w'} dy_{v,w} \\
    & \rightarrow \sv(x^*_{w,w'}) \int_{0}^{y} \int_{0}^{y_{v,w}}  e^{-k x^*_{v,w} y_{w,w'} } dy_{w,w'} dy_{v,w} \\
    & = \frac{\sv(1-x_{v,w} ) }{x_{v,w} } \left(1 - \frac{1- e^{-x_{v,w} y}}{x_{v,w} y}\right),
\end{align*}
where the last equality uses $k x^*_{v,w} = x_{v,w}$ and $x^*_{v,w} = 1 -x_{v,w}$. By combining this
with \eqref{eqn:simple_blocker_matched_vertex}, the convergence is proven, and so the proof is complete.

\CM{\subsection{Proof of \Cref{lem:triangle_free_vertex_split}}} \label{pf:lem:triangle_free_vertex_split}

Recall that $w \in N_{G}(u) \setminus \{v\}$ 
is the vertex which is copied $k \ge 1$ times in the construction $G_k$, yielding $w_1, \ldots ,w_k$ instead of $w$.
For convenience, we define $\tpartial(u): = \partial_{G}(u) \setminus \{(w,u), (u,v)\}$. 
Now, since $w \notin N_{G}(v)$, we know that $\obj_{G}(v,y) = \obj_{G_k}(v,y)$. 
Thus, the goal is to prove that
\begin{equation} \label{eqn:desired_integral}
    \int_{0}^{1} (\obj_{G}(u,y) - \lim_{k \rightarrow \infty} \obj_{G_k}(u,y)) \cdot \obj_{G}(v,y) dy \ge 0,
\end{equation}
where we've used the dominated convergence theorem in order to exchange the order of the point-wise convergence and integration. 
We first compute $\lim_{k \rightarrow \infty}\obj_{G_k}(u,y)$. Now, recalling that
$T(x,y) =\frac{\sv(1-x) }{x} \left(1 - \frac{ e^{-xy}}{x y}\right),$
we can write $\obj_{G_k}(u,y)$ as:
\begin{align*}
\sum_{f \in \tpartial(u)} T(x_f,y) y \sv(x_f) \ell(x_{w,u}/k,y)^k  \prod_{g \in \tpartial(u) \setminus \{f\}}  \ell(x_{g},y) +  k  T(x_{u,w}/k,y) y \sv(x_{u,w}/k) \ell(x_{w,u}/k,y)^{k-1}  \prod_{g \in \tpartial(u)} \ell(x_{g},y) \\
+ \ell(x_{w,u}/k,y)^k \prod_{g \in \tpartial(u)} \ell(x_{g},y).
\end{align*}
(Here the middle term groups the contributions of the edges $(u,w_i)$ for each $1 \le i \le k$).
By using the continuity of $a$, and the fact that $a(0)=1$,
\[
    \lim_{k \rightarrow \infty} \ell(x_{w,u}/k,y)^k = 
    \lim_{k \rightarrow \infty} \ell(x_{w,u}/k,y)^{k-1} = e^{-x_{w,u}y}, \, \lim_{k \rightarrow \infty} k (y \sv(x_{u,w}/k) \ell) \ell(x_{u,w}/k,y))= y x_{u,w}.
\]
Moreover, $\lim_{x \rightarrow 0^+} T(x,y) = a(1) y/2$. Thus,
$
 \lim_{k \rightarrow \infty}  k  T(x_{u,w}/k,y) y \sv(x_{u,w}/k)  = \frac{a(1)x_{u,w} y^2}{2}.
$
By combining all these expressions, $\lim_{k \rightarrow \infty} \obj_{G_k}(u,y)$
is equal to
\[
 \sum_{f \in \tpartial(u)} T(x_f ,y) \cdot \sv(x_f)y e^{-x_{w,u}y} \prod_{g \in \tpartial(u) \setminus \{f\}} \ell(x_{g},y) + \left(\frac{a(1)x_{u,w}y^2}{2} + 1 \right) e^{-x_{w,u}y} \prod_{g \in \tpartial(u)} \ell(x_{g},y).
\]
Let us now compare $\lim_{k \rightarrow \infty} \obj_{G_k}(u,y)$ to $\obj_{G}(u,y)$, the latter of which we rewrite in the following way:
\[
  \sum_{f \in \tpartial(u)} T(x_f,y) \cdot \sv(x_f)y \prod_{g \in \partial(u) \setminus \{f\}} \ell(x_{g},y) +  \prod_{g \in \partial(u)} \ell(x_{g},y) +  T(x_{u,w},y) s(x_{u,w}) y \prod_{g \in \tpartial(u)} \ell(x_{g},y).
\]
Let $D_1(y)$ be the difference of each function's first term, after multiplying each by $\obj_{G}(v,y)$. I.e.,
\[
D_1(y) := \sum_{f \in \tpartial(u)}
\left( \ell(x_{u,w},y) -  e^{-x_{u,w}y} \right) T(x_f ,y) y \sv(x_f) \cdot \obj_{G}(v,e)  \prod_{g \in \tpartial(u) \setminus \{f\}} \ell(x_{g},y). 
\]
Similarly, let $D_{2}(y)$ be the difference of each function's remaining terms, after multiplying each by
$\obj_{G}(v,y)$. I.e.,
\[
 D_{2}(y):=\left(\ell(x_{u,w},y)  + T(x_{u,w},y) \sv(x_{u,w})y -  \left( 1 +\frac{a(1) x_{u,w}y^2}{2} \right) e^{-x_{u,w}y} \right) \cdot \obj_{G}(v,e)  \prod_{g \in \tpartial(u)} \ell(x_{g},y).
\]
Our goal is to show that $\int_{0}^{1} D_i(y) dy \ge 0$ for each $i \in [2]$,
as this will establish \eqref{eqn:desired_integral} and complete the proof.
We start with $D_1$. Observe that $\ell(x_{u,w},y) -  e^{-x_{u,w}y} \ge 0$
for all $y \in [0,1]$ by the first property of Proposition \ref{property:vertex_split}. Moreover, the remaining terms in each summand of $D_1$ are non-negative, so $\int_0^1 D_1 \ge 0$. Consider now $D_2$. Observe that the function $y \rightarrow \prod_{g \in \tpartial(u)} \ell(x_{g},y)$ is non-negative and non-increasing for $y \in [0,1]$, as $\ell(x_g,y):= 1 - y \sv(x_g)$, and $\sv(x_g) \in [0,1]$ for $x_g \in [0,1]$. By \Cref{lem:decreasing_probability}, both properties are also true of the function $y \rightarrow \obj_{G}(v,y)$. Thus, $y \rightarrow \obj_{G}(v,y) \cdot \prod_{g \in \tpartial(e)} \ell(x_{g},y)$ satisfies both these properties.
Now, by the second property of \Cref{property:vertex_split}, the function 
$$y \rightarrow \ell(x_{u,w},y) + T(x_{u,w},y) \sv(x_{u,w})y -  \left( 1 +\frac{a(1)x_{u,w}y^2}{2} \right) e^{-x_{u,w}y},$$
is initially non-negative, changes sign at most once, and has a non-negative integral.
Thus, we can apply \Cref{lem:integral} with $\lambda(y):= \obj_{G}(v,y) \cdot \prod_{g \in \tpartial(e)} \ell(x_{g},y)$ as the first function, and $\phi(y) = \ell(x_{u,w},y)  + T(x_{u,w},y) \sv(x_{u,w})y -  \left( 1 +\frac{a(1) x_{u,w}y^2}{2} \right) e^{-x_{u,w}y}$ as the second,
to conclude that $\int_0^1 D_2 \ge 0$.

\subsection{Proof of \Cref{prop:de_dominance} using Lemmas \ref{lem:well_controlled_probability} and \ref{lem:matching_expected_difference}} \label{pf:prop:de_dominance}
We apply Corollary $4$ from \citet{bennett2023}, where
we have intentionally chosen the below notation
to match \citet{bennett2023}.

	Define $\scr{D}= [0,1]^2$, and set $f(z, r) := (1 - r)^2$ for $(z,r) \in \scr{D}$. Recall that $w(z)=z/(1+z)$ is the unique solution to the differential equation $w'(z) = f(z,w(z))$ with initial condition $w(0)=0$. to  $(W(i)/n^2)_{i=0}^{\eps n^2}$, the filtration $(\scr{H}_i)_{i=0}^{\eps n^2}$,
	and the above system. (Note that our \textit{scaling factor} is $n^2$). First observe that we can take $\sigma = \eps, L =2$, and $B =1$ (since $f$ has Lipschitz constant of $2$, and $|f| \le 1$). 	Recall now the sequence of events $(Q_i)_{i=0}^{\eps n^2}$ defined before \Cref{lem:well_controlled_probability},
	and let $I$ be the first $i \ge 0$ such that $Q_i$ does \textit{not} occur. 
	For the remaining parameters, set $\lambda = n^{5/3}$, $\delta = n^{-1/3}$ and $b =\beta = n$,
 Clearly, $\lambda \ge \max\{ B + \beta, \frac{L + BL + \delta n^2}{3 L} \}$ for $n$ sufficiently large.

 We now verify that these parameters satisfy the `Boundedness Hypothesis', `Trend Hypothesis' conditions of [Corollary $4$, \citet{bennett2023}] for all $0 \le i < \min\{\eps n^2, I\}$. Note that due to statement of [Corollary $4$, \citet{bennett2023}], when verifying these conditions at step $i$,
	we can assume that
	\begin{equation} \label{eqn:matching_assumed_condition}
	W(i)/n^2 \le (1 + O(n^{-1/3})) w(i/n^2).
	\end{equation}
	
	The `Boundedness Hypothesis' is satisfied since $|\Delta W(i)| \le n$, and $\Var( \Delta W(i) \mid \scr{H}_i ) \le n$.
	The `Trend Hypothesis' is satisfied due to \Cref{lem:matching_expected_difference} in conjunction
	with the above definition of $I$ and \eqref{eqn:matching_assumed_condition}. Thus, $(W(i)/n^2)_{i=0}^{\eps n^2}$ satisfies all the conditions of [Corollary $4$, \citet{bennett2023}] so as to guarantee that
		$W(i)/n^2 \le (1 + O(n^{-1/3})) w(i/n^2)$
	for all $0 \le i \le \{I,\eps n^2\}$ with probability at least $1 - o(n^{-2})$. Finally, due to \Cref{lem:well_controlled_probability}, $\mb{P}[I \le \eps n^2] = o(n^{-2})$,
	and so a union bound implies that the proposition holds.

\section{Bipartite Reduction to $1$-Regular Inputs} \label{sec:added_reduction}
In this section, we prove a reduction to $1$-regular inputs for RCRS/OCRS which preserves bipartiteness.
\begin{lemma}[Reduction to $1$-Regular Inputs for Bipartite Graphs] \label{lem:one_regular_long_bipartite}
Given a bipartite graph $G=(U,V,E)$ with fractional matching $\bm{x}=(x_e)_{e \in E}$,
there exists bipartite graph $G'=(U',V', E')$ and a fractional matching $\bm{x}' = (x'_e)_{e \in E'}$ with the following properties:
\begin{enumerate}
    \item 
    $(G', \bm{x}')$ can be computed in time polynomial in the size of $(G,\bm{x})$.
    \item $(G', \bm{x}')$ is $1$-regular. \label{property:one_regular_bip}
    \item If there is an $\alpha$-selectable RCRS (respectively, OCRS) for $G'$ and $\bm{x}'$, then there exists an $\alpha$-selectable
    RCRS (respectively, OCRS) for $G$ and $\bm{x}$. \label{property:reduction_bip}
\end{enumerate}
\end{lemma}

\begin{myproof}
The graph $G'$ is constructed as follows. First, assume without loss of generality that $|U| = |V|$
(we can do this by creating dummy vertices on the smaller side with edge
values of $0$). Let $n := |U|$. We create a biclique
$K_{n,n} = (U_{K} \cup V_{K}, E_{K})$ of \emph{dummy vertices}. Let
$E_{G,K} = (U \times V_{K}) \cup (U_{K} \cup V)$ be a set of edges connecting
every vertex in $G$ to each of the vertices of the dummy $K_{n,n}$. Let
$U' = U\cup U_{K}$, $V' = V\cup V_{K}$, and $E' = E \cup E_{K} \cup E_{G,K}$.

Then, $\mathbf{x'}$ is given by setting $x'_{e} = x_{e}$ for every $e\in E$ and
$x'_{e} = (1-x_{u})/n$ for $e = (u,v) \in E_{G,K}$, where
$x_{u} := \sum_{v \in V} x_{u,v}$. Clearly, for $u\in U$, we have that
$\sum_{v\in V'} x'_{uv} = 1$ and similarly for $v\in V$,
we have that $\sum_{u\in U'}x'_{uv}=1$.

Finally, for $e=(u,v) \in E_{K}$, set
$x'_{uv} := \frac{1}{n^{2}}\sum_{v\in V} x_{v}$. Note that by the handshaking
lemma, we have that $\sum_{v\in V} x_{v} = \sum_{u\in U}x_{u}$ so
$x'_{uv} = \frac{1}{n^{2}}\sum_{u\in U} x_{u}$. Therefore, for $u\in U_{K}$:
\[
x'_{u} = \sum_{v\in V'}x'_{uv} = \sum_{v\in V}x'_{uv} + \sum_{v\in V_{K}} x'_{uv} =
\frac{1}{n}\sum_{v\in V}(1 - x_{v}) + \sum_{v_{K}\in V_{K}} \frac{1}{n^{2}}\sum_{v\in V} x_{v}
= \frac{1}{n}\sum_{v\in V}(1 - x_{v}) + \frac{1}{n}\sum_{v\in V}x_{v} = 1
\]
and similarly, for $v\in V_{K}$:
\[
x'_{v} = \sum_{u\in U}x'_{uv} + \sum_{u\in U_{K}} x'_{uv} =
\frac{1}{n}\sum_{u\in U}(1 - x_{u}) + \sum_{u_{K}\in U_{K}} \frac{1}{n^{2}}\sum_{u\in U} x_{u}
= \frac{1}{n}\sum_{u\in U}(1 - x_{v}) + \frac{1}{n}\sum_{u\in U}x_{u} = 1.
\]
Since $|U'| = 2|U|$, $|V'| = 2|V|$, and $|E' \setminus E| = |U|^{2}|V|^{2}$, we can construct $G'$ and $\bm{x}'$ efficiently. By the same argument as in the proof of Lemma~\ref{lem:one_regular_long}, an $\alpha$-selectable RCRS (respectively, OCRS) for
$(G', \mathbf{x'})$ can be used to get an $\alpha$-selectable RCRS (respectively, OCRS) for $(G, \bm{x})$.
\end{myproof}

\end{APPENDICES}

\end{document}